\documentclass[report,rmp,nofootinbib,reprint,floatfix]{cust-revtex4-1}
\usepackage{graphics}
\usepackage{epsfig}
\usepackage{amsmath}
\usepackage{ragged2e}
\usepackage[caption=false]{subfig}
\usepackage{xspace}
\usepackage{float}
\usepackage{array, longtable}        
\usepackage{xcolor}                
\usepackage{bm}                        
\makeatletter
\let\saved@longtable\longtable
\long\def\foo#1\LT@err#2#3#4!!{\def\longtable{#1#4}}
\expandafter\foo\longtable!!
\long\def\foo#1\@outputpage#2\@outputpage#3!!{%
\def\LT@output{#1\@opcol#2\@opcol#3}}
\expandafter\foo\LT@output!!
\makeatother

\def\B{V$^2$/m$^2$Hz\xspace}                 
\def\OmRF{\Omega_{\rm rf}}                 
\def\mi{m_{\rm I}}                                   
\def\omt{\omega_{\rm t}}                           

\def\Monroe95{2}
\def\Tamm00{6}
\def\Turchette00{7}
\def\Rohde01{9}
\def\Rowe02{10}
\def\DeVoe02{11}
\def\Home06{13}
\def\Deslauriers06{16}
\def\Labaziewicz0801{19}
\def\Schulz08{20}
\def\LGL08{22}                
\def\Britton08{23}
\def\Poschinger09{24}
\def\Blakestad09{25}
\def\Amini10{28}
\def\Daniilidis11{33}
\def\AGH11{39}                
\def\Allcock12{40}
\def\Harlander12{42}
\def\Hite12{46}
\def\Poulsen12{47}
\def\Arrington13{53}
\def\DGB13{57}                
\def\McKay14{60}
\def\Goodwin14{62}

\begin{document}
\title{\hspace{13mm}{\rm {\bf Ion-trap measurements of electric-field noise near surfaces}}}

\author{{\rm M.~Brownnutt$^1$}}
\email{michael.brownnutt@uibk.ac.at}
\author{{\rm M.~Kumph$^1$}}
\author{{\rm P.~Rabl$^2$}}
\author{{\rm R.~Blatt$^{1,3}$}}
\affiliation{$^1$ Institut f\"ur Experimentalphysik Universit\"at Innsbruck Technikerstrasse 25 6020 Innsbruck Austria,
$^2$ Atominstitut TU Wien Stadionallee 2 1020 Wien Austria,
$^3$ Institut f\"ur Quantenoptik und Quanteninformation \"Osterreichische Akademie der Wissenschaften\\ Technikerstrasse 21A 6020 Innsbruck Austria}

\begin{abstract}
Electric-field noise near surfaces is a common problem in diverse areas of physics, and a limiting factor for many precision measurements. There are multiple mechanisms by which such noise is generated, many of which are poorly understood. Laser-cooled, trapped ions provide one of the most sensitive systems to probe electric-field noise at MHz frequencies and over a distance range $30-3000\,\mu$m from the surface. Over recent years numerous experiments have reported spectral densities of electric-field noise inferred from ion heating-rate measurements and several different theoretical explanations for the observed noise characteristics have been proposed. This paper provides an extensive summary and critical review of electric-field noise measurements in ion traps, and compares these experimental findings with known and conjectured mechanisms for the origin of this noise. This reveals that the presence of multiple noise sources, as well as the different scalings added by geometrical considerations, complicate the interpretation of  these results. It is thus the purpose of this review to assess which conclusions can be reasonably drawn from the existing data, and which important questions are still open. In so doing it provides a framework for future investigations of surface-noise processes.
\end{abstract}
\maketitle

\textcolor{white}{.}
\newpage
\textcolor{white}{.}
\newpage

\tableofcontents

\textcolor{white}{.}
\newpage

\section{Introduction}
\label{sec:Introduction}

Electric-field noise above surfaces provides a significant challenge to many disparate areas of physics. Drift-tube experiments attempting to measure the effects of gravity on charged particles are affected by stray fields from metallic shielding, which can be several centimetres away~\citep{Darling:1992}. The presence of static patch potentials also limits the measurement of Casimir forces at distances of around $0.1-1\,\mu$m~\citep{Speake:2003, Sushkov:2011, Garcia-Sanchez:2012}. In space-based gravitational-wave experiments, variations in the electrostatic surface potential is expected to be one of the largest contributors of noise at frequencies around 1\,mHz, over distances of 1\,mm~\citep{Pollack:2008}. Motion of nanocantilevers operating at around 1\,kHz is damped by field noise in the form of non-contact friction, acting over 10\,nm \citep{Stipe:2001}. Operating at around 1\,MHz and at distances of a few tens or hundreds of $\mu$m from metallic electrodes, laser cooled trapped ions are heated by electric-field noise.

Moving from the ubiquity of electric-field noise in general to the specific mechanism giving rise to it in each instance, however, is complicated for two reasons. Firstly, seemingly disparate effects in different systems may have a common root. For example, considering the scaling with the distance, $d$, to the electrodes, finite spatial correlation of noise can mean that the same underlying physical mechanism may present itself with a scaling between $d^{0}$ and $d^{-6}$ at different length scales \citep{Dubessy:2009,Low:2011}. Secondly, within a single system multiple effects may be significant and the measurement of electric noise within a small distance or frequency window is often not sufficient to distinguish different sources of noise. Given such a complex and interrelated picture, knowing what the cause of noise is, or is not, in one particular system can inform how that, system and the relevant noise sources, relate to other areas.

Cold, trapped atomic ions can act as immensely sensitive probes of electric-field noise at frequencies around 1\,MHz, and for distances of a few tens to a few hundred~$\mu$m from a surface. Ions, which are initially laser-cooled to close to the quantum ground state, are heated by fluctuating electric fields. The spectral density of electric-field noise at the ions' motional frequencies can be inferred by measuring the heating rate. Such heating represents a severe limitation for many trapped-ion applications, where the ion should stay near the quantum ground state. For example, trapped-ion quantum computers require ions to be near the motional ground state; even `thermal' gate operations require motional quantum numbers of $\bar{n} \lesssim 20$ \citep{Kirchmair:2009}. In practical terms, with typical gate speeds, this necessitates ion-heating rates of $\dot{\bar{n}} \lesssim 100$~s$^{-1}$. As another example, the effects of ion heating in optical clocks can contribute fractional uncertainties at the 10$^{-17}$ level due to second-order Doppler shifts \citep{Rosenband:2008}. While ions in such experiments can be sympathetically cooled, the effects of external heating significantly complicate the process \citep{Wuebbena:2012}.  In addition to investigating the effect for its own sake, there is therefore significant interest within the trapped-ion community in understanding the sources of electric-field noise so that they can be mitigated.

The first systematic studies of trapped-ion heating found that the observed heating rates (and by implication the level of electric-field noise) was very much greater than would be expected from a simple consideration of  black-body radiation or Johnson noise \citep{Turchette:2000}.  Dubbed ``anomalous heating" \citep{Monroe:1995_11} there has subsequently been an increasing interest in identifying and eliminating the noise source underlying this effect.  Questions have also been raised as to whether and how this heating could be related to the effects seen in other systems, such as non-contact friction in nanocantilever systems \citep{Stipe:2001,Volokitin:2007}, the ubiquitous $1/f$ noise encountered in many solid-state devices \citep{Dutta:1981,Paladino:2014} or diffusion-induced noise observed in emission currents \citep{Timm:1966}. Moreover, there have been suggestions that understanding the heating observed in ions could shed light on crystalline structure \citep{Dubessy:2009,Low:2011}, fluctuating patch potentials \citep{Daniilidis:2011} or surface adsorbates \citep{SafaviNaini:2011}.

To this end, much has been done to investigate the heating observed in ion traps. In most studies it is assumed that the spectral density of electric-field noise, $S_\mathrm{E}$, experienced by the trapped ion varies -- at least within a certain parameter range -- as a power-law with respect to the frequency, $\omega$, the distance of the ion from the surface, $d$, and the trap temperature, $T$, such that
\begin{equation}
\label{eqn:S_E_DependenciesIntro}
S_\mathrm{E}\propto \omega^{-\alpha} d^{-\beta} T^{+\gamma}.
\end{equation}
Such power-law behaviour is also predicted -- within certain parameter ranges -- by many theoretical models. Models for different noise mechanisms give particular predictions for the frequency-scaling exponent, $\alpha$; the distance-scaling exponent, $\beta$; and the temperature-scaling exponent, $\gamma$. Experimental measurements of these exponents can therefore provide information about likely noise sources.\\

It should be stressed at the outset that this review article does not set out to claim that the issues surrounding heating of trapped ions have all been resolved. Despite an increasing amount of experimental data, ongoing theoretical work, and strengthened engagement with insights from disciplines beyond ion trapping, the dominant sources of noise and the physical mechanisms responsible for them are still not fully understood. A review of the topic should not be taken to indicate that the subject is closed; quite the opposite. Investigation of electric-field noise with trapped ions is a vibrant and active research area, with numerous new avenues of investigation opening up. The aim of this article is therefore to take stock of the extensive literature which exists already, setting out what is known, what is conjectured, and what is unknown. Rather than being the conclusion of the matter, this review should therefore be seen as the starting point for future investigations.

The remainder of this article is structured as follows.
Section~\ref{sec:FieldNoiseAndHeating} briefly summarizes the basics of ion trapping. It describes how trapped ions are affected by electric-field noise, and how they can thus be used to measure such noise.
Section~\ref{sec:ExperimentalOverview} gives an overview of the experiments carried out to characterize electric-field noise in ion traps, particularly noting what can be deduced regarding the values of $\alpha$, $\beta$ and $\gamma$. In this section the experimental results are laid out without reference to possible heating mechanisms which may underlie the observations.
Sections~\ref{sec:Theory} and \ref{sec:Microscopic} review a variety of fundamental, technical, and surface-related sources of noise which can contribute to motional heating in ion traps. For each case the expected scaling laws, as well as the magnitude of electric-field noise expected for typical ion-trapping conditions are highlighted.
By considering what has been observed experimentally in the light of theoretical considerations, Section~\ref{sec:KnownMechanisms} then outlines a number of specific experiments in which particular heating mechanism can be clearly identified. This demonstrates that there are multiple effects to be considered which can contribute noise at similar levels.
Section~\ref{sec:OpenQuestions} then highlights some experiments in which the noise is well characterized but for which the noise source cannot yet be unambiguously identified.
Finally, Section~\ref{sec:Outlook} considers the implications of noise for ion traps and proposes ways in which the field might move forward.

\section{Field Noise and Ion Heating}
\label{sec:FieldNoiseAndHeating}

The development of ion traps more than fifty years ago opened up unique possibilities for experimenting with individual atomic and molecular ions under stable and isolated conditions \citep{Paul:1958, Fischer:1959, Paul:1990, Dehmelt:1990}. Today it is a routine procedure to  trap single or multiple ions and laser cool one or more of their motional modes to the quantum ground state. Quantum manipulations and detection schemes for internal (electronic) and external (motional) degrees of freedom can be implemented with high precision \citep{Blatt:2008, Blatt:2012}. In the context of quantum information processing the goal of building scalable ion-trap architectures has recently led to a rapid progress in the design of microtraps, where planar electrodes are fabricated on a chip to allow for smaller and more complex structures \citep{Chiaverini:2005_06, Wesenberg:2008, Amini:2010}. As traps get smaller the ions are located closer to the surfaces and become more susceptible to small voltage fluctuations on the electrodes. This leads to an increase in the rate at which ions are heated in small traps. While this is a hindrance for many experiments, it also makes trapped ions an exquisite probe to study such noise processes.

This section provides a brief summary of the general principles of ion traps which are relevant for describing noise-induced heating processes in these systems. The discussion focusses on the operation of Paul traps,  rather than Penning traps. While temperature measurements have been carried out in Penning traps \citep{Djekic:2004}, as well as heating rate measurements for large crystals \citep{Jensen:2004} and -- very recently -- for single ions \citep{Goodwin:2014}, it remains the case that the majority of noise investigations have been carried out in Paul traps. Much of the discussion in this review remains relevant to Penning traps, though it is to be expected that certain aspects of heating in Penning traps will distinct from that in Paul traps. These differences may be fundamental (arising, for example, from the different trapping methods) or more technical in nature (for example, because the ion-electrode separation in Penning traps is often larger than is typical for Paul traps). For more details of both Paul and Penning traps and a review of the state of the art of ion-trap design the reader is referred to the work of \citet{Werth:2009}.

\subsection{Introduction to ion traps}
\label{subsec:IntroductionToIonTraps}

In most experiments used for the measurement of electric-field noise induced heating rates, individual atomic ions are confined using a so-called Paul trap \citep{Paul:1958, Paul:1990}. The basic principle of a Paul trap is to use a combination of static and radio-frequency (RF) quadrupole potentials which, on time average, lead to a confinement of charged particles in all three spatial directions.

Ideally, the field null of the static and RF potentials should coincide. In this case, and restricting initial consideration to the motion along the $x$ axis and near the center of the trap, where the applied potentials are to a good approximation harmonic, the dynamics of a singly charged ion of mass $\mi$ are described by the equation of motion
    \begin{equation}
    \label{eq:MathieuClassical}
    \ddot x(t) +\frac{|e|}{\mi}\left[\Phi_{\rm dc}''+\Phi_{\rm rf}''\cos(\OmRF t) \right] x(t)=0.
    \end{equation}
Here $\Phi_{\rm dc}''$ and $\Phi_{\rm rf}''$ denote the second-order derivatives of the static electric potential and the radio-frequency potential oscillating at  $\OmRF$, respectively. By introducing the dimensionless parameters $a_x= 4|e|\Phi_{\rm dc}''/(\mi\OmRF^2)$ and $q_x= 2 |e|\Phi_{\rm rf}''/(\mi\OmRF^2)$ and a rescaled time, $\tilde{t}= t\OmRF/2$, Eq.~\eqref{eq:MathieuClassical} maps onto the well-known Mathieu equation \citep{Leibfried:2003_01}. In the limit of interest, $a_x,q_x\ll1$, this exhibits bound solutions of the form
    \begin{equation}
    \label{eq:MathieuSolution}
    x(t)= X_0 \cos(\omt t+\varphi_0) \left[ 1+ \frac{q_x}{2}\cos(\OmRF t) +\dots \right],
    \end{equation}
where  $\varphi_0$ is a phase set by the initial conditions. The ion undergoes large-amplitude oscillations at frequency $\omt$ which, for $a_x,q_x\ll1$, is approximately given by $\omt\approx (\OmRF/2) \sqrt{a_x+q_x^2/2}$. This is variously (and equivalently) termed the motional, secular or trapping frequency. Superimposed on this is a smaller motion (called `micromotion') which has an amplitude $q_x/2$ of the secular motion, and oscillates at the trap-drive frequency, $\OmRF$.  To consider some typical numbers for trapping parameters, for an applied RF voltage, $V(t)=V_{\rm rf} \cos(\OmRF t)$, in a trap with an ion-electrode spacing of $d=500\,\mu$m, values of $V_{\rm rf}\sim500$\,V and $\OmRF/2\pi\sim 20$\,MHz result in trapping frequencies around $\omt/2\pi\approx 2$\,MHz.

Corrections due to micromotion become important for large-amplitude secular oscillations, for external perturbations at frequencies close to the trap-drive frequency, $\OmRF$~\citep{Blakestad:2009, Blakestad:2011}, for static offset fields that displace the ion from the RF null at the trap center \citep{Bluemel:1988, Bluemel:1989}, and for phase differences between the RF driving fields on different electrodes \citep{Berkeland:1998, Herskind:2009}. For some experiments such regimes are unavoidable. For example, in experiments with large three-dimensional ion crystals \citep{Herskind:2009} or clouds of ions \citep{Hornekaer:2002} some ions are necessarily situated away from the RF null. Nonetheless, by careful minimization of micromotion \citep{Berkeland:1998} many experiments can be operated in a regime where micromotion can be neglected. Under such circumstances, a single trapped ion can be treated, to a good approximation, as a simple harmonic oscillator. Considering only the effective harmonic potential is termed the pseudopotential approximation.

In analogy to the classical case, a full quantum description of a trapped ion can be obtained under the assumption that near the trap center the potential is purely quadratic such that the problem can be separated into the motion along each of the three principal axes of the trap.  The dynamics of the ion along a single direction are then described by the Hamiltonian
    \begin{equation}
    \label{eq:TrapHamiltonian}
    \hat H_{\rm t}(t)=\frac{\hat p^2}{2\mi} +  \frac{|e|}{2} \left[ \Phi_{\rm dc}'' +  \Phi_{\rm rf}'' \cos(\OmRF t) \right] \hat x^2,
    \end{equation}
where $\hat x$ and $\hat p$ are the quantized position and momentum operators. From Eq.~\eqref{eq:TrapHamiltonian} it follows that the equation of motion for the position operator, $\hat x(t)$, in the Heisenberg picture is equivalent to its classical counterpart given in Eq.~\eqref{eq:MathieuClassical}. The resulting expression for $\hat x(t)$ and $\hat p(t)$ can be written as
    \begin{eqnarray}
    \label{eq:PositionOperator}
    \hat x(t)&=& \sqrt{\frac{\hbar}{2 \mi \omt}} \left[\hat a^\dag u(t) + \hat a u^*(t)\right],\label{eq:Heisenberg_x}\\
    \label{eq:MomentumOperator}
    \hat p(t)&=& \sqrt{\frac{\hbar \mi}{2\omt}} \left[\hat a^\dag \dot u(t) + \hat a \dot u^*(t)\right]\label{eq:Heisenberg_p},
    \end{eqnarray}
where the dimensionless function $u(t)$ is a solution of Eq.~\eqref{eq:MathieuClassical} satisfying $u(0)=1$ and $\dot u(0)=i\omt$. The operator $\hat a= i\sqrt{\mi/(2\hbar\omt)}[ u(t) \dot {\hat x}(t) - \dot u(t) \hat x(t)]$ is constant in time \citep{Leibfried:2003_01} and for $t=0$ it can be identified with the usual harmonic-oscillator annihilation operator $\hat a=[ \mi \omt \hat x(0)+i \hat p(0)]/\sqrt{\hbar \mi\omt}$. The annihilation operator, $\hat a$, and its adjoint creation operator, $\hat a^\dag$, satisfy $[\hat a, \hat a^\dag]=1$ and, as in the case of a standard harmonic oscillator, a complete set of phonon number states can be constructed by $|n\rangle=\frac{(\hat a^\dag)^n}{\sqrt{n}}|0\rangle$, where $|0\rangle$ is the vibrational ground state of the ion defined by $\hat a|0\rangle=0$.  The general expression for $u(t)$ is given by an infinite series,
    \begin{equation}\label{eq:u_expansion}
    u(t)= e^{i b \OmRF t/2} \sum_{j=-\infty}^\infty C_{2j} e^{i j\OmRF  t},
    \end{equation}
but under stable conditions and $q_x\ll 1$ only a few terms are relevant. In this limit, the Floquet exponent, $b$, is approximately $\sqrt{a_x+q_x^2/2}$ and $C_{\pm 2}\simeq  q_x/4$. Eqs.~\eqref{eq:Heisenberg_x} and \eqref{eq:Heisenberg_p} therefore resemble the dynamics of a quantum harmonic oscillator of frequency $\omt$, with the main corrections arising from additional sidebands at frequencies $\OmRF\pm\omt$.
It should be emphasized that the number states, $|n\rangle$, defined at a specific time $t=0$ are in general not eigenstates of $\hat H_{\rm t}(t)$ and their corresponding wave functions and energies are periodically modulated in time. This does not by itself, however, constitute a source of heating since -- assuming a stable RF source, and in the absence of other perturbations at the micromotion-sideband frequency -- the system returns to its original state after each period  $t=2\pi/\OmRF$.

While the analysis here has been one dimensional, the single-ion case is readily generalized to three dimensions, where the motional modes are orthogonal \citep{Leibfried:2003_01}. The motion of multiple ions in harmonic traps is more complicated still \citep{Steane:1997, James:1998_02}. With each additional ion, there are three additional modes of oscillation. The center of mass (COM) motion, in which all ions move in phase with each other, is the lowest-frequency mode axially and the highest-frequency mode radially. In strings of multiple trapped ions the vibrational modes are in general not strictly harmonic and are weakly coupled among each other via higher-order terms in the expansion of the Coulomb potential. Still, for sufficiently cold strings of ions a description in terms of orthogonal harmonic modes holds to a very good approximation. In practice, usually a single mode is selected for a particular operation and all other `spectator' modes are neglected. The effect of multiple ion species and anharmonic potentials add still more complexity and have been treated by \citet{Morigi:2001} and by \citet{Home:2011}.

\subsection{Typical trap geometries}
\label{subsec:Trap geometries}

\begin{figure*}[t]
\begin{center}
\includegraphics[width=\textwidth]{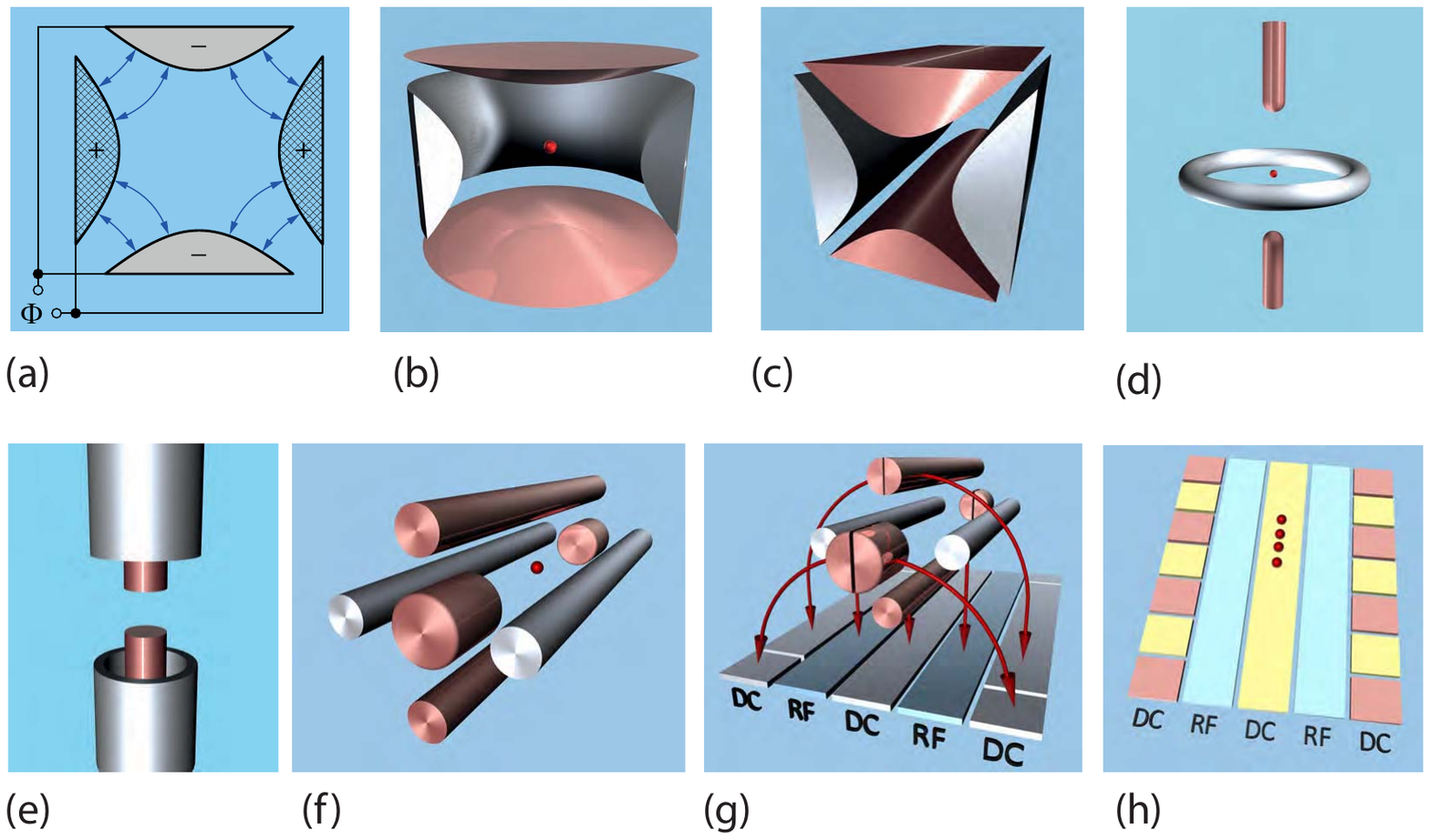}
\caption{Various possible trap geometries. The most basic geometry, of which all others are a variation, comprises perfectly hyperbolic electrodes, capable of creating a perfectly quadrupolar field (a). In generating a three-dimensional structure, a rotationally symmetric geometry can create a pseudopotential which is confining in three dimensions (b). This is the basic design of a ring trap. Alternatively, a translationally symmetric geometry can create a pseudopotential which is confining in two dimensions (c). This is the basis of the design for a linear trap, where confinement in the third dimension is provided by the addition of electrodes held at a positive DC voltage. These basic designs can be deformed to depart significantly from the idealized hyperbolae, where (d-e) are topologically ring traps and (f-g) are topologically linear traps. The axial electrodes of linear traps can also be segmented to form multiple trapping positions in a single device (h). See main text for  further discussion.}
\label{fig:TrapGeometries}
\end{center}
\end{figure*}

Having outlined the essential features of the potential used for trapping ions, we now turn to a few common trap geometries used to create such a potential. It is simple to show that a quadrupole field in two dimensions can be created by a set of hyperbolic electrodes, as depicted in Fig.~\ref{fig:TrapGeometries}(a). An oscillating quadrupole field can be created by applying an RF voltage, $(V_{\rm rf}/2) \cos(\OmRF t)$, to one pair of electrodes and an RF voltage of opposite sign, $-(V_{\rm rf}/2) \cos(\OmRF t)$, to the other pair. To create a three-dimensional trap one may either create a rotationally symmetric geometry to form a ring trap [Fig.~\ref{fig:TrapGeometries}(b)] or a translationally symmetric geometry to form the basis of a linear trap [Fig.~\ref{fig:TrapGeometries}(c)]. In the latter case the axial confinement of the ion is provided by additional DC electrodes at either end of the trap, held at a positive potential [as shown in Fig.~\ref{fig:TrapGeometries}(f)].

Rather than applying RF voltages of opposite phases, it is often technically simpler to apply an RF voltage of amplitude $V_\mathrm{rf}$ to one set of electrodes and have the other set grounded. In a linear trap where the axial electrodes are long compared to their radial separation, and in a ring trap where all other grounds are well shielded, this adds only a space-independent oscillating potential, which is of no consequence to the confining pseudopotential. If, however, the axial electrodes of a linear trap are sufficiently short that the endcaps cannot be neglected, this adds an axial component to the pseudopotential, and lifts the degeneracy of the radial secular motions.

Perfectly hyperbolic electrodes are difficult to machine and leave relatively little optical access. Fortunately, close to the RF null, the potential remains essentially harmonic even if the electrodes are deformed significantly from the ideal hyperbolic geometry. In this situation the basic physics of trapping remains unchanged, except that the voltage required to produce a given electric field at the trap center is increased by a geometry-dependent factor of typically around 1-3 \citep{Madsen:2004}.

It is common to form the electrodes of a ring trap from a simple wire loop with a pair of wires for the endcaps \citep{Raab:2000} [Fig.~\ref{fig:TrapGeometries}(d)]. The design can be deformed further so that the ring electrode is essentially bisected and moved to form a collar around each of the end caps \citep{Schrama:1993} [Fig.~\ref{fig:TrapGeometries}(e)]. Similarly, the electrodes of a linear trap are often made of simple rods \citep{Naegerl:1998} or blades \citep{SchmidtKaler:2003_11} [Fig.~\ref{fig:TrapGeometries}(f)]. Such ring traps and linear traps are typical designs for ion traps with electrode spacings of $\sim 1$\,mm. The goal of building scalable quantum information processing architectures with trapped ions has triggered substantial efforts in the development of chip-based traps. The trap electrodes can be deformed such that they all lie in a single plane \citep{Chiaverini:2005_06, Seidelin:2006}, as shown in Fig.~\ref{fig:TrapGeometries}(g). This creates a pseudopotential minimum above the surface. Such micro-fabricated planar traps prove useful for fabricating smaller and more complex structures \citep{Amini:2010, Wright:2013}, and the ions are typically held at a distance of about 100\,$\mu$m from the nearest electrode.

The electrodes of a linear trap (either a three-dimensional or a planar trap) can also be segmented \citep{Rowe:2002} as shown in Fig.~\ref{fig:TrapGeometries}(h). This allows a single trap structure to have multiple potential minima. By application of appropriate DC voltages ions can be moved between different trapping sites. This allows much greater flexibility to control the ions, though brings with it more degrees of freedom which must be controlled when considering heating.

Finally, consideration must be given to electrode structures not directly required for trapping. In order to have the lowest possible micromotion the ions should be positioned on the RF null. With a ring trap in which all electrodes are held at DC ground this would occur automatically. Fields from stray charges in the system may, however, displace the ion from the RF null and thereby cause excess micromotion \citep{Berkeland:1998}. In a linear-trap configuration the issue is compounded by the necessity of applying non-zero DC voltages to trap: any residual field at the RF null, due to applying incorrect DC voltages, or imperfections in fabrication, can displace the ions and cause excess micromotion. Excess micromotion can be minimized by use of additional compensation electrodes which, for most practical purposes, can be subsequently ignored in analyzing trap behavior. They should not, however, be neglected in the noise analysis of the system.

In summary, the ability to deform and miniaturize ion traps allows greater optical access, improved fabrication, and permits the use of lower voltages, higher trap frequencies, and novel operations. However, the close proximity of the surface makes electric-field noise one of the main challenges for such traps and illustrates the necessity of understanding and minimizing spurious electric fields from the surface.

\subsection{Electric-field noise induced heating of ions}
\label{subsec:FieldNoiseInducedHeating}

Apart from the large electric potentials which confine the ion, residual fluctuating electric fields from the environment couple to the motion of the ion and induce transitions between vibrational states. Even when cooled close to the quantum ground state, the corresponding electric transition dipole moment is $d_{\rm I} \approx  e a_0$, where $a_0=\sqrt{\hbar/(2\mi \omt)}\approx 10-50$\,nm is the extent of the ground-state wave function. This dipole moment makes trapped ions susceptible to very small electric-field fluctuations. Adopting, as above, a one-dimensional description of the ion motion along a specific trap axis, ${\bm e}_{\rm t}$, the total Hamiltonian for the ion in the presence of an additional fluctuating potential, $\Phi(t,x)$, is
\begin{equation}
\begin{split}
\hat H(t)= & \hat H_{\rm t}(t) + |e| \Phi(t,{\bm r}_{\rm I}+ {\bm e}_{\rm t} \hat x)\\
=&  \hat H_{\rm t}(t) +  |e| \Phi(t,{\bm r}_{\rm I}) -  |e| E_{\rm t} \hat x +\dots,
\end{split}
\end{equation}
where $E_{\rm t}= -{\bm e}_{\rm t}\cdot \nabla \Phi(t,{\bm r}_{\rm I})$ is the electric-field component at the position of the ion, ${\bm r}_{\rm I}$, and along the direction ${\bm e}_{\rm t}$. The global potential offset, $\Phi(t,{\bm r}_{\rm I})$, does not affect the motion of the ion. Higher-order terms in the expansion of $\Phi$ scale with additional powers of $(a_0/d)\ll 1$, and can be neglected compared to the linear coupling to the electric field. By changing into the interaction picture with respect to the bare trapping Hamiltonian, $\hat H_{\rm t}(t)$, and using the representation of the position operator given in Eq.~\eqref{eq:Heisenberg_x}, the resulting ion-field coupling is
\begin{equation}
\label{eq:IonField}
\hat H_{\rm ion-field}(t)=  d_{\rm I} \left[u(t) \hat a^\dag + u^*(t) \hat a\right] \delta E_{\rm t}(t).
\end{equation}
In general the fluctuating part of the electric field $\delta E_{\rm t}(t)= E_{\rm t}(t)-\langle E_{\rm t}\rangle$. It is assumed in Eq.~\eqref{eq:IonField} that the average stray field has been compensated, so that $\langle E_{\rm t}\rangle = 0$.

\subsubsection{Heating rate}
\label{subsec:HeatingRate}

The quantity generally measured when considering noise in ion traps is the heating rate, $\dot{\bar{n}}$, where $\bar n=\langle \hat a^\dag \hat a\rangle$ is the average phonon occupation number. We begin here by considering the special case, $\Gamma_{\rm h}$, which is defined as the rate at which an ion in the motional ground state, $|0\rangle$, is excited into the first vibrational state, $|1\rangle$.
Using Eq.~\eqref{eq:IonField} and Fermi's golden rule this rate is given by \citep{Wineland:1998_06}
\begin{equation}
\label{eq:HeatingRateGoldenRule}
\Gamma_{\rm h}=   \frac{e^2}{4\mi \hbar \omt} \sum_{j=-\infty}^\infty |C_{2j}|^2 S_{\rm E}((b/2+j)\OmRF),
\end{equation}
where
\begin{equation}\label{eq:SpectralDensity}
S_{\rm E}(\omega)= 2 \int_{-\infty}^\infty d\tau \langle \delta  E_{\rm t}(\tau) \delta E_{\rm t}(0)\rangle  e^{-i\omega \tau}
\end{equation}
is the single-sided spectral density of the electric-field noise.\footnote{Throughout this review we use this convention for the definition of the spectral density of the electric-field noise, which is generally used for reporting electric-field noise densities in ion traps. Most theoretical works use the double-sided spectral density which includes negative frequencies, and is defined without the factor of two.}
The derivation of this result is detailed in App.~\ref{Appendix_Derivation}. Due to a cross coupling between the RF and the noise fields, the heating rate in Eq.~\eqref{eq:HeatingRateGoldenRule} contains contributions from all orders of the micromotion sidebands. However, the coefficients $|C_{2j}|\sim q_x^j\sim (\omt/\OmRF)^j$ are usually small. In addition, most (though not all) known noise processes predict a spectrum which is constant or decreases at larger frequencies. In the limiting case where micromotion can be neglected, and noting that $b \OmRF/2 \simeq \omt$, the ion's heating rate then reduces to
\begin{equation}
\label{eq:Gamma(S_E)}
\Gamma_{\rm h}\simeq \frac{e^2}{4\mi \hbar \omt} S_\mathrm{E}(\omt).
\end{equation}
This result is most commonly cited in the literature and unless otherwise stated, Eq.~\eqref{eq:Gamma(S_E)} is used to relate the measured heating rate and the spectral density of the electric-field noise throughout this review.

A noteworthy deviation from the relation stated in Eq.~\eqref{eq:Gamma(S_E)}  can occur in situations where a non-compensated static off-set field, $E_{\rm stat}$, displaces the trap minimum from the RF null by an amount $\Delta x=|e|E_{\rm stat}/(\mi \omt^2)$. This shift of the trap minimum has two effects. First, the ion will undergo excess micromotion, $\hat x(t)\rightarrow  \Delta x + x_{\rm emm}(t)+ \hat x(t)$, where $x_{\rm emm}(t) \simeq \Delta x q_x/2\cos(\OmRF t)\sim \Delta x$ for small $q_x$. Excess micromotion can significantly affect the manipulation of the ion with laser light (see Sec.~\ref{subsec:HeatingRateMeasurements}), but as a coherent and periodic modulation of the mean ion position it does not  represent a source of heating per se. However, away from the RF null the electric field from the RF trapping electrodes is non-vanishing, and any noise component $\delta V(t)$ of the driving voltage can directly couple to the ion. For example,  the voltage $V(t)=V_{\rm rf}\cos(\OmRF t) + \delta V(t)$ applied to opposing electrodes produces a fluctuating electric field $\delta E(t)= \Delta x \Phi_{\rm rf}'' \delta V(t)/V_{\rm rf}$ in addition to the usual trapping potential. Since $\delta V(t)$ passes through the same filter electronics as the coherent driving field, its noise spectrum, $ S_\mathrm{V}(\omega)$,  has a dominant contribution at $\omega  \approx \OmRF$. This noise at $\OmRF$ can directly affect the ion via the higher-order, $C_{\pm 2}\simeq q_x/4$ terms in Eq.~\eqref{eq:HeatingRateGoldenRule}. In addition, the finite gradient of $\delta E(t)$ induces a mixing between $x_{\rm emm}(t)$ and $\hat x(t)$, which demodulates the RF noise components. As outlined in more detail in App.~\ref{app:ExcessMM}, both mechanisms contribute equally and in phase to the effective fluctuating force seen by the ion and give rise to  two additional heating-rate contributions \citep{Blakestad:2009}:
\begin{equation}
\Gamma_{\rm h}^{\rm rf,\pm} \simeq \frac{e^2 q_x^2 (\Phi_{\rm rf}'')^2 }{16\mi \hbar \omt}  \frac{ S_\mathrm{V}( \OmRF\pm \omt)}{V_{\rm rf}^2} (\Delta x)^2.
\end{equation}

\subsubsection{Master equation}
\label{subsec:MasterEquation}
While the initial heating rate, $\Gamma_{\rm h}$, is most commonly used to measure electric-field noise, a more general description of the ion motion in the presence of fluctuating electric fields is given in terms of the full master equation for the reduced ion density operator, $\rho_{\rm I}$. The master equation can be derived under the same conditions as has been assumed for the derivation of heating rate above (see also App.~\ref{Appendix_Derivation}) and can be written in the form \citep{Henkel:1999_11}
\begin{equation}
\label{eq:MasterEquation}
\begin{split}
\dot {\rho_{\rm I}}   = &-i (\omt+  \delta) [\hat a^\dag \hat a, \rho_{\rm I}] \\
&+ \frac{\Gamma}{2} (\bar N+1)  (2\hat a \rho_{\rm I} \hat a^\dag - \hat a^\dag \hat a \rho_{\rm I}-\rho_{\rm I} \hat a^\dag \hat a) \\
&+ \frac{\Gamma}{2} \bar N  (2\hat a \rho_{\rm I} \hat a^\dag - \hat a^\dag\hat  a \rho_{\rm I}-  \rho_{\rm I} \hat a^\dag \hat a).
\end{split}
\end{equation}
This equation describes the evolution of a damped harmonic oscillator coupled to an effective bath with mean occupation number $\bar N$. The frequency shift, $\delta$, and the damping rate, $\Gamma$, are given by
\begin{eqnarray}
\delta&=& \frac{e^2}{2\mi \hbar\omt} \, {\rm Im} \int_{0}^\infty d\tau \langle [ \delta  \hat E_{\rm t}(\tau) ,\delta \hat E_{\rm t}(0)] \rangle  e^{i\omt \tau},\label{eq:delta}\\
\Gamma&=& \frac{e^2}{\mi \hbar\omt} \, {\rm Re} \int_{0}^\infty d\tau \langle [ \delta  \hat E_{\rm t}(\tau),\delta \hat E_{\rm t}(0)] \rangle  e^{i\omt \tau}.
\end{eqnarray}
The total rate,
\begin{eqnarray}
\label{eq:GammaNbar}
\Gamma \bar N= \frac{e^2}{2\mi \hbar\omt} \,  \int_{-\infty}^\infty d\tau \langle \delta  \hat E_{\rm t}(\tau)\delta \hat E_{\rm t}(0) \rangle  e^{-i\omt \tau},
\end{eqnarray}
can be identified with the heating rate, $\Gamma_{\rm h}$, given above. In contrast to the result presented in Eq.~\eqref{eq:Gamma(S_E)}, the ion-field interaction used for the derivation of Eq.~\eqref{eq:delta}-\eqref{eq:GammaNbar} has been generalized to include quantized electric fields $\delta \hat E_{\rm t}(t)$, which do not necessarily commute at different times.
The resulting equation of motion for the average vibrational occupation number, $\bar n(t)=\langle a^\dag a\rangle(t)$, is given by
\begin{equation}
\label{eq:nbardot_t}
\dot{\bar n}(t)  =- \Gamma \bar n(t)  + \Gamma \bar N,
\end{equation}
and shows that close to the ground state $\dot{\bar n}(t)= \Gamma \bar{N} =\Gamma_\mathrm{h}$, while for long times $\bar n(t)$ approaches $\bar N$. For an equilibrium noise process one obtains $\bar N\equiv N_{\rm th}$, where $N_{\rm th}=1/(e^{\hbar\omt/k_{\rm B}T}-1)$ is the thermal equilibrium occupation number for a given bath temperature, $T$.

Typical trapping conditions of $\omt/2\pi=1$\,MHz and $T=4-300$\,K, correspond to $N_{\rm th}\approx 10^5-10^7$. Consequently, only the combined rate, $\Gamma \bar N \equiv \Gamma_{\rm h}$, is usually accessible in experiments. However, should traps be operated at millikelvin temperatures it may be possible to measure $\Gamma$ (and therefore the non-commuting part of the electric field) and $\bar N$ independently. It may be noted that work in Penning traps is not far from such low-temperature regimes \citep{Wrubel:2011}. Knowledge about the stationary occupation number, $\bar n(t\rightarrow \infty)$, provides an additional analytic tool to distinguish between equilibrium and out-of-equilibrium noise processes, or to identify whether the noise arises from a low- or high-temperature source.

\subsubsection{Heating of multiple ions}
\label{subsubsec:MultiIonHeating}

The analysis given above for a single trapped ion can be generalized to multi-ion Coulomb crystals, assuming that the ions are sufficiently cold and can be modeled as a set of coupled harmonic modes \citep{Morigi:2001, Home:2011}. In this case the coupling of $N_{\rm I}$ ions to the electric field is
\begin{equation}
H_{\rm ion-field}(t)= |e| \sum_{i=1}^{N_{\rm I}}  \delta {\bm E}(t,{\bm r}_{{\rm I},i})) \cdot \hat{\bm x}_i(t),
\end{equation}
where $\hat{\bm x}_i$ is the  position operator for an ion localized around its equilibrium position, ${\bm r}_{{\rm I},i}$. Using a normal-mode decomposition
\begin{equation}
\hat{\bm x}_i(t) =\sum_{k=0}^{3N_{\rm I}-1} \sqrt{\frac{\hbar}{2\mi \omega_k }}  {\bm c}_k(i)  \left( \hat a_k e^{-i\omega_k t} + \hat a_k^\dag e^{+i\omega_k t} \right),
\end{equation}
where $\hat a_k$ $(\hat a_k^\dag)$ is the annihilation (creation) operator for the $k$-th phonon mode with frequency $\omega_k$ and mode function ${\bm c}_k(i)$, normalized to $\sum_{i=1}^{N_{\rm I}} {\bm c}_k(i)\cdot{\bm c}_{k'}(i) =\delta_{k,k'}$. As a result the mode expansion of the electric-field coupling in the interaction picture is
\begin{equation}
\hat H_{\rm ion-field}(t) =   \sum_{k=0}^{3N_{\rm I}-1}  d_k   \left(\hat a_k e^{-i\omega_k t} + \hat a_k^\dag e^{i\omega_k t}\right) \delta E_k(t),
\end{equation}
where $d_k= |e|\sqrt{\hbar/(2\mi\omega_k)}$ is the dipole moment of the $k$-th motional mode. $\delta E_k(t)= \sum_i \delta E_{k}^i(t) $ is the projection of the electric field noise onto the $k$-th motional mode, where $\delta E_{k}^i(t)= \delta {\bm E}(t,{\bm r}_{{\rm I},i})\cdot{\bm c}_k(i)$. In analogy to the single-mode case one can define a heating rate, $\Gamma_{\rm h}^{(k)}$, for each mode,
\begin{equation}
\Gamma_{\rm h}^{(k)} =  \frac{e^2}{4\mi\hbar \omega_k}   S_{\rm E}^{(k)}(\omega_k),
\end{equation}
where
\begin{equation}
\begin{split}
S_{\rm E}^{(k)}(\omega)= & 2\sum_{i,j} \int_{-\infty}^\infty d\tau \langle \delta E_{k}^i(\tau)\delta E_{k}^j(0)\rangle e^{-i\omega \tau}.
\end{split}
\end{equation}
The heating rate of the individual modes depends on the correlation of the field noise at positions ${\bm r}_{{\rm I},i}$ and ${\bm r}_{{\rm I},j}$. In the limit of perfectly spatially correlated noise the COM mode of a linear ion string with frequency $\omega_{\rm COM}=\omega_\mathrm{t}$ and ${\bm c}_{\rm COM}={\bm e}_{\rm t}/\sqrt{N_{\rm I}}$ the heating rate is $\Gamma_{\rm h}^{({\rm COM})}=N_{\rm I} \Gamma_{\rm h}$ while $\Gamma_{\rm h}^{(k)}\approx 0$ for all the other modes. In the opposite limit of completely uncorrelated noise, the heating rates $\Gamma_{\rm h}^{(k)}\simeq (\omt/\omega_k) \Gamma_{\rm h}$ depend only on the mode frequency.

\subsubsection{Heating rates and decoherence}
\label{subsec:Decoherence}

In applications of trapped ions for quantum information processing the vibrational states, $|n\rangle$, of a single vibrational mode can be used to encode quantum information and serve as a quantum bus to communicate between qubits encoded in the internal states of two ions in a 1D crystal \citep{Cirac:2000}. For such applications it is not only the heating of the ion which is important, but also the loss of coherence of a motional superposition state. To identify the relationship between heating and decoherence, one can study the evolution of an initially prepared motional superposition state of the form $\rho_{\rm I}(t=0)=\rho_0=|\psi_0\rangle\langle \psi_0|$, where $|\psi_0\rangle = (|n_0\rangle + |m_0\rangle)/\sqrt{2}$ and $m_0\neq n_0$. In the presence of electric-field noise, i.e. under the evolution of the master equation~\eqref{eq:MasterEquation}, the coherence of this superposition, $\rho_{\rm coh}(t)=|\langle n_0| \rho_{\rm I}(t)|m_0\rangle|$, degrades (for short times) as
\begin{equation}
\dot \rho_{\rm coh}(t)\simeq - \frac{\Gamma}{2} (2\bar N+1)(n_0+m_0) \rho_{\rm coh}(0).
\end{equation}
This result shows that the overall scale of the motional decoherence rate is again set by $\Gamma_{\rm h}$ but, as usually observed in quantum mechanics, large superposition states (here meaning $n_0,m_0\gg1$) decohere even faster.

\subsection{Noise sensing with heating-rate measurements}\
\label{subsec:HeatingRateMeasurements}

The direct relation between the ion-heating rate, $\Gamma_{\rm h}$, and the spectral density of electric-field noise, $S_{\rm E}(\omt)$, established in Eq.~\eqref{eq:Gamma(S_E)} allows $S_{\rm E}(\omt)$ to be inferred by measuring the heating rate of a laser-cooled ion. In general this is done by measuring the phonon number of the ion at different times, and  calculating from this a value for  the rate of change of the phonon number, $\dot{\bar{n}}$. As detailed in Sec.~\ref{sec:ExperimentalOverview}, this has been used to infer values of the spectral density of electric-field noise ranging over ten orders of magnitude, from 10$^{-16}$ -- 10$^{-6}$\,\B. These were measured at frequencies in the range of $0.1\,\mathrm{MHz} \lesssim \omt/2\pi \lesssim 20\,\mathrm{MHz}$, and for distances from the electrodes of $30\,\mu\mathrm{m} < d < 3500\,\mu$m. Measurements have also been made on surfaces at temperatures of $6\,\mathrm{K} \lesssim T \lesssim 400\,$K. Depending on which part of this parameter space is to be measured there are various different techniques, which differ in experimental complexity and their suitability for measuring in different heating-rate regimes.

\subsubsection{Sideband spectroscopy}

One set of methods for measuring the heating rate relies on sideband-resolved cooling and manipulation techniques in the Lamb-Dicke regime~\citep{Leibfried:2003_01}. Sideband-resolved methods require the ion's motional frequency to be greater than the linewidth of the transition between two internal atomic states, $|g\rangle$ and $|e\rangle$, being probed. Given $\sim$MHz trap frequencies, this is not generally possible on dipole-allowed transitions [though see \citet{Jefferts:1995}]. Instead, either quadrupole \citep{Diedrich:1989} or Raman \citep{Monroe:1995_11} transitions are generally used.

\begin{figure}[t]
\begin{center}
\includegraphics[width=\columnwidth]{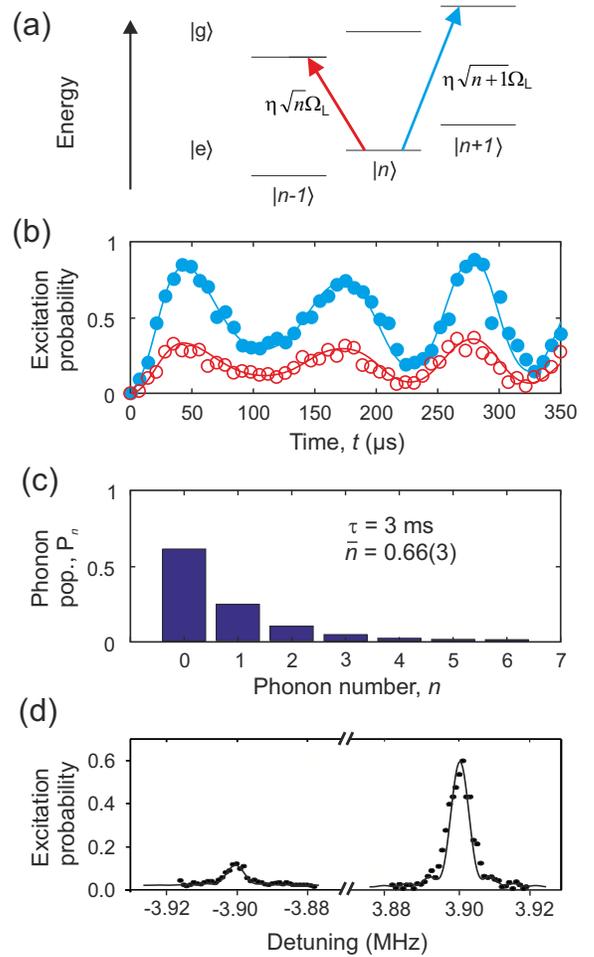}
\caption{Heating-rate measurements by sideband-resolved methods.
(a) Excitations of the ion to states with different phonon numbers have phonon-number-dependent coupling strengths. In the Lamb-Dicke limit, transitions of $|\Delta n| > 1$ can be neglected.
(b) Rabi flops driven on the red sideband (open circles) and blue sideband (filled circles). The solid lines are fits to the data from Eqs.~\eqref{eq:RSBRabiFlops}-\eqref{eq:BSBRabiFlops}.
From these fits the populations, $P_n$, can be calculated (c).
(d) Rather than driving Rabi flops, the mean phonon number, $\bar{n}$, can be calculated from the relative heights of  the red and blue sidebands.
[Figures (b-c) adapted from \citet{Harlander:2012PhD}. Figure (d) adapted from \citet{Brownnutt:2007PhD}. Used with permission.]}
\label{fig:HeatingRateMeasurement}
\end{center}
\end{figure}

In the Lamb-Dicke regime the atomic wave packet is confined to  a region much smaller than the wavelength of the transition being addressed. The Lamb-Dicke parameter, $\eta$, is defined equivalently as
\begin{equation}
\eta = k_x a_\mathrm{0} = \sqrt{\frac{\hbar k_x^2}{2 m_\mathrm{I} \omt}} = \sqrt{\frac{\omega_\mathrm{rec}}{\omt}},
\end{equation}
where $k_x$ is the projection of the incident light's wave vector into the $x$-direction and $\omega_\mathrm{rec}$ is the recoil frequency  of the ion. In the Lamb-Dicke regime, defined as $\eta^2 (2\bar{n}+1) \ll 1$, the coupling between internal and vibrational states of the ion is weak and the dominant contributions are phonon-assisted transitions $|g\rangle|n\rangle \leftrightarrow |e\rangle |n\pm 1\rangle$, which either add or subtract a single vibrational quantum. As illustrated in Fig.~\ref{fig:HeatingRateMeasurement}(a) the ion can be driven on either the red- or blue-detuned sidebands by appropriately choosing the laser frequency.  The corresponding Rabi-frequencies depend explicitly on the vibrational state, $|n\rangle$. They are respectively given by $\Omega_{n,n-1}=\eta\sqrt{n}\Omega_\mathrm{L}$ and $\Omega_{n,n+1}=\eta\sqrt{n+1}\Omega_\mathrm{L}$,  where $\Omega_\mathrm{L}$ denotes the bare Rabi frequency of the atomic transition driven by the incident light, neglecting the effect of the motion.

When driving Rabi flops on these transitions the excitation probability after driving red or blue sidebands for a time $t$ is \citep{Leibfried:2003_01}
\begin{eqnarray}
\label{eq:RSBRabiFlops}
p_{|e\rangle}^{\rm RSB}(t)&=&\frac{1}{2}\left(1-\sum_{n=0}^\infty P_n \cos(\Omega_{n,n-1} t) \right),        \\
\label{eq:BSBRabiFlops}
p_{|e\rangle}^{\rm BSB}(t)&=&\frac{1}{2}\left(1-\sum_{n=0}^\infty P_n \cos(\Omega_{n,n+1} t) \right),
\end{eqnarray}
where $P_n$ is the population of the motional state $|n\rangle$ at time $t=0$. Example data are given in  Fig.~\ref{fig:HeatingRateMeasurement}(b).
From this signal the $P_n$ can be extracted from a Fourier transform of $p_{|e\rangle}(t)$ [Fig.~\ref{fig:HeatingRateMeasurement}(c)]. This method does not need to assume thermal states but can rather infer the values of $P_n$ independently \citep{Meekhof:1996}. By repeating the measurement of $P_n$ for different waiting times, $t_{\rm w}$, after ground-state cooling, the heating rates, $\Gamma_\mathrm{h}$ and $\dot{\bar{n}}$, can be found. This method works well for $\bar n\lesssim 2$ and low heating rates of up to about $\Gamma_{\rm h} \sim 100$\,s$^{-1}$.

If it is sufficient to only measure $\bar{n}$, and assuming that the distribution of $n$ is approximately thermal, a similar method, which requires less data taking, uses the asymmetry between red and blue sidebands \citep{Monroe:1995_11}. This is illustrated in  Fig.~\ref{fig:HeatingRateMeasurement}(d). Taking the ratio of the excitation probabilities on the red and blue sidebands [Eqs.~\eqref{eq:RSBRabiFlops} and  \eqref{eq:BSBRabiFlops}] the mean phonon number is given by  $p_{|e\rangle}^{\rm RSB}/p_{|e\rangle}^{\rm BSB}=\bar{n}/(1+\bar{n})$. By repeating this measurement for different values of $t_\mathrm{w}$ the heating rate, $\dot{\bar{n}}$, can be inferred. Again, this method works well for $\bar n\lesssim 2$ and low heating rates  of up to about $\Gamma_{\rm h} \sim 100$\,s$^{-1}$.

It may be noted that, because it assumes a thermal distribution of $n$, this method can give incorrect results when the ions are not - or not always - in thermal states. Such non-thermal states can occur if the ion is coherently excited, such as occurs with noise from Digital to Analog Converter (DAC) cards \citep{Blakestad:2010PhD}. Alternatively, the ions may sometimes be excited to non-thermal states by collisions with background gas atoms. When a collision occurs the ion becomes very hot, so that the red and blue sidebands are approximately the same heights. Such events may not be immediately obvious, as they are averaged with many collision-free events. However, for collision rates of a few per second or greater, this can lead to a significantly higher apparent heating rate than would be measured without the collisions \citep{Chiaverini:2014}.

For higher phonon numbers and higher heating rates, one can compare higher-order sidebands. Generalizing to the  $k$-th-order red and blue sidebands, the ratio of the  corresponding excitation probabilities, $p_{|e\rangle}^{{\rm RSB}_k}$ and $p_{|e\rangle}^{{\rm BSB}_k}$, is
\begin{equation}
\label{eq:kthOrderSideband}
\frac{p_{|e\rangle}^{{\rm RSB}_k}}{p_{|e\rangle}^{{\rm BSB}_k}}=\left(\frac{\bar{n}}{1+\bar{n}}\right)^k.
\end{equation}
This method is most sensitive for the sideband order, $k$, nearest to the value of $\bar{n}$ \citep{Turchette:2000}.

As the sidebands of the different motional modes are generally well resolved from each other, all of these sideband-spectroscopic methods allow independent measurement of the heating rates on different modes. The methods have the drawback, however, that they require access to the resolved-sideband and Lamb-Dicke regimes. This requires a narrow-linewidth laser and ground-state cooling.

For larger Lamb-Dicke parameters or higher phonon numbers, $\eta^2\bar n>1$,  increasing numbers of phonon sidebands contribute to the ion-laser interaction, leading to a corresponding suppression of the oscillator strength for the carrier transition. The reduced Rabi frequency for the carrier is
\begin{equation}
\Omega_c=\Omega_L \sum_{n=0}^\infty P_n \langle n |e^{i\eta(\hat{a}+\hat{a}^\dagger)}|n\rangle = \Omega_L e^{-\eta^2 (\bar n+1/2)},
\end{equation}
where, for the second equality, a thermal phonon distribution has been assumed. This can then be used to infer a value of $\bar{n}$ and thence $\dot{\bar{n}}$ \citep{Rowe:2002}.

\subsubsection{Doppler recooling}

Another method for heating-rate measurements uses Doppler recooling in the weak-binding regime (i.e. in a regime where the motional sidebands are not resolved) \citep{Epstein:2007, Wesenberg:2007}. In the first step the ion is cooled. It is not, however, necessary that the ion be cooled to the motional ground state, or even to the Doppler-cooling limit. The cooling laser is turned off and, after a waiting period of duration $t_{\rm w}$ during which the ion can heat up, the ion is Doppler cooled again. Evaluating the heating rate is based on the fact that hot ions, being Doppler shifted further from resonance with the cooling laser, scatter less light than cold ions.  At the end of the waiting time, the hot ion initially scatters fewer cooling photons, but scatters progressively more as it is cooled, until the Doppler-limited temperature is reached. During this recooling process the time-resolved fluorescence of the ion is recorded. From this, the initial mean phonon occupation number of the ion before recooling, $\bar{n}(t_\mathrm{w})$, can be inferred. By repeating the measurement for different waiting times, $t_\mathrm{w}$, the heating rate, $\bar{n}$, can be found.

This method has several distinct advantages over sideband-resolved methods: it works well at much larger phonon numbers, around $\bar{n}\sim 10^4$; it does not require ground-state cooling; and it does not require narrow-linewidth lasers. However, it also suffers a number of limitations: The waiting times required to measure low heating rates can become very long. It cannot provide information about the populations, only about the average phonon number. This is normally not a problem as, to a good approximation, the states are thermal. More significantly, as it is unable to resolve the separate motional sidebands, it cannot provide information regarding the heating rates of different motional modes. The heating rate for a particular mode must be estimated by making certain assumptions about the way different modes might be heated. One option is to assume that one mode is heated predominantly and that only that one mode contributes to the change in fluorescence. In many experiments using linear ion traps the axial mode has a much lower frequency than the radial modes, and usually the noise spectrum is such that low frequency modes are heated faster than high frequency modes (see Sec.~\ref{subsec:FrequencyScaling}). Under these conditions, neglecting heating of radial modes is not an unreasonable starting point, though it would be expected to slightly overestimate the heating rate \citep{Wesenberg:2007}. At the opposite extreme, one may assume that all modes are heated equally. Simulations using some typical trap parameters have seen that this can give a result smaller by around  a factor of two \citep{Wesenberg:2007}. While there is not a consistent usage of a particular set of assumptions across the literature \citep{Daniilidis:2011, Allcock:2012}, papers using this method generally state what assumption they have made regarding this issue.

\subsubsection{Other methods}

Sideband spectroscopy and  Doppler recooling are the most common methods for measuring heating rates, though not the only ones. \citet{Tamm:2000} compared the linewidth-broadening of the secular-vibration sidebands relative to the carrier resonance. In their case, however, the heating rate was below the sensitivity of the method and thus only an upper bound to the heating could be inferred. Similar problems are encountered with simple ion-loss measurements \citep{Splatt:2009}, where the heating rate is estimated from the ion-loss rate in a shallow trap. This method works for very high temperatures and large heating rates. However, away from the trap center the harmonic potential approximation as well as the neglected influence of micromotion assumed in Eq.~\eqref{eq:Gamma(S_E)} are no longer justified and a relation between heating rate and $S_{\rm E}(\omt)$ is questionable.

Using a phase Fresnel lens with a working distance of only 3\,mm, \citet{Norton:2011}  were able to image a trapped ion with a resolution of $\sim$ 400\,nm. They could then use the spatial extent of the ion's position to estimate its temperature, over the range $20 \lesssim \bar{n} \lesssim 10^5$. A similar scheme has been used by \citet{Knuenz:2012} using standard optics. This method has correspondingly lower spatial resolution, but can still be effective for low trap frequencies (tens of kHz). Such  spatial thermometry has not yet been used to measure heating rates, though it may be a useful technique in certain regimes.

Interference effects from light scattered by trapped ions can be used to infer the ion temperature. This was first demonstrated by \citet{Eichmann:1993}, using the fringe patterns of light emitted from a two-ion crystal to infer a temperature of 2.5\,mK, equivalent to $\bar{n}\sim50$. Using a single ion, it is possible to create an interferometer by retroreflecting light scattered by the ion so that the reflected light interferes with directly scattered light from the ion. As the length of the interferometer arm is varied any motion of the ion will reduce the visibility of the fringes. \citet{Slodicka:2012} used this effect to measure ion phonon numbers in the range $1 < \bar{n} < 40$. These methods are, however, rather involved and have not yet been used to measure heating rates.

\subsection{Limits to the measurement sensitivity}
\label{subsec:HeatingIonTraps}

Given enough experimental time, measuring the heating rate of a trapped ion provides a very sensitive way to measure the absolute electric-field noise strength at the location of the ion.  For typical parameters, a heating rate of $\Gamma_{\rm h}=1$ s$^{-1}$ corresponds to an electric-field noise sensitivity of $S_{\rm E}(\omt)\approx  10^{-14}$\,\B, and the detection of lower heating rates leads to a correspondingly greater sensitivity. In practice, a limit on the sensitivity can arise from other sources of heating,  which are discussed in detail by \citet{Wineland:1998_05} and outlined briefly here.

The ion can be heated by collisions with background-gas atoms. At the typical pressures used in ion-trap experiments ($\sim$10$^{-11}$\,mbar) and for single ions these can be expected to occur every few minutes; sufficiently infrequently that they can be neglected as a heating mechanism at the levels currently observed. The effects of such collisions are even less concerning in cryogenic experiments which are typically carried out at $\sim$4\,K: while the heating rates from fluctuating fields are generally lower in such systems, the background pressures expected [$\sim$10$^{-17}$\,mbar \citep{Antohi:2009}] rule out any significant effect from background collisions. One caveat to this is to note that the local pressure near the trap may be significantly higher than the measured background pressure in the chamber \citep{Chiaverini:2014}. This  is because the trap itself is a source of particles. These may desorb naturally from the trap surface. Additionally, particles may be emitted when photoelectrons -- initially ejected by scattered laser light -- are accelerated by the RF fields and impact on the trap electrodes.

The motional mode of interest can also be heated by other spectator modes. For a single ion in an ideal harmonic potential the three modes of motion are decoupled. However, in any real trap, higher-order terms in the trapping potential can couple these motions. If the spectator modes are not initially cold they can heat the mode of interest. In most experiments, however, the traps can be considered harmonic, and $\omega_\mathrm{x}$, $\omega_\mathrm{y}$, $\omega_\mathrm{z}$ are chosen to be separated so that that parametric interconversion is negligible \citep{Wineland:1998_05}.

A further issue arises with multiple ions in a single trap. The number of spectator modes increases with the number of ions in the crystal, $N_\mathrm{I}$, scaling as 3$N_\mathrm{I}$-1. This increases the possibility that one of the modes is not perfectly cooled and can couple to the mode of interest. This is all the more significant as radial modes in a chain can couple to the axial motion even in a perfectly harmonic trap. Nonetheless, most experiments select mode spacings such that this is not a problem. Most importantly for the purposes of this paper, almost all heating-rate measurements discussed here were taken with single ions.

Finally, when dealing with multiple ions in the presence of the RF confining potential, collisions are not conservative, but rather take energy from the RF field. This leads to RF heating \citep{Ryjkov:2005}. This effect is most pronounced when the ions form a cloud, rather than a crystal. However, even for 14 ions in a linear crystal, under normal conditions, it is estimated that this mechanism could lead to sufficient heating that the crystal melts after $\sim$15\,minutes without cooling \citep{Chen:2013}.

In summary, the measurement of electric-field fluctuations in ion traps is usually based on a number of assumptions. Firstly, it is assumed that the additional heating mechanisms listed above can be neglected. Secondly, it is assumed that heating from coupling to the micromotion sidebands can be neglected. Finally, it is assumed that the ion is initially sufficiently cold, implying that $\dot{\bar{n}}\approx \Gamma_{\rm h}$ and also that anharmonic terms in the trapping potential or the Coulomb interaction can be ignored. Given these conditions, it is straightforward to convert between the heating rate, $\dot{\bar{n}}$, and the spectral density of electric-field noise, $S_\mathrm{E}$, using the relationship in Eq.~\eqref{eq:Gamma(S_E)}. It should be noted that the $\omt$ term in the denominator of Eq.~\eqref{eq:Gamma(S_E)} means that the frequency exponents of $\dot{\bar{n}}$ and of $S_\mathrm{E}$ differ by one.

Being able to measure noise in this way, the main limitation for ions as electric-field sensors comes simply from the fact that the ion measures only the total field noise (at the trapping frequency) and cannot directly distinguish between noise from different sources. The interpretive step consists of measuring how the noise scales under different conditions, how it varies from day to day, or how it can be suppressed. Various models of possible heating mechanisms can then be compared to the observations and an attempt made to unpick which noise sources are likely to have been observed in any given experiment. This review follows such a line: First considering in Sec.~\ref{sec:ExperimentalOverview} the total levels of noise observed and how they vary under various conditions. Secs.~\ref{sec:Theory} and \ref{sec:Microscopic} outline various models of heating mechanisms and what noise characteristics they would predict. Sec.~\ref{sec:KnownMechanisms} then attempts to interpret the experimental results by seeing which, if any, of the models considered describe the observed behavior.

\section{Experimental Overview}
\label{sec:ExperimentalOverview}

The spectral density of electric-field noise varies with a range of physical parameters. Some of these lend themselves relatively easily to quantitative consideration. A systematic investigation of the observed variations of the noise over a large range of these parameters could provide a promising method for identifying the dominant or limiting noise sources in trapped-ion systems. To this end, experimental data is often analyzed using the ansatz\footnote{While $\alpha$, $\beta$, and $\gamma$ are standardly used to denote the scaling exponents, there is currently no standardized convention in the literature regarding which of these letters represents the exponent for which variable. For clarity, the convention stated here will be used throughout this review, and results from publications using different conventions will be `translated' without further comment.}
\begin{equation}
\label{eqn:S_E_Dependencies}
S_\mathrm{E} \propto \omega ^{-\alpha}d^{-\beta}T^{+\gamma}.
\end{equation}
It should be noted that there is no a-priori reason for this ansatz to hold. Moreover, even in systems exhibiting such power-law behavior, it is not necessarily that the behavior should be consistent over the entire accessible parameter space. Indeed, several models predict non-power-law behaviors and in general $\alpha$, $\beta$ and $\gamma$ must be interpreted as `local' scaling coefficients, which may themselves depend on the frequency-, distance- and temperature regime under consideration. Given the variety of behaviors which may be expected, the characterization of the noise scaling with respect to these basic parameters can already provide significant information about the source of the noise. That being noted, the diversity has the potential to make the picture rather complicated. The various scaling predictions of different theories are given in Secs.~\ref{sec:Theory} and \ref{sec:Microscopic}. At this point, however, we consider the scalings observed in experiments, unencumbered by theoretical models.

\begin{center}
\begin{table*}
\begin{tabular}{l l l l l l}
\hline
       &                             &             &                             &             &                         \\
$[1]$  & \citet{Diedrich:1989}       & [22]        & \citet{Labaziewicz:2008_10} & [43]        & \citet{Wang:2012PhD}    \\
$[2]$  & \citet{Monroe:1995_11}      & [23]        & \citet{Britton:2008PhD}     & [44]        & \citet{Doret:2012}      \\
$[3]$  & \citet{King:1998}           & [24]        & \citet{Poschinger:2009}     & [45]        & \citet{Wilpers:2012}    \\
$[4]$  & \citet{Roos:1999}           & [25]        & \citet{Blakestad:2009}      & [46]        & \citet{Hite:2012}       \\
$[5]$  & \citet{Myatt:2000}          & [26]        & \citet{Britton:2009}        & [47]        & \citet{Poulsen:2012}    \\
$[6]$  & \citet{Tamm:2000}           & [27]        & \citet{Leibrandt:2009}      & [48]        & \citet{Brama:2012}      \\
$[7]$  & \citet{Turchette:2000}      & [28]        & \citet{Amini:2010}          & [49]        & \citet{Steiner:2013}    \\
$[8]$  & \citet{SchmidtKaler:2000}   & [29]        & \citet{Allcock:2010}        & [50]        & \citet{Allcock:2013}    \\
$[9]$  & \citet{Rohde:2001}          & [30]        & \citet{Wang:2010_06}        & [51]        & \citet{Warring:2013}    \\
$[10]$ & \citet{Rowe:2002}           & [31]        & \citet{Wang:2010_12}        & [52]        & \citet{Vittorini:2013}  \\
$[11]$ & \citet{DeVoe:2002}          & [32]        & \citet{McLoughlin:2011}     & [53]        & \citet{Arrington:2013}  \\
$[12]$ & \citet{Deslauriers:2004}    & [33]        & \citet{Daniilidis:2011}     & [54]        & \citet{Mount:2013}      \\
$[13]$ & \citet{Home:2006PhD}        & [34]        & \citet{Brown:2011_03}       & [55]        & \citet{Hite:2013}       \\
$[14]$ & \citet{Stick:2006}          & [35]        & \citet{Harlander:2011}      & [56]        & \citet{Chiaverini:2014} \\
$[15]$ & \citet{Seidelin:2006}       & [36]        & \citet{Ospelkaus:2011}      & [57]        & \citet{Daniilidis:2014} \\
$[16]$ & \citet{Deslauriers:2006_09} & [37]        & \citet{Herskind:2011}       & [58]        & \citet{Niedermayr:2014} \\
$[17]$ & \citet{Letchumanan:2007}    & [38]        & \citet{Blakestad:2011}      & [59]        & \citet{Kumph:2014}      \\
$[18]$ & \citet{Epstein:2007}        & [39]        & \citet{Allcock:2011}        & [60]        & \citet{McKay:2014}      \\
$[19]$ & \citet{Labaziewicz:2008_01} & [40]        & \citet{Allcock:2012}        & [61]        & \citet{Mehta:2014}      \\
$[20]$ & \citet{Schulz:2008}         & [41]        & \citet{Akerman:2012}        & [62]        & \citet{Goodwin:2014}    \\
$[21]$ & \citet{Benhelm:2008_06}     & [42]        & \citet{Harlander:2012PhD}   &             &                         \\
       &                             &             &                             &             &                         \\
\hline
\end{tabular}
\caption[Reference numbers]
{Comprehensive list of references for trapped-ion heating-rate measurements. The figures and text of this paper denote the traps using the reference numbers here. Where a single publication reports heating rates for several traps, or for one trap under markedly different conditions, these different results are labeled with a letter following the reference number.}
\label{tab:References}
\end{table*}
\end{center}

\subsection{Frequency scaling}
\label{subsec:FrequencyScaling}

\begin{figure}[t]
\includegraphics[width=\columnwidth]{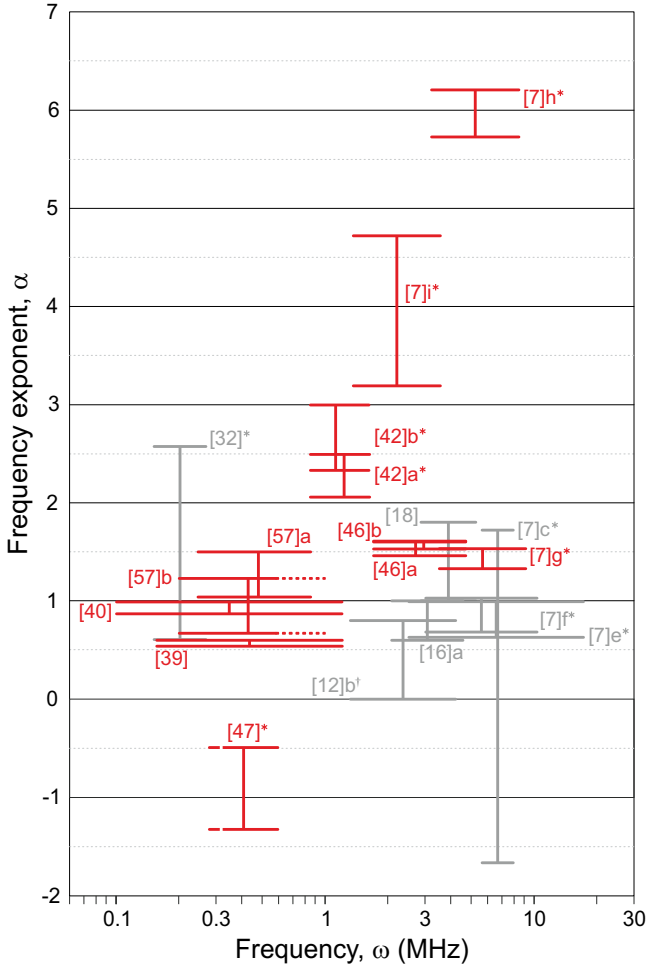}
\caption{\label{fig:alpha_omega}
Frequency exponent, $\alpha$, as a function of the ion motional frequency, $\omt$, in room-temperature traps. The lateral extent of the bars gives the frequency range over which the noise was measured. The vertical extent of the bars indicates the uncertainty in $\alpha$. Data are taken from the relevant references in table~\ref{tab:References}. For points marked with $^\dagger$, the uncertainty on $\alpha$ was not stated in the original paper, and has been estimated from the published data. For points marked with $^\ast$, neither the value of $\alpha$ nor its uncertainty were explicitly stated in the original paper, and have been estimated from the published data. For experiments in which resonance peaks were observed (points [\Poulsen12] and [\DGB13]b) the regions over which the resonances occurred are indicated by a dotted line. Selected data are discussed in detail in the text, and are highlighted here in red.}
\end{figure}

Many sources of noise are expected to follow a power law scaling and give rise to 1/$f$ noise. Strictly, this would mean that $S_\mathrm{E} \propto \omega ^{-\alpha}$, where $\alpha=1$. It should, however, be noted that ``1/$f$ noise" is something of a flexible term and is often used for any scalings in the range $0.9 \lesssim \alpha \lesssim 1.4$ \citep{Dutta:1981}. Even allowing for the flexible terminology, there are several possible noise sources which exhibit scalings significantly different to 1/$f$. There are still other noise sources for which the spectrum does not even follow a power law. By measuring the ion-heating rate for different motional frequencies it is possible to see whether, for a particular experiment and over a particular frequency range, the noise does indeed follow a power law and, if so, what the scaling exponent,  $\alpha$, is in that particular instance. Given the fact that different noise mechanisms can give rise to various values of $\alpha$ (as discussed in Secs.~\ref{sec:Theory} and~\ref{sec:Microscopic}) such measurements are of interest because the determination of $\alpha$ provides information about the possible mechanisms underlying the noise. Many experiments have deduced a value of $\alpha$ for the electric-field noise they observe, and these are shown in Fig.~\ref{fig:alpha_omega}. It is immediately clear from Fig.~\ref{fig:alpha_omega} that there is not an overall frequency-scaling law which holds for all systems. We now consider several specific cases for which power-law scaling behavior has been observed, and for which the measurements provide a relatively tightly constrained value of $\alpha$ for a particular experiment.

Data points [$\AGH11$] \citep{Allcock:2011} and [$\Allcock12$] \citep{Allcock:2012} shown in Fig.~\ref{fig:alpha_omega} were measured in a single trap at Oxford University.\footnote{The references from which all experimental data points in this review were taken are summarized in table~\ref{tab:References}.}
The Doppler recooling method used for the heating-rate measurement \citep{Allcock:2010} cannot distinguish between signals arising from heating of different modes. However as $\omega_\mathrm{ax} \ll \omega_\mathrm{r}$ (by at least a factor of 3.5) it was assumed that any heating of the radial modes could be neglected. The trap was a linear, SiO$_2$-on-Si surface trap with aluminium electrodes, in which the ion was 84\,$\mu$m above the trap surface (98\,$\mu$m from the nearest electrode). The electrodes were 2.4\,$\mu$m thick, with a 2-3\,nm native oxide and 8\,nm RMS (root mean square) roughness. Any exposed surfaces on the silicon substrate were coated with 114\,nm of gold. Any bare surfaces of the silica supporting the electrodes were recessed well below the electrode plane. Points [$\Allcock12$] and [$\AGH11$] were measured before and after laser cleaning respectively. Prior to cleaning the noise scaling was consistent with a $1/f$ source [$\alpha = 0.93(6)$]. Following laser cleaning the heating rate was around a factor of two lower, and the scaling had changed to $\alpha=0.57(3)$. These results are discussed further in Sec.~\ref{subsec:LaserCleaning}.

Data points [$\Turchette00$]g and [$\Hite12$]a,b were measured in different traps at NIST.
Point [$\Turchette00$]g \citep{Turchette:2000} shows the scaling for the axial component of the electric-field noise in a linear, three-dimensional, gold-on-alumina trap. The ion was 365\,$\mu$m from the nearest electrodes, which were made of 0.75\,$\mu$m of evaporated gold. The value of $\alpha=1.43(1)$ is significantly different from 1, though consistent with 1.5.

Points [$\Hite12$]a,b \citep{Hite:2012} show the scaling for the axial component of the electric-field noise in a surface trap made of gold on crystalline quartz. The ion was 40\,$\mu$m above the electrodes, which were made of 10\,$\mu$m of electroplated gold. Point [$\Hite12$]a was measured in a trap for which the surface of the gold electrodes had been exposed to air. Point [$\Hite12$]b was measured in the same trap after it had been cleaned in-vacuo by an argon-ion beam. The freshly revealed surface of the electrodes had therefore never been exposed to air. While the absolute value of the heating rate subsequent to cleaning was lower by two orders of magnitude, the frequency scaling of the electric-field noise did not change significantly between these two experiments; both exhibit $\alpha \approx 1.5$. This scaling is consistent with that seen in the other NIST trap considered above ([$\Turchette00$]g). These results are discussed further in Sec.~\ref{subsec:PlasmaCleaning}.

Data points [$\DGB13$]a,b were measured in a single trap at the University of California, Berkeley \citep{Daniilidis:2014}. They show the scaling for the axial component of the electric-field noise in a linear surface trap. The ion was 100\,$\mu$m above the electrodes, which were made of a copper-aluminium alloy on a fused-quartz substrate. Point [$\DGB13$]a was measured in a trap for which the copper surface of the electrodes had been exposed to air. The measured frequency scaling was $\alpha=1.27(23)$ over the range 246\,kHz - 852\,kHz. Point [$\DGB13$]b was measured in the same trap after it had been cleaned in-vacuo by an argon-ion beam. The freshly revealed surface of the electrodes had therefore never been exposed to air.  Like the work of \citet{Hite:2012}, the absolute value of the heating was reduced by around two orders of magnitude.  The frequency scaling was measured to be $\alpha=0.95(28)$ over the range 200\,kHz - 580\,kHz, not significantly different from the pre-cleaning value. This work additionally indicates presence of a broad resonance peak centered at around 800\,kHz, which was clearly visible following cleaning. These results are discussed further in Sec.~\ref{subsec:PlasmaCleaning}.

Data points [$\Harlander12$]a,b were measured in a single trap at the University of Innsbruck \citep{Harlander:2012PhD} and show the scaling for the axial component of the electric-field noise in a linear, three-dimensional, gold-on-alumina trap. The ion was 257\,$\mu$m from the nearest electrodes, which were made of 10-15\,$\mu$m of electroplated gold. Data point [$\Harlander12$]a was measured when the DC lines to the trap were low-pass filtered using an RC filter with a cutoff frequency of 1~MHz, and the noise exhibited a frequency-scaling exponent of $\alpha$=2.3(2). For point [$\Harlander12$]b, an RC filter with a lower cutoff frequency of 370~kHz was used. In this configuration the spectral density of electric-field noise was a factor of 1.6(8) lower, and the frequency scaling was measured to be $\alpha$=2.7(3). The frequency scaling if the noise in both configurations implies that the mechanisms for the dominant sources of heating in these instances do not display a $1/f$ spectrum. These results are discussed further in Sec.~\ref{subsec:Filters}.

Data point [$\Poulsen12$] was measured at Aarhus University \citep{Poulsen:2012}
and shows the axial component of the electric-field noise in a linear trap. The trap was made of stainless-steel rods at a distance of 3500\,$\mu$m from the ion and plated with 5\,$\mu$m gold. The heating rate was observed to be independent of the trap frequency, except for a narrow resonance at $\omega=295$\,kHz. The resonance was found to be caused by  a switch-mode power supply in a neighboring experiment. The remaining noise would require a source for which $S_\mathrm{E}\propto f$, which is significantly different from the oft-assumed $1/f$ spectrum. These results are discussed further in Sec.~\ref{subsec:NoiseAtAarhus}.

Data points [$\Turchette00$]c,h were measured in a single trap at NIST \citep{Turchette:2000}. They show the scaling for the radial component of the electric-field noise in a molybdenum ring trap. The ion was 125\,$\mu$m away from the nearest electrode, which was a fork-shaped endcap. The trap initially had a frequency scaling of $\alpha=0(2)$. The trap was removed from vacuum and cleaned with HCl to remove the Be coating deposited by the atomic source. It was then electropolished in phosphoric acid, and rinsed in distilled water followed by methanol. Following cleaning, the spectral density of electric-field noise was significantly reduced (over the frequency range measured)  and the frequency scaling had changed to $\alpha=6.0(2)$ (point [\Turchette00]h). This is significantly different from $1/f$. These results are discussed further in Sec.~\ref{subsec:TurchettesWeirdTraps}.

Finally, data point [$\Turchette00$]i was measured in a ring trap in NIST \citep{Turchette:2000} in which the ion was 280\,$\mu$m away from the nearest electrode, which was a fork-shaped endcap. The trap was made in the same piece of metal as trap  [$\Turchette00$]h, and separated from it by 1.7\,mm. The traps had therefore undergone exactly the same treatment and handling. Point [$\Turchette00$]i in Fig.~\ref{fig:alpha_omega} shows the scaling for the radial component of the electric-field noise after the trap had been cleaned. The trap initially had a heating rate that was relatively `typical' (see point [$\Turchette00$]d in Fig.~\ref{fig:S_E_d}), though the frequency scaling was not measured. Following the same cleaning procedure already described for trap [$\Turchette00$]h, above, the spectral density of electric-field noise was significantly reduced (over the frequency range measured). From the published data the frequency scaling is estimated to be $\alpha=4.0(8)$. This is significantly different from $1/f$. These results are discussed further in Sec.~\ref{subsec:TurchettesWeirdTraps}.

While each of these cases is discussed in more detail in Secs.~\ref{sec:KnownMechanisms} and \ref{sec:OpenQuestions} it suffices at this point to note that it is simply not possible to assume a single universal scaling behavior which holds for all traps. Some experimental results tightly constrain $\alpha$ and support a value of $\alpha=1$; still other experiments have scalings markedly different from $\alpha=1$. This lack of a universal frequency-scaling behavior suggests that the higher-than-expected level of heating observed in ion-trap experiments does not necessarily come from one single, universal effect, but rather has multiple sources, which may be different for different experiments.\\

\begin{figure}[t]
\includegraphics[width=\columnwidth]{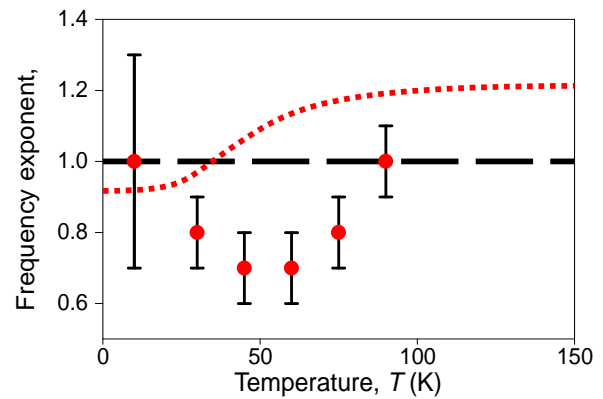}
\caption{Frequency exponent, $\alpha$, as a function of temperature for a gold-on-quartz surface trap.
The values for the data points and error bars are taken from  Fig.\,4 of \citet{Labaziewicz:2008_10}.
The dotted line shows the theoretically expected dependence from the model proposed
[\citet{Labaziewicz:2008_10} drawing on \citet{Dutta:1981}]
with the dashed line showing temperature-independent 1/$f$ scaling.}
\label{fig:alpha_T}
\end{figure}

All of the measurements discussed so far were made at room temperature. It has been suggested \citep{Labaziewicz:2008_10} that, in cases where the source of electric-field noise for ion traps is due to thermally activated processes, one could expect such sources to freeze out at low temperatures, leading to a temperature-dependent value of $\alpha$.  \citeauthor{Labaziewicz:2008_10} measured the value of $\alpha$ as a function of trap temperature for the lowest-frequency vibrational mode in a linear surface trap (trap [$\LGL08$]c~i). The ion was 75\,$\mu$m above the electrodes, which were made of gold on a crystal-quartz substrate. The results are shown in Fig.~\ref{fig:alpha_T}. These results suggest that in this experiments there may be some variation of the frequency scaling with temperature. The theory underlying possible temperature-dependent changes in $\alpha$ is outlined in Sec.~\ref{subsubsec:NonUniformTLFDistribution}, and the experiment is discussed further in Sec.~\ref{subsec:CryoNoiseFloor}.

\subsection{Distance scaling}
\label{subsec:DistanceScaling}

It is intuitively reasonable that the closer an ion is held to a source of electric-field noise, the more  perturbed it would be by that source. By measuring the ion-heating rate for different trap sizes it should be possible to see how the heating rate scales with the distance from the electrodes and, potentially, to infer a scaling exponent, $\beta$. As different noise mechanisms can give rise to various values of $\beta$ (as discussed in Secs.~\ref{sec:Theory} and~\ref{sec:Microscopic}) such measurements are of interest because the determination of $\beta$ provides information about the possible mechanisms underlying the noise. This may at first seem similar to the procedure used in Sec.~\ref{subsec:FrequencyScaling} to measure $\alpha$. However, while it is relatively simple to vary the trap frequency and leave all else unchanged, varying the trap size in a controlled way is very much more difficult. This section reviews some experiments which have varied the trap size while minimizing other uncontrolled differences in the set up. It also reviews  what can and cannot be learned from a compilation of heating-rate data from many different experiments.

\begin{figure}[t]
\includegraphics[width=\columnwidth]{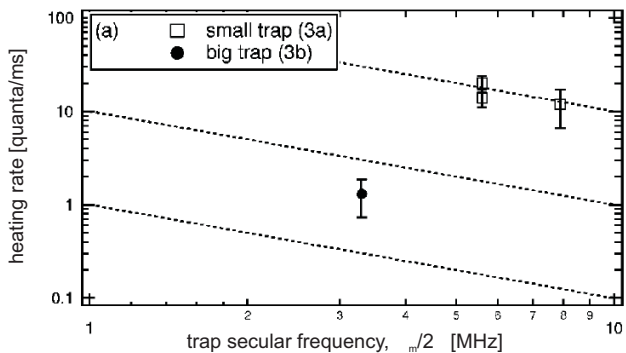}
\caption{Heating rate, $\dot{\bar{n}}$, as a function of trap secular frequency, $\omega_\mathrm{m}$ ($\equiv \omt$), for ring traps of two different sizes. The small trap (called 3a by \citeauthor{Turchette:2000}, and [$\Turchette00$]c in this review) had $d=125$\,$\mu$m. The large trap (called 3b, or [$\Turchette00$]d) had $d=280$\,$\mu$m. The dashed lines show scaling of $\dot{\bar{n}}\propto \omega^{-1}$ (i.e. $\alpha=0$). Assuming this frequency scaling, the data imply a distance scaling of $\beta=3.8(6)$. [Reprinted figure from \citet{Turchette:2000} and used with permission. \copyright 2000 by the American Physical Society.]}
\label{fig:Turchette00Fig6a}
\end{figure}

The distance scaling of noise in ion traps was first discussed by \citet{Turchette:2000}. They considered nine traps in total (six ring traps and three linear traps) ranging in size from $d=170$\,$\mu$m to $d=600$\,$\mu$m. For a controlled investigation of distance scaling they fabricated a pair of ring traps for which the RF electrodes of both traps were fabricated in a single piece molybdenum material. The traps' endcaps were made using a second, shared, piece of molybdenum. The two traps could be operated simultaneously, driven by the same electronics, and held in a single vacuum apparatus. They had ion-electrode separations of 125\,$\mu$m and 280\,$\mu$m respectively. (The initial results from these traps are referred to as [$\Turchette00$]c,d in this review.) The heating-rate results obtained by \citeauthor{Turchette:2000} for $^9$Be$^+$ ions in these two traps are shown in Fig.~\ref{fig:Turchette00Fig6a}. The traps were not operated at the same frequency but, assuming $\alpha=0$, \citeauthor{Turchette:2000} extrapolated the  heating rate to deduce a distance scaling of $\beta=3.8(6)$. The value of $\alpha=0$ was taken from measurements made in [$\Turchette00$]c, and it seems reasonable that [$\Turchette00$]d -- being fabricated in the same piece of metal and run by the same electronics -- might display the same behavior. While the uncertainty on the measurement of $\alpha$ is rather large, the general conclusion of \citeauthor{Turchette:2000} does not sensitively depend on assumptions about the frequency scaling: a value of $\alpha=1.5$, as observed for trap [$\Turchette00$]g, would imply a distance scaling (for the data shown in Fig.~\ref{fig:Turchette00Fig6a}) of $\beta\approx 5$. Considering the magnitude of the heating rates and the results of the size-scaling measurements they concluded that the observed heating was inconsistent with a thermal electronic noise source.

\begin{figure}[t]
\includegraphics[width=\columnwidth]{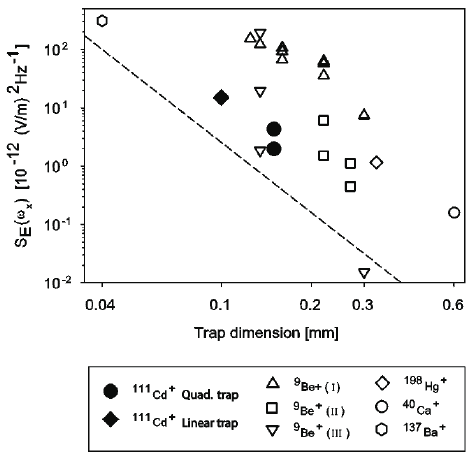}
\caption{Spectral density of electric-field noise, $S_\mathrm{E}$, at the position of the ion as a function of the distance, $d$, from the ion to the nearest electrode. Data are taken from \citet{Deslauriers:2004} ($^{111}$Cd$^+$), \citet{Turchette:2000} ($^{9}$Be$^+$ I and III), \citet{Rowe:2002} ($^{9}$Be$^+$ II), \citet{Diedrich:1989} ($^{198}$Hg$^+$), \citet{Roos:1999} ($^{40}$Ca$^+$), and \citet{DeVoe:2002} ($^{137}$Ba$^+$). The dashed line shows the $1/d^4$ scaling predicted by a simple patch potential model. [Reprinted figure  from \citet{Deslauriers:2004} and used with permission. \copyright 2004 by the American Physical Society.]}
\label{fig:Deslauriers04Fig6b}
\end{figure}

\citet{Deslauriers:2004} measured the heating rates of $^{111}$Cd$^+$ ions in a ring (quadrupole) trap and a linear trap. They concluded that, compared to the noise expected from a thermal (black-body or Johnson) source, the absolute heating rates observed were higher than expected by several orders of magnitude for the ring trap, and by a factor of twenty for the linear trap. Because of this higher-than-expected absolute value they inferred that the heating they saw was from a source other than thermal noise. In order to compare their results to other published heating rates they plotted their results along with those from thirteen other traps (seven ring traps and six linear traps) ranging from $d=80$\,$\mu$m to $d=600$\,$\mu$m in size.\footnote{The ring-trap result taken from \citet{DeVoe:2002} had an ion-electrode separation of of 80\,$\mu$m. The measurement provides only an upper limit on the heating rate.} The plot is reproduced here in Fig.~\ref{fig:Deslauriers04Fig6b}. From this data they concluded that the heating rates which they had measured using trapped $^{111}$Cd$^+$ ions were significantly lower than those measured in other traps of similar dimension and trap frequency. It may be noted that at no point did \citeauthor{Deslauriers:2004} demonstrate -- or claim to demonstrate -- that the combined data could be used to support a distance scaling of $d^{-4}$.

\begin{figure}[t]
\includegraphics[width=\columnwidth]{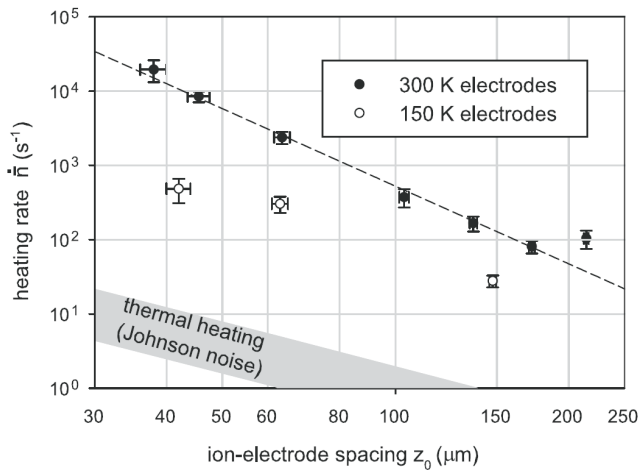}
\caption{Scaling of heating rates in a needle trap. Axial heating rate, $\dot{\bar{n}}$, as a function of the ion-electrode separation, $z_0$ ($\equiv d$). The dashed line, to which the measured data fit well, has a gradient of -3.5, corresponding to $\beta=3.5$. [Reprinted figure from \citet{Deslauriers:2006_09} and used with permission. \copyright 2006 by the American Physical Society.]}
\label{fig:Deslauriers06_09Fig4}
\end{figure}

To make a highly controlled investigation of distance scaling, \citet{Deslauriers:2006_09} used a trap which consisted of two needle-like, tungsten, RF electrodes with grounded sleeves. These were attached to axial translation stages, allowing the tip-to-tip separation (and consequently the ion-electrode separation) to be controllably varied over a wide range with micrometer resolution. They measured the heating rate for seven different ion-electrode spacings in the range $38\,\mu\mathrm{m}<d<216\,\mu$m. As they were able to vary the position of the trap electrodes in situ, this provides the only controlled measurement to date of heating as a function of electrode proximity which did not require a comparison across different trap structures, electrode materials or surface qualities. Their plot is reproduced here in Fig.~\ref{fig:Deslauriers06_09Fig4}. The measurements fit well to a scaling of heating rate with ion-electrode separation of $\dot{\bar{n}} \sim d^{-3.5\pm0.1}$. This corresponds to a scaling exponent of $\beta=3.5(1)$.
The distance scalings expected with a needle geometry for various noise sources are considered in Secs.~\ref{subsubsec:Johnson_DistanceScaling} and \ref{subsubsec:PlaneAndNeedleDiffusion}. Likely types of sources are discussed in Sec.~\ref{subsec:NeedleTrap}.

\begin{figure*}[t]
\includegraphics[width=\textwidth]{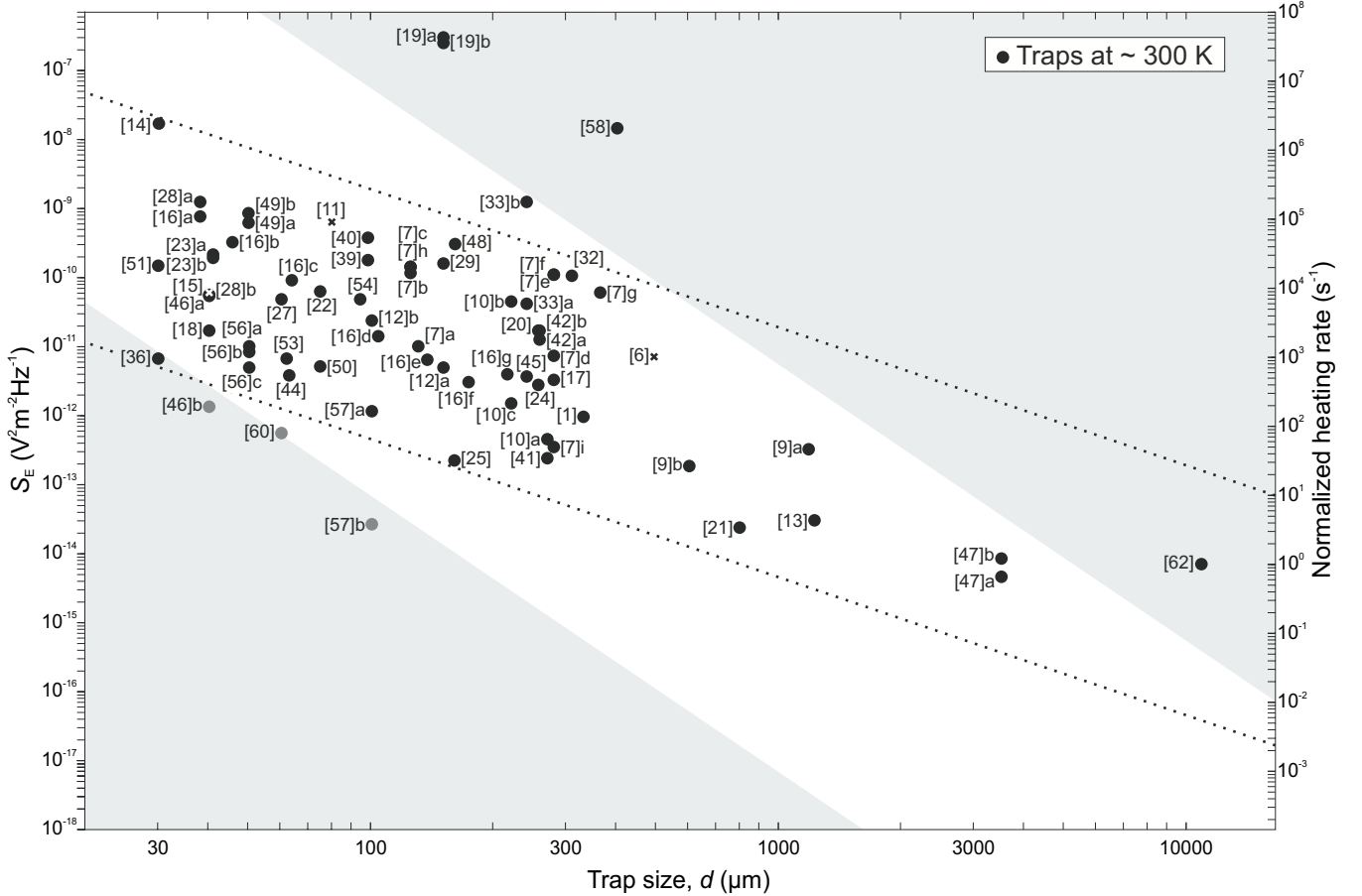}
\caption{Spectral density of electric-field noise, $S_\mathrm{E}$, as a function of the distance, $d$, from the ion to the nearest electrode, for traps operated nominally at room temperature. Data points are taken from the relevant references in table~\ref{tab:References}. On the right, the ordinate scale is given as the equivalent heating rate of a $^{40}$Ca$^+$ ion with a motional frequency of $\omt =2 \pi \times$1\,MHz. The shaded regions indicate an envelope scaling with $d^{-4}$. The dotted lines indicate an envelope scaling with $d^{-2}$. See Sec.~\ref{subsec:DistanceScaling} for discussion, including the uses and significant limitations of plotting such data on a single graph.}
\label{fig:S_E_d}
\end{figure*}

Following these investigations, several papers \citep[e.g.][]{Epstein:2007, Hughes:2011, Daniilidis:2011} compiled the results of multiple experiments into a single plot, and indicated that the combined data lent support to $d^{-4}$ scaling models of heating. Such plots have subsequently been widely invoked to claim that the experimental data show a $d^{-4}$ scaling \citep[e.g.][]{Dubessy:2009, SafaviNaini:2011, Herskind:2011, Herschbach:2012, Poulsen:2012, Pyka:2014}. The remainder of this section will construct comprehensive plots for all of the ion-trap heating-rate data currently available, highlighting the uses and the limitations of such plots. In particular, regarding the limitations of the plots, we observe that the full data set cannot be used to draw general conclusions about distance-scaling laws in ion traps. As a specific corollary to this, the full data set cannot be used to support the claim that, in general, heating in trapped-ion systems follows a $d^{-4}$ scaling.

Despite the claimed limitations, such plots can be useful in that they provide a map by which to orient discussions, and give a visual representation to aid comparisons of where different experiments lie relative to one another. To this end, Fig.~\ref{fig:S_E_d} shows the inferred spectral density of electric-field noise, $S_\mathrm{E}$, as a function of ion-electrode separation, $d$, in traps nominally operated at room temperature, for all available published data to date. To make it clear what is being displayed here, what the plot can be used to show, and what it cannot be used to show, we shall now discuss various aspects of the data compilation.\\

In order to best compare like with like, when measurements were made of several modes of motion in a single trap, the axial COM heating rate has been plotted here. When heating rates were taken over a range of frequencies, with all else being constant, the result for $\omt= 2\pi \times 1$\,MHz has been plotted if the frequency range included this value or, if it did not, the result for the frequency closest to $2 \pi \times 1$\,MHz has been plotted.

To keep the presentation as clear as possible, for the instances in which a paper reports repeated measurements taken in a single trap under essentially similar conditions, only a single data point (the median) has been included here. Similarly, if multiple publications include heating rates from the same trap under broadly similar conditions, only one result is included. Specifically, \citet{Monroe:1995_11, King:1998} and \citet{Myatt:2000} report heating rates for three traps which are also described by \citet{Turchette:2000} (plotted as points [\Turchette00]a,b,e here). \citet{SchmidtKaler:2000} and \citet{Roos:1999} report heating rates for traps which are also described in \citet{Rohde:2001} ([\Rohde01]a,b here). \citet{Blakestad:2009} and \citet{Blakestad:2011} discuss the same results ([\Blakestad09] here).  Finally, the results of \citet{Britton:2009} and \citet{Harlander:2011} were also reported in the theses of \citet{Britton:2008PhD} (trap [\Britton08]) and \citet{Harlander:2012PhD} (trap [\Harlander12]) respectively. In each case, only one result has been shown on the graph.

If heating rates were measured in a single trap under markedly different conditions, a data point is shown for each of the various sets of conditions. For instance, \citet{Turchette:2000} measured heating rates in two molybdenum ring traps which initially gave results which are plotted in Fig.~\ref{fig:S_E_d} as [$\Turchette00$]c and [$\Turchette00$]d. (The original data are also reproduced here in Fig.~\ref{fig:Turchette00Fig6a}). Following venting of the chamber and recleaning of the traps the behavior of both traps changed dramatically. The frequency scaling in trap [\Turchette00]c changed from $\alpha \approx 0$ to $\alpha = 6.0(2)$. The heating rate over the frequency range measured also dropped, by around two orders of magnitude at 10\,MHz. This is plotted as point [\Turchette00]h in Fig.~\ref{fig:S_E_d}. (The values of the electric-field noise plotted for points [\Turchette00]c and [\Turchette00]h in Fig.~\ref{fig:S_E_d} are very similar due to the difference in $\alpha$ between the two traps and the different frequencies at which the measurements were taken.) Trap [\Turchette00]d also underwent a significant reduction in heating rate, by around three orders of magnitude at 3\,MHz. The spectral density of electric-field noise is plotted as point [\Turchette00]i in Fig.~\ref{fig:S_E_d}.  After the change it was measured to have a frequency scaling of $\alpha=4.0(8)$. Given such unusual behavior, even without any clear understanding of exactly what it was which caused the behavior to change, two points have been plotted for each trap: before and after the change in behavior respectively. Despite plotting these points on a single graph, we note and stress the caveat, stated initially by \citeauthor{Turchette:2000} that it is difficult to draw general conclusions from the data for this particular trap.  These data points are discussed in more detail in Sec.~\ref{subsec:TurchettesWeirdTraps}.

Data points [$\Schulz08$] and [$\Poschinger09$] were taken for heating rates in the same trap, with the difference being attributed to an improved voltage supply \citep{Poschinger:2009}. (These two measurements are discussed in detail in Sec.~\ref{subsec:PowerSupply}.) Data points [$\Poulsen12$]a,b were also measured in a single trap, with [$\Poulsen12$]b being on resonance with interference for other lab equipment, while [$\Poulsen12$]a was away from any such resonances. It has been shown that surface treatment can make heating rates in a single trap higher ([$\Daniilidis11$]a,b) or lower ([$\AGH11$], [\Allcock12]). (These are discussed in detail in  Secs.~\ref{subsec:SurfaceEffects} and \ref{subsec:LaserCleaning} respectively.) Finally, \citet{Hite:2012}, \citet{Daniilidis:2014} and \citet{McKay:2014} have shown that removal of surface contamination by ion-beam cleaning of the electrodes can greatly reduce the heating rate. Where this has been done in a single trap, Fig.~\ref{fig:S_E_d} displays points for both the pre-cleaning ([$\Hite12$]a, [$\DGB13$]a) and post-cleaning ([$\Hite12$]b, [$\DGB13$]b) measurements. It may be considered that ion-beam cleaning is a sufficiently distinctive procedure that the heating rates from such traps should be analyzed separately. All points measured after some form of ion-beam cleaning are colored grey in Fig.~\ref{fig:S_E_d}. These measurements are then discussed in detail in Sec.~\ref{subsec:PlasmaCleaning}.

It is worth additionally highlighting data point [$\Goodwin14$] \citep{Goodwin:2014}. As well as being notable for the large trap size, the result is interesting as it is the only heating rate to date of a single ion in a Penning trap. This complicates any comparison with the other data shown here; while considerations for some heating mechanisms (such as Johnson noise and adatom diffusion) are essentially the same in both Penning and Paul traps, other mechanisms (such as issues concerning coupling to micromotion) become moot for Penning traps. Additionally there may be effects (such as coupling between magnetron and cyclotron motions\citep{Djekic:2004}) which have not been considered in this review, but which become important for Penning traps.

A final comment should be made regarding data points which have not been plotted. Measurements by \citet{Tamm:2000}, \citet{DeVoe:2002} and \citet{Amini:2010} are often included in compilations of heating rates. In fact these results provide only an upper limit to the observed heating rate, not the heating rate itself. These are plotted in Fig.~\ref{fig:S_E_d} as [$\Tamm00$], [$\DeVoe02$], and [$\Amini10$]b respectively, but are shown as crosses. Any heating rates for which the trap electrodes were at temperatures significantly below 300~K have been excluded from this plot. Results for cryogenic heating-rate measurements are shown in Fig.~\ref{fig:CryoS_E_d}.  \\

Having accounted for the selection of the data displayed, it is worth discussing how the data should most clearly be presented. The data shown have been measured in many different experiments, under many different sets of conditions. In what follows we discuss how the data might be standardized, both to best facilitate clear presentation and to allow -- as far as possible -- a meaningful comparison of like with like. Firstly, given the expected dependencies of the noise, rather than plotting $S_\mathrm{E} (d)$ it might be considered more appropriate to plot $\omega^{+\alpha} T^{-\gamma} S_\mathrm{E} (D)$. We therefore consider (Secs.~\ref{Subsubsec:NormalisingForFrequency} - \ref{subsubsec:NormalisingForDistance}) how the data might be best be standardized and displayed to properly account for frequency scaling, for the geometry dependence of distance scaling, and for temperature scaling. We then consider  (Sec.~\ref{subsubsec:OtherSystematicErrors}) a number of systematic effects which are not taken into account in a simple consideration of scalings. These may introduce apparent correlations where none exist. Alternatively they may mask or skew genuine correlations.

\subsubsection{Normalizing for frequency}
\label{Subsubsec:NormalisingForFrequency}

Given that traps are operated at different frequencies, and given that $S_\mathrm{E}$ is frequency dependent, it may be desirable to take this dependence into account. This was first done by \citet{Epstein:2007} who noted that there were several cases where the values of $\omega S_\mathrm{E}$ for a given trap were bunched together, and concluded that $S_\mathrm{E}\propto 1/\omega$ was a better assumption than $S_\mathrm{E}$ being independent of $\omega$. From the frequency dependencies shown in Fig.~\ref{fig:alpha_omega}, it can be seen that this is not unreasonable.

For the purposes of this review, however, we note that a significant number of data sets exist for which $\alpha \neq 1$. Additionally, for the majority of experiments, $\alpha$ has not been measured and is essentially unknown. In order to plot what is known, the figures throughout this paper show $S_\mathrm{E}(d)$, for the measured value closest to 1~MHz, but without any further normalization for frequency. It may be noted that there is no strong correlation between a trap's size and the motional frequency at which it is operated. Consequently, the different frequencies at which the traps are run does not systematically change any global trends in the data.

\subsubsection{Normalizing for temperature}
\label{subsubsec:NormalisingForTemperature}

There are two broad temperature regimes in which ion-trap experiments operate: `room temperature' and cryogenic temperature. `Room-temperature' traps are not deliberately temperature controlled. However, the presence of high-voltage RF applied to the electrodes can deposit large amounts of energy into the trap. Exactly how much energy is deposited will vary as a function of the applied voltage, the construction of the trap, and the attendant electronics. In vacuum this energy cannot be convected away. The degree to which it can be conducted away depends strongly on the way the trap is mounted. The efficiency of radiative transport will vary with, for example, the geometry, and the surface finish of the electrodes.

Given the differences in such factors between different experiments, the exact operating temperatures might be expected to vary significantly. For most `room-temperature' experiments there is no data regarding the trap temperature under normal operation. However, a number of experiments in which the electrode temperature has been measured can inform considerations. In a macroscopic trap ($d=800\,\mu$m) with stainless-steel electrodes held in a Macor mount, a temperature rise of $\Delta T \sim$\,100\,K was observed for typical operating conditions \citep{Chwalla:2009PhD}. In a trap of nominally identical geometry, but made using titanium electrodes held in a sapphire mount a temperature rise of $\Delta T \sim$\,2\,K was observed for typical operating conditions.\footnote{Personal communication, P. Schmidt, PTB, 2012.} In a gold-coated, SiO$_2$-on-Si, microfabricated trap ($d=240\,\mu$m) a temperature rise of $\Delta T \sim$\,3\,K was observed for typical operating conditions \citep{Wilpers:2012}. Considering these three data points, there is not a particular correlation between a trap's size and its operating temperature. For the sake of discussion, one may nonetheless wish to consider to what extent the data may be skewed if -- the above results notwithstanding -- there really were a strong correlation such that large traps (requiring high RF voltages) were systematically very hot, and small traps (with lower RF voltages and better thermal contact) were systematically closer to room temperature. Assuming a temperature scaling of $\gamma=2$ (see discussion in Sec.~\ref{subsec:TemperatureScaling}) even a rise of $\sim$140\,K would only lead to a factor of two increase in the heating rate. This is comparable with the trap-to-trap variation seen between nominally identical traps, and can be neglected for the purposes of most comparisons. For all intent and purposes `room-temperature' traps can therefore be reasonably considered to be at around 300\,K.

Cryogenic traps are typically held at around liquid-helium temperatures. [The exception to this being the first ever cryogenic experiment, for which the electrodes were estimated to be at 150\,K \citep{Deslauriers:2006_09}.] The effect of changing the temperature by around two orders of magnitude is significant (see discussion in Sec.~\ref{subsec:TemperatureScaling}). Heating-rate measurements at cryogenic temperatures are thus treated entirely separately from room-temperature measurements, and are not included in Fig.~\ref{fig:S_E_d}.

\subsubsection{Normalizing for geometry}
\label{subsubsec:NormalisingForDistance}

In the first paper considering a comparison of heating rates in multiple traps, \citet{Turchette:2000} correctly stated that the resistances, $R$, in the trap electrodes and connecting circuits give rise to an electric-field noise spectral density $S_\mathrm{E}(\omega)=4k_\mathrm{B}TR(\omega)/D^2$, where $D$ is the characteristic distance from the trap electrodes to the ion. They went on to consider an idealized trap geometry in which the electrodes form a spherical conducting shell of radius $d$ around the ion. Making the association $D\sim d$, they showed that a thermal electronic noise model gives a scaling $\dot{\bar{n}}\propto D^{-2}$ while a patch-potential model gives $\dot{\bar{n}}\propto D^{-4}$.

The association $D\sim d$ is reasonable for a spherical ion trap. However, as discussed in detail in Sec.~\ref{subsubsec:Johnson_DistanceScaling}, it cannot be assumed for an arbitrary trap geometry. Nonetheless, in the subsequent literature it has often been claimed that thermal electronic noise models predict a scaling $\dot{\bar{n}}\propto d^{-2}$, while the patch-potential model predict $\dot{\bar{n}}\propto d^{-4}$,  where $d$ is the distance between the ion and the closest trap electrode. The change from considering the characteristic distance, $D$, between the ion and the trap electrodes, to considering the distance, $d$, between the ion and the closest trap electrode is a very significant one, and unwarranted in the general case. While it happens to be true for a spherical trap that $D \propto d$, such a relationship does not generally hold for an arbitrary geometry.

One may wonder, given the importance of the characteristic distance for scaling considerations, why $S_\mathrm{E}(d)$, rather than $S_\mathrm{E}(D)$, is predominantly discussed in the literature. One reason for considering the ion-electrode separation is that it can be easily measured or calculated. By contrast, a trap's characteristic distance cannot generally be analytically calculated but must rather be simulated; moreover it can vary sensitively on the exact details of the geometry such as the curvature of the electrode edges which are -- in practice -- difficult to know accurately. As discussed in detail in Sec.~\ref{subsubsec:Johnson_DistanceScaling}, the characteristic length scale, $D_{i,j}$, is obtained by solving for the electric-field component along the $i$ principal axis at the equilibrium ion position, ${\bm r}_{\rm I}$, due to a voltage $V_j$ on the $j$-th trap electrode and using $E^{(j)}_i({\bm r}_{\rm I})=V_j/D_{i,j}$ \citep{Leibrandt:2007_01}. It is clear that if the heating originates from a particular feature in the trap -- for example the RF electrodes or the DC endcaps -- then the important figure is the characteristic distance \textit{of that electrode}. Thus, for example, if the dominant noise source is a thin layer of thermally driven adsorbates evenly distributed across a surface trap then the $j$-th electrode can be considered to be a uniform, infinite plane; the exact layout of the individual electrodes is irrelevant. This is in contrast to a situation in which the dominant heating is due to technical noise from voltage supplies for different electrodes of a segmented trap. Under such circumstances, at the position of the ion, the noise contributions originating from different places on a single electrode will be correlated,  while the noise originating from different electrodes will be uncorrelated. In this instance the characteristic distance will therefore not only depend on the electrode geometry, but will even be different for different positions in the same trap \citep{Leibrandt:2007_01}. Furthermore, even at a particular position, for a given geometry and specified noise model, the characteristic distance may still not be uniquely constrained: in the case of fluctuating patches of finite size, for example, the characteristic distance is dependent on the size of the patch \citep{Low:2011}. It can thus be seen that the value of the characteristic distance, $D$, for any given trap depends both on the type and details of the noise model considered. Consequently, any attempt to infer a noise model from the scaling of $S_\mathrm{E}(D)$ is, to a certain extent, circular: a particular noise model is necessarily already implicit in the calculation of $D$.

Some papers include a trap's characteristic distances (assuming a certain noise model), or sufficient information to calculate them. However, the vast majority give only the ion-electrode separation. For comparisons aimed at elucidating heating mechanisms, simply stating the ion-electrode separation has significant limitations. In an ideal world, the characteristic distance of all historic traps would be calculated for a number of noise models, and plotted. In practice this is not feasible, though the contribution of future work to the discussion would be greatly enhanced by including the characteristic distance of each new trap for a selection of likely noise models.

Aside from the practical and fundamental difficulties with plotting $S_\mathrm{E}(D)$, there are very good pragmatic reasons in favor of plotting $S_\mathrm{E}(d)$. When considering such issues as what fabrication techniques might be suitable for making a trap, how tightly a laser beam must be focussed to avoid being clipped by electrodes, or what additional apparatus can be integrated near the trap structure, the ion-electrode separation provides a very useful measure of the trap size. If one wishes to compare the usefulness of various traps for a particular application, and neglecting any attempt to understand the physics underlying the noise, then the ion-electrode separation is, pragmatically speaking, a sensible parameter to consider.

For these reasons, and accepting the necessary limitations attendant with such a decision, the plots in this review show $S_\mathrm{E}(d)$.

\subsubsection{Other systematic effects}
\label{subsubsec:OtherSystematicErrors}

In the considerations outlined above, the additional variables for which one is attempting to control can be stated as some parameter. There exist, however, a number of other considerations which are not so simple to quantify. These are discussed in turn here.\\

\noindent $\bullet$ Systematic differences by group\\
        Some groups -- for a variety of reasons -- build systematically smaller traps than others. Different groups also see different sources of noise (as suggested in Sec.~\ref{subsec:FrequencyScaling} and elaborated upon in Sec.~\ref{sec:KnownMechanisms}). If the absolute levels of heating from the various sources are different (even taking all other scalings into account) then a systematic connection between the size of trap that a group uses and the type of noise they observe could give rise to, or alter, any apparent trends in the scaling. Without knowing in advance what the different absolute levels are, it is difficult to even estimate the magnitude of the effect this may have. \\

\noindent $\bullet$ Material differences\\
        It is possible, even likely, that different materials have different noise properties. This may be because the dominant noise mechanism varies between different materials. Alternatively it may be because a single noise mechanism presents very differently in different materials. For example, considering noise sources related to crystal grain boundaries in the electrodes, the grain structure in gold electrodes will be different to the grain structure in stainless steel electrodes. Gold is commonly used as an electrode material for traps of $d\sim100$\,$\mu$m, while stainless steel is commonly used for traps of $d\sim1000$\,$\mu$m. A systematic connection between the trap size and trap material, coupled with a connection between trap material and noise, would potentially cause systematic shifts in the observed relationship between trap size and noise level. Without knowing in advance what the sources of noise for each material are, it is difficult to even estimate the magnitude of the effect this may have. \\

\noindent $\bullet$ Different reporting conventions\\
        There are many different problems which could cause an increase in trap heating rates. In order to compare what is ultimately limiting the experiment, it can therefore be argued that one should compare the lowest heating rate from each experiment \citep{Wang:2012PhD}. Alternatively, it can be argued that given some distribution of results one should quote the average value \citep{Britton:2008PhD}. Given that some experiments observe a variation of several orders of magnitude from day to day in a single trap \citep{Wang:2012PhD} this difference in reporting methods is not necessarily insignificant. Without knowing the exact reporting method used for each experiment it is difficult to predict how this might affect the overall result, if at all. It would be helpful if future contributions to the literature would state which reporting convention was used, or give both the lowest and the average (arguably the median) results.\\

\noindent $\bullet$ The importance of the absolute heating rate\\
        While it may be interesting to say that one trap is better or worse than another `scaled for size', in most applications, such as quantum information processing, it is the absolute heating rate that is important. For example, the implementation of a M\o lmer-S\o rensen gate made by \citet{Kirchmair:2009} would require a heating rate below $\sim$100 s$^{-1}$ in order to obtain a gate infidelity of less than 10$^{-3}$. This requirement is independent of the size of the trap in which the gate is implemented. Groups using small traps may therefore be motivated to achieve heating rates which are as low as possible, while groups using larger traps may have little need to improve over rates which are \textit{low enough}. This may skew the reported heating rates to make any apparent scaling with distance exhibited by the full data set smaller than the actual scaling of the physical mechanisms involved.\\

\noindent $\bullet$ Selection bias\\
        It is often reported that a trap's heating rate is `typical for traps of this size.' It is reasonable that a group whose central interest is not heating-rate studies will work to improve a trap's operation until it is `typical', and then stop working so hard to improve it. However, such data may then be indicative of what was expected, rather than normative for what is possible. This does not rule out the eventuality that various groups may have reached fundamental limits; however, it is a reminder that not all reported results are necessarily at such a limit, nor do they all claim to be. A selection bias would tend to skew the reported data towards showing whatever scaling trend is expected.\\

\subsubsection{Summary}

Two experiments have been performed which attempted controlled comparisons of heating rates as a function of trap size: \citet{Turchette:2000} compared the heating rates of two similar traps which had different ion-electrode separations. These two traps exhibited heating rates which suggested $\beta$=3.8(6). \citet{Deslauriers:2006_09} measured heating rates in a single trap for which the ion-electrode separation could be varied in situ. This allowed $d$ to be varied, while ensuring that other aspects of the experiment which may have otherwise influenced the result, such as the electrode material, the electrode surface finish, driving electronics and so forth, were unchanged. This experiment gave results fitting well to a distance-scaling exponent of $\beta=3.5(1)$.

Compilations of heating rates from other experiments provide a useful visual overview of the topic. However, they have a number of practical and fundamental limitations regarding their possible use to infer any scaling laws that would shed light on the mechanisms underlying the observed heating. As different experiments may be limited by different sources of noise, it is not necessarily expected, and has not been experimentally demonstrated, that there should be a generally applicable distance-scaling law for heating in ion traps.

\subsection{Temperature scaling}
\label{subsec:TemperatureScaling}

Assuming that trapped ions are heated by some thermal, or thermally activated, effect in the trap, it seems reasonable that reducing the temperature of the trap would reduce the ion-heating rate. This is of interest for two reasons. Firstly, because the change of the observed level of noise with temperature would shed light on the underlying mechanism (as discussed in Secs.~\ref{sec:Theory} and~\ref{sec:Microscopic}). Secondly, because a reduction in the heating rate -- without having to increase the trap frequency or the trap dimensions -- would be useful in itself. This section considers a number of experiments which have measured the ion-heating rate over a range of temperatures, and thereby determined values of the temperature-scaling exponent, $\gamma$, for particular traps. It then considers the difference in the heating rate observed between `typical' traps operated at room temperature and `typical' cryogenic traps.

The first experiment to cool down the trap electrodes \citep{Deslauriers:2006_09} showed that reduced electrode temperatures did indeed lead to reduced ion-heating rates. The trap consisted of two needle-like RF electrodes with grounded sleeves. The needles were cooled to approximately 150\,K through contact with a liquid-nitrogen reservoir. This reduced the heating rate of a single $^{111}$Cd$^+$ ion by around an order of magnitude compared to room temperature. This would suggest a temperature scaling of $\gamma \sim 3$. Since then, numerous experiments have been carried out at several institutions with traps operated at around 6\,K. The traps are held in liquid $^4$He cryostats, and the power dissipated by the trap typically raises the temperature of the trap itself a few degrees above 4\,K.

\begin{figure}[t]
\begin{center}
\includegraphics[width=\columnwidth]{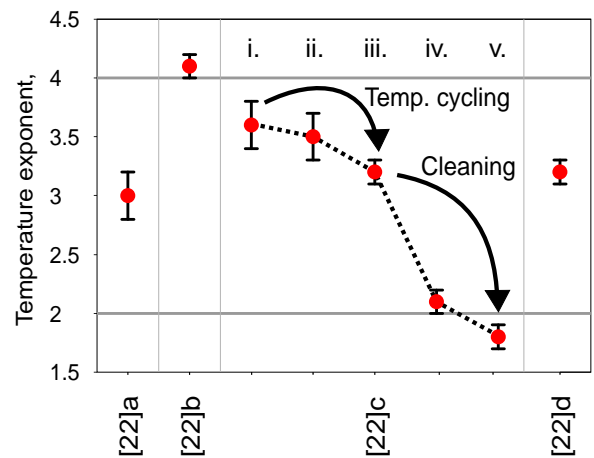}
\caption{Measured values of the temperature exponent, $\gamma$, for four different traps \citep{Labaziewicz:2008_10}. Traps [$\LGL08$]a,b,d were measured only once at cryogenic temperatures. For trap [$\LGL08$]c five measurements were made. Following measurement i the trap was cycled from 7\,K to 130\,K, without breaking vacuum. Following measurement ii the trap was cycled from 7\,K to 340\,K without breaking vacuum. Following measurements iii, and iv the trap was brought to room temperature, cleaned with lab solvents in air, and then recooled.}
\label{fig:gamma}
\end{center}
\end{figure}

\begin{figure}[t]
\begin{center}
\includegraphics[width=\columnwidth]{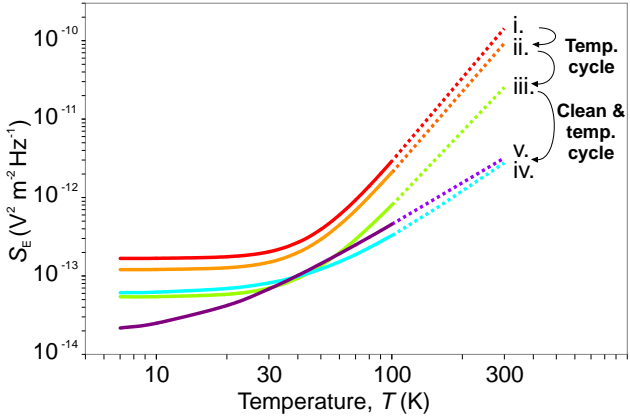}
\caption{Spectral density of electric-field noise, $S_\mathrm{E}$, as a function of temperature, $T$, for a single trap ([$\LGL08$]c) following different treatments. $S_\mathrm{E}(T)$ was inferred from ion-heating rates taken between 7\,K and 100\,K \citep{Labaziewicz:2008_10}. The published fit parameters [Table I of \citet{Labaziewicz:2008_10}] have additionally been extrapolated here to room temperature (dotted lines).}
\label{fig:S_E_T}
\end{center}
\end{figure}

\citet{Labaziewicz:2008_10} studied four different traps in detail. All four traps were made of gold electrodes on crystal-quartz substrates and had an ion-electrode separation of 75~$\mu$m. For one of the traps they also took repeated measurements following various cleaning and handling procedures. In each case they measured both the temperature-scaling exponent, $\gamma$, and the absolute heating rate as a function of temperature over the range 7\,K-100\,K. They found that the spectral density of electric-field noise was not simply proportional to some exponential increase of the form $S_\mathrm{E}\propto T^{+\gamma}$ as assumed in the naive ansatz proposed in Eq.~\eqref{eqn:S_E_Dependencies} above. Rather, it (usually) took the form $S_\mathrm{E} \propto S_\mathrm{0}[1+(T/T_\mathrm{0})^{+\gamma}]$. The eight different measured values of $\gamma$ are plotted in Fig.~\ref{fig:gamma}. The five fit curves for trap [$\LGL08$]c are also plotted in Fig.~\ref{fig:S_E_T}. The four traps, prepared and operated in nominally identical fashions, initially had temperature exponents in the range  $3 \lesssim \gamma \lesssim 4$. These are similar to the results obtained by \citet{Deslauriers:2006_09}. Beyond this, there are several points worth noting.

Firstly, $\gamma$ varied as a function of processing. Not only did $\gamma$ change following different processing steps for trap [$\LGL08$]c, but initially measured values $\gamma$ for the four different traps were different at a level significantly greater than the uncertainty on each individual point. This suggests that some -- as yet unidentified -- factors in trap preparation play a role.

Secondly, rather than a simple exponential decrease with decreasing temperature, the noise level plateaued below around 30\,K, at a value of $\sim 0.1\times 10^{-12}$\,\B. All of the traps (with the exception of [$\LGL08$]c-v) reached this floor at similar temperatures. They also all (with the exception of [$\LGL08$]d) exhibited similar absolute noise levels having reached this floor [in the range $0.02 \times 10^{-12} < S_\mathrm{E}/($\B$)< 0.2 \times 10^{-12}$]. It can thus be seen, at least in the instance of these traps, that the variation of several orders of magnitude in the heating rate at room temperature is less strongly reflected in the cold-temperature behavior. Possible causes of such a noise floor are discussed in Sec.~\ref{subsec:CryoNoiseFloor}. Here it is simply noted that such a noise floor is observed.

\begin{figure*}[t]
\begin{center}
\includegraphics[width=\textwidth]{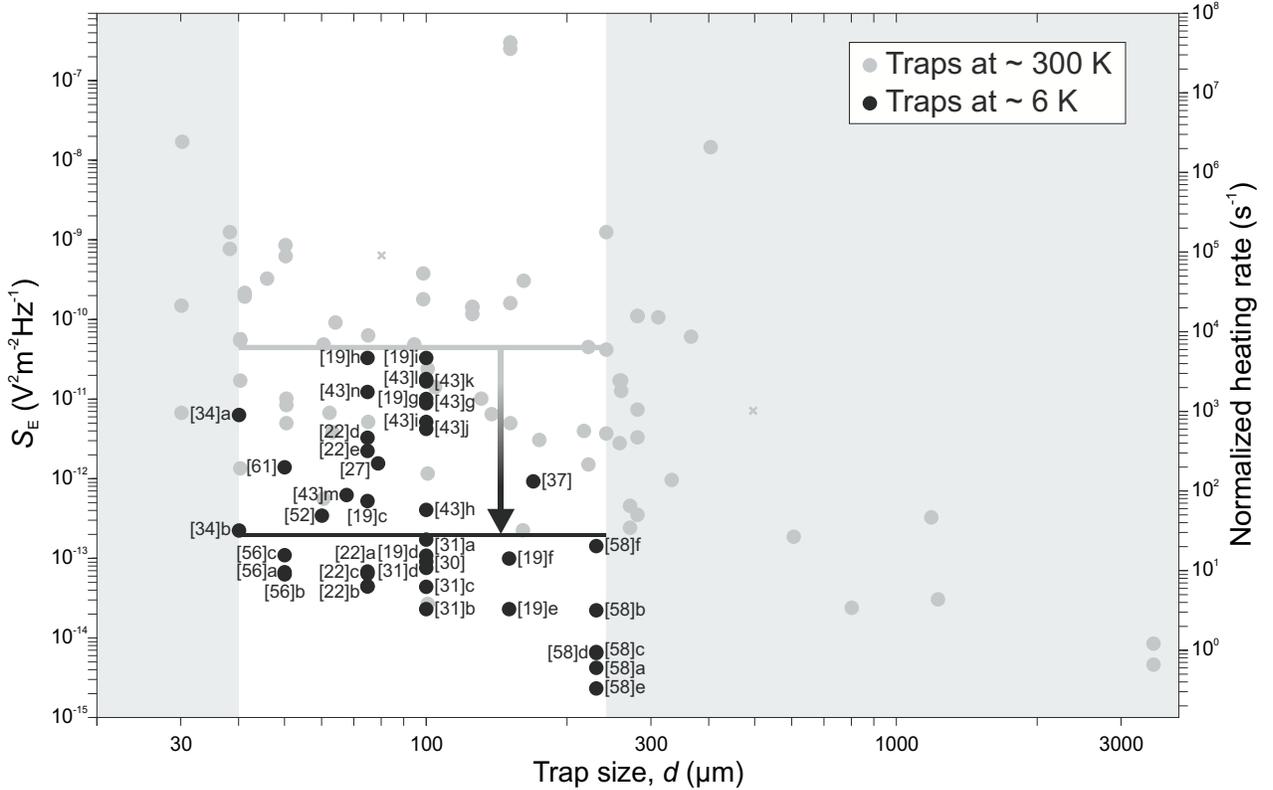}
\caption{Spectral density of electric-field noise, $S_\mathrm{E}$, as a function of the distance, $d$, from the ion to the nearest electrode, for traps at cryogenic temperatures (all nominally around 6\,K). Data points are taken from the relevant references in table~\ref{tab:References}. The gray points are the same room-temperature data as shown in Fig.~\ref{fig:S_E_d}, given here for comparison. On the right, the ordinate scale is given as the equivalent heating rate of a $^{40}$Ca$^+$ ion with a motional frequency of $\omt =2 \pi \times 1\,$MHz. The white area ($30\mu\mathrm{m}\leq d \leq 230\mu$m) shows the size range over which cryogenic measurements have been made. The solid black line marks the median spectral density of electric-field noise measured in  cryogenic traps. The solid gray line marks the median spectral density of electric-field noise measured on room-temperature traps over the same size range.}
\label{fig:CryoS_E_d}
\end{center}
\end{figure*}

Thirdly, traps which started with higher heating rates also had higher values of $\gamma$. Whatever effect caused the reduction in heating rate at room temperature, there is a concomitant reduction in the strength with which heating rates scale as a function of temperature. This means that one cannot necessarily expect to independently combine the expected improvements of cleaning with that of cryogenic temperatures. Consequently, while traps with very high heating rates at room temperature (e.g. [$\Labaziewicz0801$]a,b in Fig.~\ref{fig:S_E_d}) see a very large reduction in heating rates when cooled to cryogenic temperatures (see [$\Labaziewicz0801$]e,f in Fig.~\ref{fig:CryoS_E_d}), claims of improvement by seven \citep{Labaziewicz:2008_01} or even eight \citep{Antohi:2009} orders of magnitude are not likely to be the rule for most traps.\\

\begin{figure}[t]
\begin{center}
\includegraphics[width=\columnwidth]{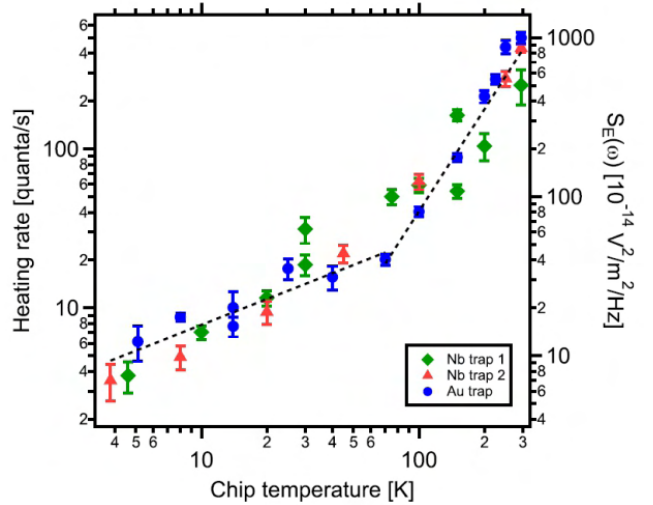}
\caption{Spectral density of electric-field noise, $S_\mathrm{E}$, as a function of temperature, $T$, for traps made of different materials. The levels of heating are similar in the three traps, despite the different trap-electrode materials. The dotted lines are fits to power laws in temperature to all data. Above 70~K the scaling exponent is $\gamma$=2.13(5). Below 70~K it is 0.54(4). [Reprinted figure from \citet{Chiaverini:2014} and used with permission. \copyright 2014 by the American Physical Society.]}
\label{fig:S_E_Tii}
\end{center}
\end{figure}

\citet{Chiaverini:2014} studied three traps over an even larger temperature range of 4~K$<T<$295~K. They measured the axial component of the electric-field noise, acting on single $^{88}$Sr$^+$ ions. All three traps had ion-electrode separation of 50~$\mu$m and sapphire substrates. The electrodes of two traps were made of 2~$\mu$m of sputtered niobium, while those of the third were made of 500~nm of thermally evaporated gold. The results are reproduced in Fig.~\ref{fig:S_E_Tii}.

Like \citet{Labaziewicz:2008_10} they observed a reduction of around two orders of magnitude in the noise between room temperature and cryogenic temperatures. Despite this, the details of the scaling behavior over the temperature range was qualitatively different from that seen by \citet{Labaziewicz:2008_10}. Assuming that the similar absolute levels of heating indicated a single source of heating in the three traps, despite the different materials, and fitting the behavior of the three traps by the same scaling parameters, they inferred a value of $\gamma$=2.13(5) above 70~K, and $\gamma$=0.54 below that temperature.

Both \citet{Labaziewicz:2008_10} and \citet{Chiaverini:2014} observed temperature scaling behavior in which the scaling exponent was smaller at lower temperatures. However, the plateau seen in most of the traps of \citeauthor{Labaziewicz:2008_10} was not reproduced by Chiaverini and Sage. Additionally, the critical temperature at which the scaling behavior changed was different between the two groups. \\

It is not clear at this stage that the dominant source of heating -- either in the high- or low-temperature regimes -- is the same for the two sets of experiments. Despite this, it would be interesting to know, even without necessarily identifying the cause of heating in each instance, what degree of heating-rate reduction might be expected by operating a trap at cryogenic temperatures. Pragmatically speaking, such a number is particularly interesting when setting up an experiment: without knowing in advance what source of noise might limit the experiment at any given stage of development, it would be useful to estimate what order of magnitude improvement might one expect to gain by investing in a cryostat.

The observed levels of noise for cryogenic traps as a function of ion-surface distance are shown in Fig.~\ref{fig:CryoS_E_d}. The data were selected using similar criteria as for the room-temperature data shown in Fig.~\ref{fig:S_E_d}, as described in Sec.~\ref{subsec:DistanceScaling}. If a publication reports repeated measurements taken in a single trap under essentially similar conditions, only a single data point (the median) has been plotted here. Similarly, if multiple publications include heating rates from the same trap under broadly similar conditions, only one result is included. If measurements were taken at different frequencies, the measurement result taken closest to 1~MHz has been shown. If measurements were taken at different temperatures, the measurement taken closest to 6~K has been shown. All of the caveats laid out in Sec.~\ref{subsec:DistanceScaling} regarding Fig.~\ref{fig:S_E_d} also hold here for Fig.~\ref{fig:CryoS_E_d}.

The possibility of different physical mechanisms underlying the noise in the different traps, whereby each experiment might therefore be expected exhibit a scaling which is essentially unknown, means that any fit to the data is broadly meaningless. In order to obtain a quantitative comparison between the data sets for the two different temperatures, a heuristic way of identifying the `typical' level of noise seen in a cryogenic experiment must be used. To this end the median spectral density of the electric-field noise, $\tilde{S}_\mathrm{E}$, is considered. Given the traps are reasonably symmetrically distributed across the size range, this is not an unreasonable measure. Considering all measurements so far made, this is $\tilde{S}_\mathrm{E}(T=6~\mathrm{K})=0.2\times 10^{-12}$~\B, and is indicated in Fig.~\ref{fig:CryoS_E_d} by a solid black horizontal line. To identify a typical level of noise for room-temperature traps of a similar size to cryogenic traps, the median value of the spectral density of electric-field noise  for all room-temperature measurements made on traps in the range $30\mu\mathrm{m}\leq d \leq 230\mu$m was calculated, and found to be $\tilde{S}_\mathrm{E}(T=300~\mathrm{K})=40\times 10^{-12}$~\B.

This change in noise level is in line with the estimate of two orders of magnitude for good traps which was made previously by considering the data of \citet{Labaziewicz:2008_10} and \citet{Chiaverini:2014}. It should, however, be noted that there are numerous possible noise sources which may limit experiments, and that different sources may scale differently with temperature. It is therefore not necessarily the case that the same reduction will be seen in all experiments. Very early work on cryogenic traps \citep{Labaziewicz:2008_01} saw that, in two traps with very high heating rates at room temperature, the rates are suppressed by 7 orders of magnitude by cooling them to 6\,K. The absolute heating rates were then 2 orders of magnitude lower than the heating rates reported for typical traps of similar size operated at room temperature. This initial result is shown by subsequent data to be reasonably typical: bad traps may be improved dramatically, though good traps might only be expected to have an improvement of around two orders of magnitude. Other specific noise sources may, however, behave differently again.

\section{Selected Sources of Electric-Field Noise}
\label{sec:Theory}

Having presented, in Sec.~\ref{sec:ExperimentalOverview}, an overview of the main results and trends observed experimentally, the following two sections (\ref{sec:Theory} and \ref{sec:Microscopic}) consider various sources of electric-field noise which may contribute to the heating of trapped ions.

The current section considers several common sources of electric-field noise which are related to the electromagnetic environment and the trapping circuitry. Starting with arguably the most fundamental source of noise, Sec.~\ref{subsec:Blackbody} addresses the effect of black-body radiation; firstly in free space and then above surfaces. The actual level of noise observed in free space is generally very much higher than this black-body value, due to electromagnetic interference from both man-made and natural sources. Direct coupling of the ion to such fields is considered in Sec.~\ref{subsec:EMI}. Background electromagnetic radiation can also couple to parts of the experimental apparatus, and from there perturb the ion. This can potentially couple noise into the system much more efficiently than by simple free-space coupling directly to the ion. This effect is termed electromagnetic (EM) pickup and is discussed in Sec.~\ref{subsec:EMPickup}. Moving on to noise originating in the experimental components themselves, Sec.~\ref{subsec:JohnsonNoise} considers Johnson-Nyquist noise. In the simplest case, this is the thermal noise of the electrons in resistive elements in the experiment, and this section considers a number of often-overlooked resistive elements. Additionally, Sec.~\ref{subsubsec:TechnicalNoise} considers the effects of technical noise, taken here to mean noise from  equipment such as power supplies, and coupled to the experiment though the wiring. While this is often not considered strictly Johnson noise, it is included here due to the similar distance scaling that it shares with Johnson noise. Finally, Sec.~\ref{subsec:SpaceCharge} considers the noise that can arise when the vacuum is no longer neutral.  Specifically, it considers the effect of  surface-emitted electrons which perturb the ion as they move through free space towards an anode.

Inspired by the observations of \citet{Turchette:2000} much recent work has been carried out on a class of models in which such electric-field noise arises from variations of the electrode potential on a microscopic scale. These are considered in Sec.~\ref{sec:Microscopic}, together with a discussion of several microscopic mechanisms which have been suggested in the literature to explain the physical origin of this noise.

The major results of both Sec.~\ref{sec:Theory} and Sec.~\ref{sec:Microscopic} are summarized in  App.~\ref{Appendix_Summary}. The appendix is intended to help maintain an overview and shows indicative values only. The reader should always refer back to the discussions provided in the main paper.

\subsection{Black-body radiation}
\label{subsec:Blackbody}

A charged particle in a trap is subject to heating by fluctuating electromagnetic (EM) fields and, even in the absence of nearby electrodes or noisy circuits, the ion is affected by freely propagating electromagnetic radiation. A fundamental source for a finite electromagnetic background is black-body radiation. While the level of this noise is far too low to play a significant role in ion-trap experiments, it is important to calculate as it provides the baseline used for calculating and discussing the level of noise due to  other sources of electromagnetic interference. This section first considers  black-body radiation in free space, and then considers how this noise is modified in the vicinity of conducting surfaces such as trap electrodes.

\subsubsection{Black-body radiation in free space}
\label{subsubsec:FreeSpaceBB}

If the frequency of the electromagnetic radiation, $\omega\ll k_\mathrm{B} T/\hbar$,  then the single-sided spectral energy density of black-body electromagnetic radiation in vacuum is $u_\mathrm{em}(\omega)=2 k_\mathrm{B} T \omega^2/( \pi c^3)$ \citep{LLStatPhys:1980}, where $k_\mathrm{B}$ is Boltzmann's constant, $T$ is the temperature, and $c$ is the speed of light. The total energy of the field is equally shared between the electric and magnetic energy densities, $\epsilon_0 \langle \bm{E}^2\rangle/2$ and $ \langle \bm{B}^2\rangle/(2\mu_0)$, where $\epsilon_0$ and $\mu_0$ are the  the electric permittivity and magnetic permeability in free space, respectively.  By taking into account the fact that, for an isotropic system, each spatial component of the electric field provides 1/3 of the total power the (single sided) electric-field noise spectrum along a specific trap axis is~\citep{Henkel:1999_11}
    \begin{equation}
    \label{eqn:BB}
    S^{\mathrm{(BB)}}_\mathrm{E}=\frac{2k_\mathrm{B} T\omega^2}{3 \epsilon_0 \pi c^3}.
    \end{equation}
At room temperature and frequencies of $2 \pi \times 1$\,MHz, this gives a spectral density of electric-field noise of approximately $10^{-22}$\,\B.  This is far below the observed level of noise in ion-trap experiments (cf. Fig.~\ref{fig:S_E_d}).  It is nonetheless important to know this level, as black-body radiation at room temperature is the reference point for EMI or ambient RF noise, as explained below.

\subsubsection{Black-body radiation above surfaces}
\label{subsubsec:SurfaceFields}

Compared to black-body radiation in free space, the presence of a nearby surface can substantially enhance the electric-field background, due to finite losses of electromagnetic radiation in dielectrics or metals. According to the dissipation-fluctuation theorem these losses are associated with an spectral density of electric-field noise, $S_\mathrm{E}$, which, in thermal equilibrium and for the conventions used in this review, can be expressed as \citep{Agarwal:1975}
    \begin{equation}
    S^{({\rm BBS})}_\mathrm{E}= \frac{4\hbar}{e^{\hbar\omega/k_{\rm B}T}-1} {\rm Im} \, G_{\rm E}({\bm r}_{\rm I},{\bm r}_{\rm I},-\omega).
    \end{equation}
In this expression $G_{\rm E}({\bm r},{\bm r}',\omega) = \sum_{ij} {\bm e}_{{\rm t},i} {\bm e}_{{\rm t},j}  G_{\rm E}^{ij}({\bm r},{\bm r}',\omega)$, where the electric Green's function, $G_{\rm E}^{ij}({\bm r},{\bm r}',\omega)$, determines the electric-field components ${\bm E}_i({\bm r},\omega)$ emitted from an oscillating dipole at ${\bm r}'$, i.e. ${\bm E}_i({\bm r},\omega) = G_{\rm E}^{ij}({\bm r},{\bm r}',\omega){\boldsymbol \mu}_j({\bm r}^\prime,\omega)$. For simple geometries the Green's function can be evaluated analytically~\citep{Agarwal:1975}. For an ion trapped at a distance $d$ above an infinite plane with dielectric constant $\epsilon(\omega)$ and in the quasi-static limit, the imaginary part of the Green's function is approximately given by \citep{Henkel:1999_06}
    \begin{equation}
    \label{eq:ImGiiStatic}
    {\rm Im} \, G_{\rm E}(-\omega)= \frac{\omega^3}{6\epsilon_0 \pi c^3} + \frac{s_{ij}}{16\pi\epsilon_0 d^3} {\rm Im} \, \frac{\epsilon(\omega)-1}{\epsilon(\omega)+1},
    \end{equation}
where $s_{ij}$ is a diagonal tensor with $s_{xx}=s_{yy}=1/2$ and $s_{zz}=1$. The first term in this expression corresponds to the free-space black-body noise spectrum. For a metallic surface the dielectric function at low frequencies can be approximated by $\epsilon(\omega)\simeq 2ic^2/(\omega^2\delta_s^2)$, where $\delta_s(\omega)=\sqrt{2\epsilon_0 c^2 \rho_{\rm e}/\omega}$ is the skin depth for a metal with resistivity $\rho_{\rm e}$. Under such conditions, ${\rm Im} \frac{\epsilon(\omega)-1}{\epsilon(\omega)+1}\simeq 2\omega \rho_{\rm e} \epsilon_0$ and the second term in Eq.~\eqref{eq:ImGiiStatic} can be interpreted as the damping of the ion's image dipole.
The quasi-static limit given in Eq.~\eqref{eq:ImGiiStatic} above is only valid for distances $d<\delta_s(\omega)$ which, for gold and at MHz frequencies, corresponds to about $80$\,$\mu$m. An approximate result for the spectral density of electric-field noise which applies also for larger distances is given by \citep{Henkel:1999_11}
    \begin{equation}
    S^{({\rm BBS})}_\mathrm{E}= S^{({\rm BB})}_\mathrm{E}  + s_{ii}\frac{k_{\rm B}T\rho_{\rm e}}{2\pi d^3}\left( 1+ \frac{d}{\delta_s}\right),
    \end{equation}
where already the high-temperature limit, $k_{\rm B}T \gg \hbar \omega$, has been assumed. At a distance of $d=100$\,$\mu$m and $T=300$\,K, where $\rho_{\rm e}=2.44\times 10^{-8}\,\Omega$m, the level of electric-field noise is around $S^{({\rm BBS})}_\mathrm{E}\approx  10^{-17}$\,\B, scaling as $d^{-2}$ for larger distances and as $d^{-3}$ for an ion approaching the surface. Although the presence of a surface significantly enhances black-body radiation, the overall noise level is still not enough to explain the observed heating rates (cf. Fig.~\ref{fig:S_E_d}).

\subsection{Electromagnetic interference}
\label{subsec:EMI}

Radio-frequency engineers have long known that at low- (30-300\,kHz), medium- (0.3-3\,MHz) and high-  (3-30\,MHz) frequency bands, the level of black-body radiation is too low to be responsible for the noise level observed on antennas.  The ion motional frequency of most ion-trap experiments is in the medium-frequency (MF) band, and such radiation is pervasive: sources as diverse as lightning strikes in the tropics \citep{Parrot:2009}, the solar wind hitting the ionosphere \citep{Bunch:2009} and data being sent along power transmission lines \citep{Hansen:2003} cause noise which is near the secular frequency of many ion-trap experiments. Moreover, MF radiation is hardly attenuated by the atmosphere, natural materials, or buildings, and not efficiently radiated into space as it is reflected by the ionosphere.

The standard measure of this EMI is called the external noise factor,  $F_\mathrm{a}$, which is a dimensionless factor, defined as the extra noise that a perfect antenna receives above the black-body radiation noise at a temperature of $T=300$\,K. The power spectral density of the electromagnetic interference is then $S^{\mathrm{(EMI)}}_\mathrm{E}=F_\mathrm{a} \times S^{\mathrm{(BB)}}_\mathrm{E}$. Radio engineers have measured the external noise factor's statistics.  It varies greatly, both spatially and temporally.  In the MF band, outdoors, away from man-made structures, $F_\mathrm{a}$ is about 60\,dB \citep{ITU-R:2009}, and in cities it is approximately 80\,dB, decreasing with frequency as $\omega^{-3}$.  At 1\,MHz, natural sources have an $F_a$ range of 0-100\,dB, 99.5\% of the time.  i.e. 0.5\% of the time natural sources exceed 100 dB.  The electric-field noise levels for $F_\mathrm{a}=80$\,dB at 1\,MHz is then $S^{\mathrm{(EMI)}}_\mathrm{E} \sim 10^{-14}$\,\B, decreasing as $\omega^{-1}$. This level of noise is seen in some ion-trap experiments (cf. Fig.~\ref{fig:S_E_d}).

Worse yet, inside commercial buildings, where most laboratory experiments would be carried out, there is even more electrical noise at these frequencies than is found outdoors \citep{Fernandez:2010}. The external noise factor in commercial buildings has been measured to be approximately 120\,dB at 1\,MHz and falls off with $\omega^{-5}$ \citep{Landa:2011}. An unshielded ion trap exposed to `typical' levels of indoor noise would therefore see $S^{\mathrm{(EMI)}}_\mathrm{E}  \sim 10^{-10}$\,\B falling off with $\omega^{-3}$.

Of course, trapped ions are surrounded by electrodes,  in a vacuum vessel made of metal, which shields them to some extent.  Nonetheless, EMI from ambient noise in buildings cannot be ruled out as a source of direct ion heating  and care must be taken to shield the ion trap so that EMI does not affect the experiment.

$F_\mathrm{a}$ is measured relative to the black-body background, which scales as $\omega^{2}$ [see Eq.~\eqref{eqn:BB}]. This means that the frequency-scaling exponents of  $F_\mathrm{a}$ and $S^{\mathrm{(EMI)}}_\mathrm{E}$ differ by~2. For an indoor experiment, where the external noise factor, $F_\mathrm{a}$, falls as $\omega^{-5}$, one might expect the spectral density of electric-field noise, $S^{\mathrm{(EMI)}}_\mathrm{E}$, to scale in free space as $\omega^{-3}$. However, the ion is partly shielded by the electrodes, and also shielded by the vacuum chamber and other conductors [such as heat shields in cryogenic systems \citep{Brown:2011_09}]. Electric-field shielding via conductors works best at lower frequencies (the opposite being true for magnetic-field shielding) and so, while shielding can provide substantial reductions in the expected heating rate, it would also change the expected frequency scaling of the electric-field noise. The effect of this EMI shielding should be taken into account \citep{Miller:1966, Bridges:1988, Rajawat:1995}.  Furthermore, any strong sources at specific frequencies in the noise spectrum have not been included in this analysis: radio stations and electronic devices can emit strong signals which would be further above the levels mentioned above. These would manifest themselves as peaks in the power spectrum and would be ephemeral and/or geography dependent.

The heating due to direct electromagnetic interference would be expected to be independent of the trap size provided the shielding did not change as a function of trap geometry. In practice, changing the geometry of the electrodes will almost certainly change the shielding at the ion, unless care is taken to shield the entire experiment from EMI.  Some non-zero value of the distance-scaling exponent, $\beta$, would therefore be expected, but this would have to be calculated for each specific apparatus.

Obviously, the background black-body radiation level scales linearly with temperature [see Eq.~\eqref{eqn:BB}]. However, the level of environmental EMI impinging on an experiment would not be expected to vary with temperature. That said, if the ion trap has metallic shielding which is cooled, the shielding's effectiveness will change significantly as its conductance increases with decreasing temperature. At low temperatures this will eventually reach a plateau when the conductance no longer changes.

\subsection{Electromagnetic pickup}
\label{subsec:EMPickup}

As explained in Sec.~\ref{subsec:EMI}, electric fields from EMI can directly interact with an ion in a trap. As well as interacting with the ion directly, it is also possible that EMI causes voltage fluctuations on the wires and electronics of the experiment, which then carry this noise to the trap. These voltage fluctuations caused by EMI coupling to wires are referred to as electromagnetic (EM) pickup, and are considered in detail here. The electric component of EMI can be well shielded by metallic conductors. However, the magnetic-field noise from EMI is not as easily shielded and can be picked up by any conductive loops in the trap wiring. An example of where such loops may occur in ion traps is the wiring leading to the RF electrodes: in a linear-trap configuration, a single RF source is usually split to drive two trap electrodes. The capacitance between the two RF electrodes then completes the loop.  According to Faraday's law, any changing magnetic flux perpendicular to the surface formed by a loop will induce an electromotive force (EMF) around the loop.  Since the small capacitance between the two trap electrodes has a very large impedance, this is where the full voltage drop of the induced EMF will present itself.

The frequency scaling, $\alpha$, of noise due to EM pickup would be dependent upon the scaling of the underlying EMI radiation. Beyond this, it must be taken into account that the voltage $V_\mathrm{L}$, at frequency $\omega$, induced in a loop of wire from an incident magnetic field, $B$, normal to the loop, also at frequency $\omega$, is given by \citep{Kanda:1993}
    \begin{equation}
    V_\mathrm{L}=\omega B A_\mathrm{L},
    \end{equation}
where $A_\mathrm{L}$ is the loop area.  The spectral densities of magnetic- and electric-field noise are related by $S_\mathrm{B}=\epsilon_0\mu_0 S_\mathrm{E}$. Using the results discussed above, the spectral density of the EM-pickup voltage noise $S^{\mathrm{(PU)}}_\mathrm{V}$ on the electrode is
    \begin{equation}
    S^{\mathrm{(PU)}}_\mathrm{V}=F_{\rm a}\frac{2 \mu_0 k_\mathrm{B} T A^2_\mathrm{L} \omega^4}{3 \pi c^3}.
    \end{equation}
This means that, for the same free-field noise, $\alpha$ for EM pickup would be smaller by two than for direct coupling of the free fields to the ion [cf. Eq.~\eqref{eqn:BB}]. Thus, if the EMI scaled with  $\alpha=3$ (as is expected for noise indoors, see Sec.~\ref{subsec:EMI}) then the EM pickup would scale with $\alpha=1$.

At a given frequency and temperature, the noise on the electrodes caused by EM pickup is indistinguishable -- from the ion's point of view -- from Johnson noise. Consequently, the spectral density of electric-field noise above the surface scales as $D^{-2}$ where $D$ is the characteristic length scale of the electrode. This is discussed more fully under Johnson noise in Sec.~\ref{subsec:JohnsonNoise}.

As EM pickup is caused by the EMI interaction with the electronics of the ion trap, temperature might, at first, not seem to be relevant.  However, if there is any shielding in the experiment, the shielding's effectiveness is dependent upon temperature.  Typically, as temperature is lowered, the resistance is lowered and the shield becomes more effective at damping magnetic fluctuations, so that for a typical experiment $\gamma>0$.

The above discussion of EM pickup explains what sort of power spectral density of voltage noise can be expected on a trap electrode.  The electric-field noise, to which this will give rise at the position of the ion, is explained below under Johnson noise in Sec.~\ref{subsec:JohnsonNoise}.  For completeness, we give an estimate here of the absolute noise level expected from EM pickup: Considering an environment (typical office building) with $F_\mathrm{a}=120$\,dB, the voltage noise due to pickup on an unshielded 10~cm diameter loop of wire will be $S^{\rm (PU)}_\mathrm{V}\sim10^{-18}$\,V$^2$/Hz.  This is equivalent to the Johnson noise of a 200\,$\Omega$ resistor at room temperature. If this voltage noise is on a trap electrode with a characteristic length scale of $D=100$\,$\mu$m, then the electric-field noise will be $S^{\rm (PU)}_\mathrm{E}\sim10^{-10}$\,\B.  This level will scale in proportion to the square of the area enclosed by the loop.

\subsection{Johnson-Nyquist noise}
\label{subsec:JohnsonNoise}

Johnson-Nyquist noise (or Johnson noise) is the electrical noise generated by thermal motion of charge carriers in a conductor.  It has been shown \citep{Johnson:1928, Nyquist:1928} that the power spectral density of such voltage noise is given by
    \begin{equation}
    S^{(\mathrm{JN})}_ \mathrm{V}=4 k_\mathrm{B} TR(\omega,T),
    \label{eq:JohnsonNoise}
    \end{equation}
where $R(\omega,T)$ is the effective real resistance, at frequency $\omega$, of the whole circuit from the two terminals across which the voltage noise is observed. What constitutes the \textit{whole circuit} is broadly a matter of convention. Some analyses consider only the trap electrodes, others consider the associated passive electronics, and still others the full system up to and including active components such as power supplies. The Johnson noise due to the bulk resistance of the trap electrodes themselves is almost always negligibly low. However, as shown below, Johnson noise can appear due to losses in other elements in the experiment that are often overlooked.

For the purposes of analysis, some authors lump several noise sources together, modeling both technical noise (discussed here in Sec.~\ref{subsubsec:TechnicalNoise}) and EM pickup (discussed here in Sec.~\ref{subsec:EMPickup}) as Johnson noise. This can be done by considering Eq.~\eqref{eq:JohnsonNoise} and replacing the resistance, $R$, by a (much) larger effective resistance \citep{Lamoreaux:1997}, or alternatively, a resistance at a (much) higher effective temperature \citep{Wineland:1998_05}.  Here we shall consider Johnson noise only from electrical elements on the circuit which leads to the trap electrodes.

The spectral density of the electric-field noise due to Johnson-Nyquist noise is
    \begin{equation}
    S^{(\mathrm{JN})}_\mathrm{E}=\frac{S^{(\mathrm{JN})}_\mathrm{V}}{D^2}=\frac{4 k_\mathrm{B} TR(\omega,T)}{D^2},
    \label{eqn:char_len}
    \end{equation}
where $D$ is the characteristic length scale of a particular electrode in the ion trap.

A simplistic interpretation of Johnson noise may assume that it exhibits an essentially flat noise spectrum ($\alpha=0$) and that the $D^2$ term in the denominator of Eq.~\eqref{eqn:char_len} would imply $\beta=2$. Given many metals have $R \propto T$, one might also anticipate $\gamma=2$. Finally, assuming trap electrodes to have a resistance of $\sim 0.1\,\Omega$ one might expect the electric-field noise due to Johnson noise at 300\,K, in a trap of $D=100\,\mu$m to be around $S^{(\mathrm{JN})}_\mathrm{E}\sim10^{-13}$\,\B. Such an analysis is, however, rather too simplistic to describe the situation in realistic ion-trap experiments.

In what follows we consider a number of physical effects which cause the Johnson-noise analysis to be more complex than the simple picture just presented. First, the frequency dependence of Johnson noise is considered (Sec.~\ref{subsubsec:Johnson_FrequencyScaling}). We then consider the distance scaling (Sec.~\ref{subsubsec:Johnson_DistanceScaling}) - specifically how the characteristic length, $D$, relates to the ion-electrode separation, $d$, for various geometries. This is followed by sections on the temperature scaling  (\ref{subsubsec:Johnson_TemperatureScaling}) and absolute levels (\ref{subsubsec:Johnson_Absolute}) of Johnson noise. Finally, consideration will be given to technical noise (Sec.~\ref{subsubsec:TechnicalNoise}).

\subsubsection{Frequency-dependent resistance}
\label{subsubsec:Johnson_FrequencyScaling}

It can immediately be seen from Eq.~\eqref{eqn:char_len} that the frequency scaling, $\alpha$, of Johnson noise depends on the frequency scaling of $R(\omega)$ which is flat for standard ideal resistors, i.e. $\alpha = 0$. However, electrical circuits connected to the trap almost always contain frequency-dependent impedance elements such as filters. Consequently, such frequency dependencies must be considered carefully as it is, in principle, possible to achieve any value of $\alpha$.

Since we are considering ion heating due to the electric-field noise at the trap's secular frequency, the Johnson-noise-induced heating would be dependent upon the real part of the impedance seen from the trap electrodes at that frequency \citep{Leibrandt:2007_01}.  It is easy to underestimate this resistance and it is nearly always much higher than the direct current (DC) resistance of the circuit.  This is because frequency-dependent elements are always lossy; even superconductors have AC (alternating current) resistance \citep{Zar:1963}.

Idealized capacitors, inductors and transmission lines (wires) have, as their main characteristics, only reactances (imaginary impedances). However, the materials which make up these devices are not perfect and have losses. The impedance of a wirewound inductor is an obvious example: a wirewound inductor typically has a very low DC resistance, but if the current is kept constant while the frequency of the current is increased, the inductor will heat up.

Real dielectrics are also lossy, characterized by a loss tangent, $\tan{\theta}$.  Any capacitance incorporating a dielectric has an equivalent parallel resistance (EPR) given by \citep{Vandamme:2010}
            \begin{equation}
            \mathrm{EPR}=\frac{1}{\tan{(\theta)} \omega C},
            \end{equation}
where $\omega$ is the angular frequency and $C$ is the capacitance of the capacitor. A large EPR would be considered characteristic of a high-quality capacitor and have little associated noise.  Only by modeling the rest of the circuit connected to the capacitor, can the equivalent series resistance (ESR) be computed.  The ESR is then easily understood as the resistance which goes into Eq.~\eqref{eqn:char_len} to determine the Johnson noise. We now consider the ESR that might reasonably be expected from a number of components.

While capacitors used in the trap electronics may be the most obvious instance of such loss in dielectrics, the dielectrics which hold the trap must also be considered. Macroscopic traps often have their electrodes mounted on Macor \citep{Sinclair:2001, Rohde:2001} which has relatively high dielectric losses. This may be a possible cause of the significant dissipation observed in some ion-trap systems \citep{Chwalla:2009PhD}. Surface traps usually consist of thin ($\sim$1\,$\mu$m) metallic electrodes on dielectric substrates such as quartz \citep{Seidelin:2006}, silica \citep{Allcock:2011} or sapphire \citep{Daniilidis:2011}. Excellent (low-loss) dielectric materials have loss tangents of $\sim10^{-5}$. Typically the ion-trap electrodes have a capacitance of just a few picofarads, which means that at trap frequencies of $2 \pi \times$1\,MHz, even in the best case,  the equivalent parallel resistance will not be higher than 10\,G$\Omega$. If such  a trap electrode were to be isolated from its voltage source at the trap frequency, this would give rise to an ESR of $\simeq1$\,$\Omega$.   To keep Johnson noise below such a level requires careful attention to the drive electronics and filters so that they absorb this noise from the electrodes, while filtering out the noise invariably coming from the power supplies.

Losses from magnetic components can be treated in a very similar manner (again, characterized by a loss tangent). This can be seen, for instance, in the case of a layer of nickel, which can be used as a barrier layer between gold and copper. Being magnetic it can cause substantial losses at microwave frequencies \citep{Shlepnev:2011}. This is less likely to be a significant issue for ion traps as, in most ion-trap systems, magnetic materials are generally avoided due to concerns over magnetic-field instability affecting the ion's electronic states.  More commonly, magnetic components in ion-trap systems include wirewound inductors inside filters or resonators which lead to the trap electrodes.  The quality factor, $Q$, relates an inductor's real and imaginary impedances as
    \begin{equation}
    \label{eq:inductor-Q}
    Q=\frac{\omega L}{R_\mathrm{L}},
    \end{equation}
where $L$ is the inductance of the inductor and $R_\mathrm{L}$ is the real part of the inductor's impedance.  The loss tangent is related to the quality factor by $\tan{\theta}=1/Q$.  If a filter inductor, $L=100$\,$\mu$H, is intended to work at $\omega=2\pi\times1$\,MHz, and has a quality factor of $Q=20$, then it has an ESR of $\sim30$\,$\Omega$ at this frequency.  This resistance would then need to be correctly accounted for in any Johnson noise analysis.

Eddy-current losses in conductors near a wire carrying RF and microwave frequencies can also cause the effective resistance of the wire to increase substantially above that expected from the bulk DC resistance \citep{Poritsky:1954}.  At medium RF frequencies,  and at   a distance $\simeq100$\,$\mu$m away from a non-magnetic conductor, this effect would give an ESR on the order of only 1\,m$\Omega$ per cm of wire.  This effective resistance could be enhanced by several factors.  It increases roughly with the logarithm of the magnetic permeability of the nearby conductor.  Multiple turns of the current-carrying wire would increase the effect roughly with the logarithm of the number of turns squared.  At 1\,MHz the effect saturates at around 0.1\,$\Omega$/cm.  However, as the frequency is increased, the effective resistance due to eddy-current losses scales proportionally without limit.  Details are given by \citet{Poritsky:1954}.  One situation in which this effect is of particular relevance to ion-trap experiments is when a shield is placed over a helical resonator or wirewound inductor.  The resulting drop in the inductor's $Q$ is due to these eddy-current losses.

\subsubsection{Characteristic length scale}
\label{subsubsec:Johnson_DistanceScaling}

The characteristic length of a system, $D$, determines the conversion of voltage fluctuations to electric-field fluctuations, as defined in Eq.~\eqref{eqn:char_len}. Electric-field-noise measurements in ion traps usually report the minimum ion-electrode distance, $d$. While this is often much simpler to determine, it is generally quite different from the characteristic length, $D$.  Indeed, for some noise sources and geometrical symmetries the characteristic length can tend to infinity. This would, for instance, occur due to common-mode rejection in the case of perfectly correlated noise on the RF electrodes, provided the ion was perfectly localized on the RF null.

For an arbitrary trap geometry, the electric field from voltages on electrodes in vacuum can be computed using the Laplace equation,
    \begin{equation}
    \nabla^2\Phi=0,
    \end{equation}
where $\Phi$ represents the potential everywhere in space, and the voltages on the electrodes represent the boundary conditions for solving this differential equation. The electric field is then given by $\bm{E}=-\nabla\Phi$.

In general, there can be several characteristic length scales, corresponding to the different components of the electric field and to the particular trap electrode under consideration.  If a voltage is applied to the $j$-th electrode, and the $i$-th component of the resulting electric-field $\bm{E}^{(j)}_i$  is calculated, then the corresponding characteristic length scale is
    \begin{equation}
    D_{i,j}=V_j/\bm{E}_{i}^{(j)}.
    \end{equation}
Closely related to the characteristic dimension is a dimensionless quantity called the dipole geometrical efficiency factor  [sometimes simply called a geometrical efficiency factor, \citep{Deslauriers:2006_09}], $\kappa_{i,j}$. This relates the ion-electrode separation to the characteristic distance as
    \begin{equation}
    \label{eq:kappa}
    \kappa_{i,j}=\frac{2d_j}{D_{i,j}}=\frac{2d_j \bm{E}_{i}^{(j)}}{V_j},
    \end{equation}
where $d_j$ is the distance of the ion from the $j$-th electrode.  The factor of 2 in Eq.~\eqref{eq:kappa} ensures that the efficiency is equal to 1 for a pair of parallel plates, where the ion is trapped halfway between.

For the simple geometries considered in this section, symmetry provides that there is only one direction in which the ion can be heated.  The heating is also entirely due to just one voltage (or voltage difference) of interest.  For simplicity, the subscripts for $\kappa$, $d$, $D$, $V$, or $E$ will therefore be omitted for the remainder of this section. The results, however, can be readily generalised to more complicated structures. If the geometry of a trap is uniformly scaled then $\kappa$ is constant, $d$ is proportional to $D$,  and $\beta=2$. If, however, some dimensions of a trap are changed while others remain constant, then one must solve for the electric field as a function of the changing geometry to calculate $\kappa$, $D$ and $\beta$.  There exist a limited number of geometries for which the characteristic distances can be analytically calculated; in general it must be found numerically. Three geometries are considered here: surface, spherical and needle geometries. In the first instance they are all calculated analytically. A more complicated -- and realistic -- needle geometry is then calculated numerically.\\

\noindent $\bullet$ Planar geometry\\
An often-used idealization for determining the effects of voltage noise on the electrodes is to approximate the trap by two parallel and infinitely extended plates \citep{Lamoreaux:1997, Wineland:1998_05, Wineland:1998_06, James:1998_02, Henkel:1999_11}.  This configuration approximates the situation for an ion trapped above a planar microtrap, or more generally describes the limit in which the ion-surface distance is much smaller than  any features on the trap such as the extent or curvature of the electrode. The electric field normal to two plates separated by a distance $s$, and due to a voltage difference $V$, is simply $V/s$.   In any real geometry consisting of large but finite planes there exists some far-field ground, such as the vacuum chamber. As the theoretical model considers infinite planes there is no quadrupole term and no pseudopotential minimum in which the ion might be trapped. The position of the ion between the plates is thus arbitrary, though we can choose to place the ion at $d=s/2$, so that the characteristic length scale is $D=s=2d$ and $\kappa=1$.  The distance scaling of the electric-field noise is then $\beta=2$.\\

\noindent $\bullet$ Spherical geometry\\
Another idealized geometry consists of two spherical electrodes of equal radius, $R_\mathrm{el}$, held a distance $s$ apart, and a far-field ground. In this geometry the ion is trapped halfway between the two spheres. Provided $s \gg R_\mathrm{el}$, the charge distribution on the surface of the spheres will give the same electric field half way between the two spheres as that of two point charges located at the spheres' centers.  This configuration is sometimes used to approximate a needle Paul trap or the endcaps of a linear Paul trap.  The electric-field  at the center of the ion trap would then be
    \begin{equation}
    E=\frac{C V}{4\pi\epsilon_0 d^2},
    \end{equation}
where $C$ is the capacitance of the two conducting spheres, $V$ is the voltage difference between them, and $d=s/2$.  The capacitance of one sphere in such a two sphere system, $C\simeq2\pi\epsilon_0(R_\mathrm{el}+R_\mathrm{el}^2/s)$ \citep{Lekner:2012}, allows for an analytical solution of the field to be found in the limit $s \gg R_\mathrm{el}$. The characteristic length scale is then $D=2d^2/R_\mathrm{el}$. The dipole geometrical efficiency factor is $\kappa=R_\mathrm{el}/d$ and the distance-scaling exponent  is $\beta=4$.\\

\begin{figure}[t]
\begin{center}
\includegraphics[width=\columnwidth]{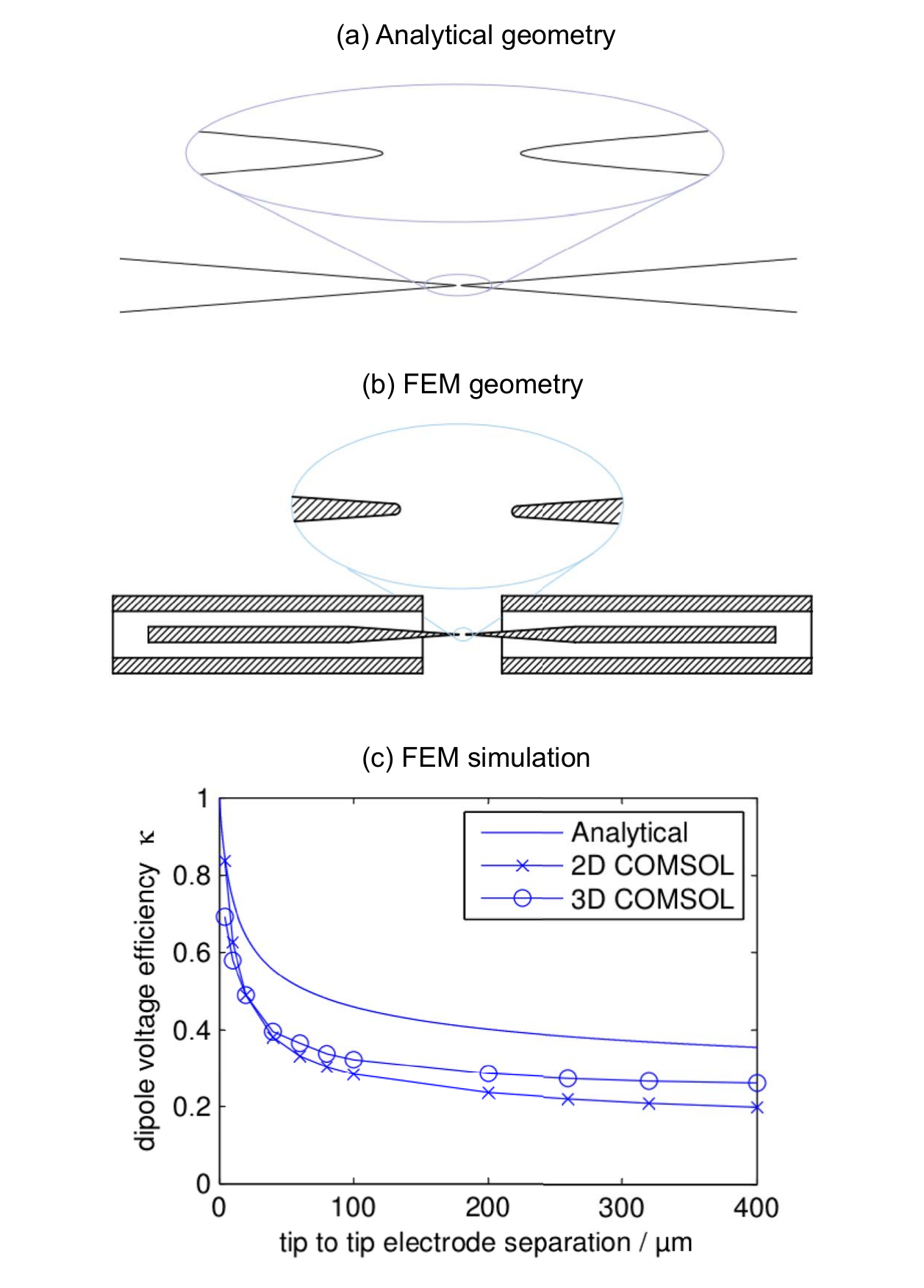}
\caption{Distance scaling of Johnson noise for a needle trap.
(a) Electrode geometry considered for the analytic result given in Eq.~\eqref{eq:needle-analytic}.
(b) Electrode geometry used for the numerical calculation of the electric field at the center of a needle trap. The taper angle is 4$^\circ$ and a conducting ground sleeve, 3\,mm inside diameter, is recessed 2.3\,mm from the tips. This geometry was chosen to approximate the experiment of  \citet{Deslauriers:2006_09}.
(c) The dipole efficiency factor $\kappa$ defined in Eq.~\eqref{eq:kappa} is plotted as a function of the tip separation.}
\label{fig:needle-sims}
\end{center}
\end{figure}

\noindent $\bullet$ Needle geometry\\
Another geometry which has been studied in ion-trap experiments is that of two needle electrodes \citep{Deslauriers:2006_09}. The ion is trapped halfway between the two needle tips.  The distance, $d$, between the ion and the needle tip can be varied, and the expected electric-field noise due to voltage noise on the electrodes can be calculated by solving the Laplace equation. If the needle electrodes are approximated by hyperbolic surfaces of revolution [see Fig.~\ref{fig:needle-sims}(a)], then the Laplace equation for the electric field can be solved in prolate spheroidal coordinates \citep{Chen:1996}.  This approximation has the advantage that it can be solved analytically, though it suffers from the limitation that the radius of curvature of the needle tips cannot be independently specified of the taper angle of the needle tips. As with the planar geometry above, the absence of a far-field ground in the mathematical model means that no trap is formed with the application of RF voltage to the tips. However, for the purposes of calculating the heating rate due to voltage noise on the needle tips, the model does provide some insight.

Consider two hyperbolic needle tips with radius of curvature $R_\mathrm{el}$, and separated by a distance $2d$. The electric field half-way between them is given by
    \begin{equation}
    \label{eq:needle-analytic}
    E=\frac{V}{d}\frac{v_0}{\ln\left(\frac{1+v_0}{1-v_0}\right)},
    \end{equation}
where $v_0=\sqrt{1/(1+R_\mathrm{el}/d)}$, and $V$ is the potential difference between the two tips.  This simplified needle geometry has a characteristic length scale $D=d \ln(\frac{1+v_0}{1-v_0})/v_0$, which can be fitted to a power law, in order to estimate the distance scaling, $\beta$,  for a particular trap geometry. Taking $R_\mathrm{el}=3$\,$\mu$m and approximating Eq.~\eqref{eq:needle-analytic} over the range 30\,$\mu$m $< d <$ 200\,$\mu$m with a power law of the form $d^{-\beta}$ gives $\beta\cong2.4$.  This is sensitively dependent on parameters such as tip radius of curvature, $R_\mathrm{el}$, and the exact range of distances, $d$.  For the above example, if $d$ has an uncertainty of $\pm$5\,$\mu$m and $R_\mathrm{el}$ has an uncertainty of $\pm$1\,$\mu$m, then $\beta=2.4(2)$.

For more realistic electrode geometries the electric field due to a voltage difference between the two tips must be calculated numerically. This can be done using, for example, finite-element modeling (FEM) software such as COMSOL. Figure~\ref{fig:needle-sims}(b) shows a geometry similar to that described by \citet{Deslauriers:2006_09}. The needle electrodes have a tip radius of curvature $R_\mathrm{el}=3$\,$\mu$m and a taper angle of 4$^{\circ}$. Also included are ground sleeves of 3\,mm inside diameter, recessed 2.3\,mm from the needle tips and electrically isolated from the needles with an alumina tube. Considering the dipole efficiency factor, $\kappa$, Fig.~\ref{fig:needle-sims}(c) compares the numerically evaluated results\footnote{Using COMSOL Multiphysics v3.4.} to those obtained analytically from Eq.~\eqref{eq:needle-analytic}. This numerical example confirms the overall trend of $\kappa$ also for a realistic electrode configurations and, at least for this specific geometry, the absolute value deviates only by a factor of two from the analytic prediction. By fitting the numerically calculated characteristic distance, $D$, to a power law and assuming that the measured ion-electrode separation has an uncertainty of $\pm$5\,$\mu$m [which is the uncertainty claimed by \citet{Deslauriers:2006_09}] then the predicted value of $\beta$ is 2.5(2). The comparison of these simulation results with experimental results, are discussed in detail in Sec.~\ref{subsec:NeedleTrap}.

\noindent $\bullet$ Surface roughness\\
A final point worth noting in connection to the distance scaling of Johnson noise concerns the effect of the electrode surface not being geometrically smooth. Small features on the electrode could give rise to a local enhancement of Johnson noise. For instance, if there is a small cylindrical cone with a spherical end-tip protruding above the surface then, assuming its tip radius, $r_\mathrm{cc}$, is much smaller than its height, its resistance is given by \citep{Deslauriers:2006_09}
\begin{equation}
R=\frac{\rho_{\rm e}}{\pi r_{\rm cc} \tan{\theta}_{\rm cc}},
\end{equation}
where $\rho_{\rm e}$ is the bulk resistivity and $\theta_{\rm cc}$ is the angle of the cone. Therefore, a small crystal of gold with a radius of a few nanometers at the tip, growing from the surface of an ion trap, could be equivalent to a nanometer-sized electrode having a resistance of several ohms. If the surface is populated with patches of a poor conductor, such as an oxide \citep{Harlander:2010, Wang:2011}, then small patches with orders-of-magnitude higher resistances could be expected. The noise due to such patch potentials -- while having its origin in Johnson noise -- would have a very different distance scaling. The distance scaling of patch potentials is discussed in Sec.~\ref{sec:Microscopic}.

\subsubsection{Temperature scaling of Johnson noise}
\label{subsubsec:Johnson_TemperatureScaling}

From Eq.~\eqref{eqn:char_len} it should be expected that the level of Johnson noise varies with temperature. This picture is complicated by the temperature dependence of the various resistances involved in the system.  Many metals' resistance varies as $R\propto T$. This would suggest, in the simplest case, that $\gamma=2$. That being said, at low temperatures, materials can depart significantly from such simple behavior \citep{Ekin:2007}. This applies both to the simple materials such as the trap electrodes and also to any electronic components which may be held at cryogenic temperatures. Finally, it should be noted that even if the trap itself is cooled to very low temperatures, not all of the attendant electronics may be at the same temperature. This will complicate the picture of any temperature scaling.
Ultimately, each individual experiment must be modeled properly to check what value of $\gamma$ would be expected, and how that value might change in different temperature regimes.

\subsubsection{Absolute values of Johnson noise}
\label{subsubsec:Johnson_Absolute}

The calculation of the expected field noise above a surface has been extensively modeled by \citet{Leibrandt:2007_01}.  The main results of which are plotted in Fig.~\ref{fig:KnownMechanisms} as points [A] and [B].  Point [A]a at $\sim10 \times 10^{-12}$\,\B corresponds to the expected axial electric-field noise 60\,$\mu$m above a surface-electrode segmented linear ion trap, due to the Johnson noise of the electrode material and attendant filter electronics [see Fig. \ref{fig:TrapGeometries}(h)]. The calculation was made for an ion situated directly above the gap between two electrode segments, such that it is sensitive to the uncorrelated nature of the noise between the two electrodes. The calculated level of noise in this instance is comparable to the level of noise observed in a number of traps of similar size. Point [A]b, at $\sim1 \times 10^{-12}$\,\B corresponds to the expected axial electric-field noise in the same trap as before, but at a position above the center of a segment. As the ion is above the center of an electrode and Johnson noise on the electrode is correlated the axial heating rate is lower.

An instance where Johnson noise is less dominant is indicated by point [B] in Fig.~\ref{fig:KnownMechanisms} at $\sim2 \times 10^{-14}$\,\B.  This point corresponds to the expected level of axial electric-field noise due to the Johnson noise of the electrode material and filter electronics in a two-layer segmented linear ion trap [similar to a segmented version of the trap shown in Fig.~\ref{fig:TrapGeometries}(f)], with an ion-electrode separation of $\sim300$\,$\mu$m. The calculation was modeled after the geometry and electronics used by \citet{Rowe:2002}. The noise is around a factor of 20 below the level measured experimentally by \citet{Rowe:2002} (point [$\Rowe02$]a in Fig.~\ref{fig:KnownMechanisms}.

\subsubsection{Technical noise}
\label{subsubsec:TechnicalNoise}

Technical noise is defined for the purposes of this review as noise coming from power supplies and other voltage sources such as digital-to-analog  (DAC) cards. It arises from the imperfect nature of the power supplies in laboratory equipment. While the physical mechanism of the technical noise from power supplies is usually Johnson-Nyquist or EM pickup in nature and amplified up by electronics, it is considered separately here.   Technical noise can be modeled as a resistor which is very hot \citep{Wineland:1998_05} or very large \citep{Lamoreaux:1997}.

Considering the likely range of absolute values for the noise, many DC power supplies and DACs used for experiments have specifications which allow an upper limit to be put on the level of any such technical noise.  A typical power supply might have 5\,mV of noise spread across 20\,MHz (1 $\mu$V/Hz$^{1/2}$) which, using Eq.~\eqref{eq:JohnsonNoise}, is equivalent to the Johnson noise on a 75\,M$\Omega$ resistor.  One would need to use roughly 80\,dB of filtering at the trap frequency to lower the electric-field noise so that this technical noise becomes roughly equal to the Johnson noise expected from the bulk resistance ($\sim 1\,\Omega$) of the trap electrodes.  With this level of filtering and a characteristic distance $D=100$~$\mu$m, the expected spectral density of electric-field noise at the position of the ion would be of $S_\mathrm{E} \sim 10^{-12}$\,\B. Such aggressive filtering is possible, but not trivial. Moreover, it is easy for such filter electronics to have low-Q inductors [cf. Eq.~\eqref{eq:inductor-Q}] or loops of wire subject to EM pickup (see Sec.~\ref{subsec:EMPickup}) which can themselves become more significant source of noise than the filtered technical noise.

The frequency scaling exponent, $\alpha$, of technical noise could in principle take any value as the device could exhibit resonances.  Distance scaling, $\beta$, of technical noise, would be $1/D^2$ since it is proportional to the voltage noise on the electrodes. The temperature scaling, $\gamma$, might be expected to be flat. However, if the filters change their response as the temperature changes, one would expect a non-zero $\gamma$.

\subsection{Space charge}
\label{subsec:SpaceCharge}

Most models of ion-trap experiments assume that the vacuum is neutral. However, if some mechanism allows for charges within the vacuum then these charges would cause electric-field noise at the position of the trapped ion.  A common source of charge in vacuum, which has been documented in ion traps \citep{Wineland:1998_05}, is electron emission from electrode surfaces.  The mechanism of electron emission in any instance depends on the details of the experiments: Field emission is made more likely by sharp points, rough electrode surfaces or high voltages \citep{Murphy:1956}. Thermal emission is made more likely by high voltages or locally hot electrodes \citep{Murphy:1956}. Photoelectric emission is made more likely by use of  short-wavelength laser light \citep{Linford:1933}. Regardless of the emission mechanism, the effect on the trapped ion is  similar. Electrons escaping the surface of a cathodic electrode follow the field lines created by the high-voltage trap drive and terminate at the anode. These moving charges create electric-field noise at the position if the ion, leading to heating. Effects due to secondary charges which might be ejected by electron bombardment of the anode are not considered here.

Under typical ion-trap conditions it takes $\lesssim 100$~ps for an emitted electron to traverse the trapping region. The RF trap drive is therefore essentially static during the process. As the emitted electron follows the arc of the electric-field line from the cathodic to the anodic electrode its distance from the trapped ion varies during its flight. Even without solving for the exact electrostatic field of a particular trap geometry, a quantitative analysis of a single electron flying through an ion trap is instructive.

The Coulomb force between an emitted electron and a trapped ion is assumed to be a Gaussian pulse of temporal width $\tau_\mathrm{e}=100$~ps. Fourier analysis shows that the spectral density of electric-field noise, $S^{\mathrm{(SC)}}_\mathrm{E}$, due to this Coulomb force is frequency independent for $\omega < 1/\tau_\mathrm{e} =10$\,GHz, and has a value of
    \begin{equation}
    \label{eq:SpaceChargeNoise}
    S^{\mathrm{(SC)}}_\mathrm{E}= \frac{e^2 \tau_\mathrm{e}^3}{16 \epsilon_0^2 d^4}.
    \end{equation}
Here, $d$ is the average distance from the ion to the emitted electron as the electron travels from the cathode to the anode. It is assumed to be equal to the distance from the ion to the nearest electrode. Using the values stated above, each emitted electron contributes a spectral density of electric-field noise of order $10^{-28}$\,\B for $d=40$\,$\mu$m.

If the electrons being emitted from the surface are uncorrelated in time then, at an emission current of 1\,$\mu$A, a value of $S^{\mathrm{(SC)}}_\mathrm{E}\sim 10^{-15}$\,$\mathrm{V^2/m^2 Hz}$ might be expected. Such a large field-emission current would not be difficult to detect and perhaps engineer away. However, many electron-emission experiments have seen that electron-emission noise follows a periodic or oscillatory nature, which is not white noise in character \citep{Tringides:1985, Dharmadhikari:1991}.  Depending on the frequency of these oscillations, currents as low as a few thousand electrons per second could generate electric-field noise of order 10$^{-15}$\,\B at 1~MHz. $S^{\mathrm{(SC)}}_\mathrm{E}$ would then fall-off with frequency, instead of exhibiting the white-noise behavior described above. In all cases, these point charges emanating from the surface of the electrodes would exhibit $\beta=4$, as seen from Eq.~\eqref{eq:SpaceChargeNoise}, provided all other operating parameters remain constant. The frequency and distance scalings which are expected from the space charge model considered here are the same as those observed in some experiments. However, the absolute values of noise expected for all but the highest foreseeable emission currents are below those observed in most experiments.

\section{Microscopic Models for Noise Above Non-Ideal Surfaces}
\label{sec:Microscopic}

In the analysis of Sec.~\ref{sec:Theory} it was usually assumed that in the relevant frequency range of $\omt\sim$ MHz, the whole trap electrode can be described as an ideal equipotential. This assumption is not in general true for real metallic surfaces, where regions of different crystal orientation, surface roughness or adsorbed atoms and compounds lead to local variations of the potential \citep{Herring:1949}. These so-called patch potentials play an important role in many areas of physics and represent, for example, an experimental limitation for precision measurements of the Casimir-Polder force between closely spaced metallic plates \citep{Sandoghdar:1992,Harber:2005} or gravity tests with charged elementary particles \citep{Camp:1991,Darling:1992}. It was first suggested by \citet{Turchette:2000} that fluctuating patch potentials on the electrodes could also be the source of the unexpectedly large heating rates observed in ion traps. In their original work \citet{Turchette:2000} showed that, for a simplified spherical trap geometry, local rather than extended voltage fluctuations lead to a $d^{-4}$ scaling and therefore a strong enhancement of heating rates for small trap dimensions. Subsequent studies have investigated in more detail how the distance scaling is affected by finite patch sizes \citep{Dubessy:2009,Low:2011} and by the electrode geometry \citep{Low:2011}.

The general theoretical framework and the main theoretical predictions for patch-potential heating are summarized in Sec.~\ref{subsec:PatchPotentials}. A class of models which may also be invoked in consideration of the noise seen in ion traps  involve two-level fluctuators (TLFs), in which a particle can occupy, and fluctuate between, one of two states. The instance in which TLFs are distributed in a thin layer on the electrode surface provides a special case of patch potentials, which is considered in Sec.~\ref{subsec:TwoLevelFluctuators}. These models provide valuable predictions for distinguishing between, for example, technical noise sources which lead to global fluctuations of the electrode voltage and noise sources related to microscopic processes on the electrode surface which lead to local fluctuations. However, the patch-potential model itself does not make any predictions on the origin of these fluctuations. Sections \ref{subsec:AdatomDipoles}-\ref{subsec:AdatomDiffusion} of this review describe different microscopic processes, namely fluctuating adatomic dipoles and adatom diffusion, which have been suggested as potential underlying mechanisms for localized field fluctuations. The analysis of these processes provides additional predictions for the frequency and temperature scaling of the noise.

\subsection{Patch-potential models}
\label{subsec:PatchPotentials}

\subsubsection{Origin of patch potentials}

The term ``patch-potential" refers quite generally to a local variation of the potential on an otherwise homogeneous, biased electrode surface. Different mechanisms are predicted to produce such microscopic potential variations. Most commonly patch potentials are attributed to regions of different crystal orientation and surface adsorbates \citep{Herring:1949}. For a clean and regular surface the otherwise homogeneous density of the electrons inside the metal is distorted at the surface, which creates an effective dipole layer at the metal-air interface. This dipole layer changes the work function, $W$, of the electrode by $\Delta W=e\Phi_{\rm p}$. Here $e$ is the charge of the electron and $\Phi_{\rm p}$ is the patch potential, which is related to the dipole moment per unit area, $\mathcal{P}$, by \citep{Jackson:1999}
    \begin{equation}
    \Phi_{\rm p}= \mathcal{P}/{\epsilon_0}.
    \end{equation}
The value of $\mathcal{P}$ depends on the material's surface properties, in particular on the relative orientation of the crystal lattice and the surface. Consequently, small regions of different crystal orientation can lead to variations of $\Phi_{\rm p}$ over microscopic distances. A similar effect arises from adsorbed atoms and molecules, which are polarized when approaching the surface and form additional dipole layers.

The static potentials of metallic surfaces have been measured using various methods. From thermionic-emission-current experiments, it is known that the work function of metal surfaces can vary by several tens of millivolts, depending on the crystal orientation \citep{Herring:1949}. On gold surfaces patch potentials with sizes ranging from 10\,nm to 10\,$\mu$m, and $\Phi_{\rm p}$ $\sim$ meV have been measured using Kelvin probes \citep{Camp:1992,Rossi:1992}. The surface dipoles created by adsorbates can be directly observed on the level of single atoms in precision experiments with cold trapped atoms. For example, \citet{Obrecht:2007} utilized a magnetically trapped Bose-Einstein condensate to measure the electric-field distribution emanating from a cluster of Rb atoms adsorbed on various surfaces. The measured induced dipole moment of $\mu\sim 5$\,D (debye) per atom is consistent with theoretical predictions for alkaline atoms absorbed on metallic surfaces. ($1\,\mathrm{D} \approx 3.33 \times 10^{-30}$\,Cm.) Finally, trapped ions have been used to investigate laser-induced surface dipoles \citep{Harlander:2010} and the long-term variations of stray electric fields over several months \citep{Haerter:2014}.

While static patch fields on metal surfaces are relatively well understood, little is known about their fluctuations, in particular in the MHz frequency regime of interest. Sections \ref{subsec:AdatomDipoles}-\ref{subsec:AdatomDiffusion} of this review describe several mechanisms which have been suggested in the literature to explain the observed heating rates in ion traps in terms of microscopic atomic processes on the surface.

\subsubsection{Electric-field noise from fluctuating patches}

\begin{figure}[t]
\begin{center}
\includegraphics[width=\columnwidth]{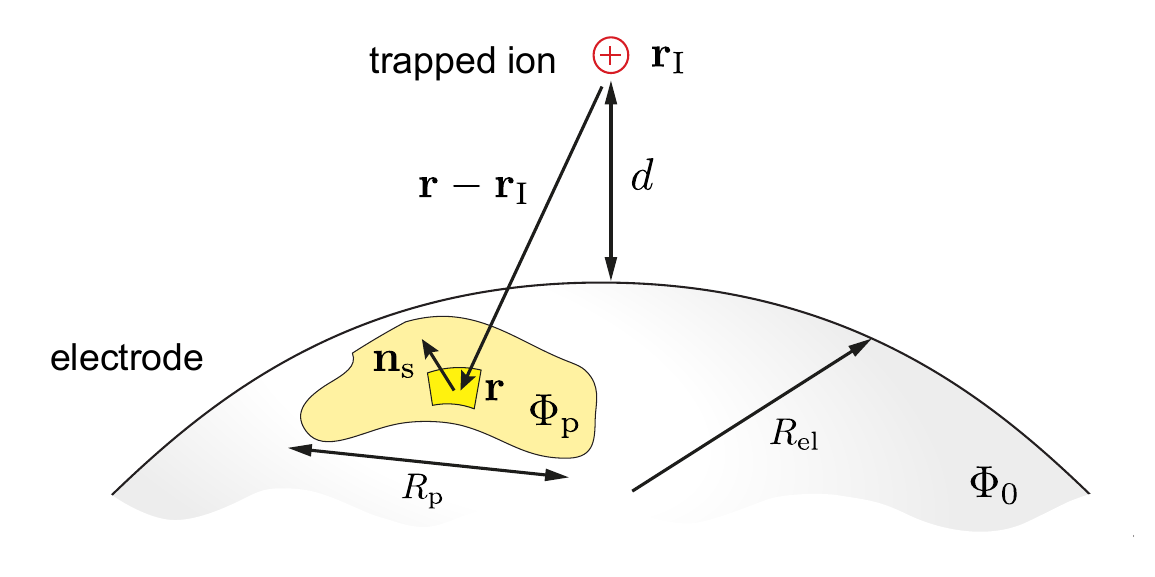}
\caption{Relevant length scales considered in the patch-potential model. The ion is trapped at a distance $d$ above an electrode of characteristic dimension $R_{\rm el}$. Patch fields on the electrode produce local potential fluctuations $\Phi_{\rm p}({\bm r},t)$, which are correlated over a length scale $r_{\rm c}\approx R_{\rm p}$ set by the typical radius of the patches, $R_{\rm p}$. }
\label{fig:PatchPotentialModel}
\end{center}
\end{figure}

Fig.~\ref{fig:PatchPotentialModel} illustrates the general setting of an ion, which is trapped a distance $d$ above an electrode with a radius of curvature $R_{\rm el}$. The electrode is considered in general to be an equipotential at $\Phi_0$, while surface imperfections cause the electrode potential within different areas on the surface (``patches") to be enhanced or reduced by an amount $\Phi_{\rm p}({\bm r},t)$.

For the following discussion the physical origin of these voltage variations is left unspecified, but it is assumed that potential fluctuations are correlated only over a length scale $r_{\rm c}\approx R_{\rm p}$, which corresponds to the characteristic patch radius, $R_{\rm p}$. In the absence of additional free charges\footnote{This is typically the case for metallic electrodes. Here surface charges are compensated by their image charges and form effective dipoles. However, this is in general not the case for charges located on insulating parts of the trap, where a different field distribution can arise.} and by assuming that the potential difference $\Phi_{\rm p}({\bm r},t)$ occurs within a layer that is thin compared to the ion distance, $d$, the electrostatic potential at the position of the ion, $\Phi({\bm r}_{\rm I},t)$, is the solution of the Laplace equation with boundary conditions given by $\Phi_0+\Phi_{\rm p}({\bm r},t)$ on the electrode surface, $S$. By setting $\Phi_0=0$ for simplicity, the potential can in general be expressed as \citep{Jackson:1999}
    \begin{equation}
    \label{eq:Phi_Patch}
    \Phi({\bm r}_{\rm I},t)= - \int_S d^2 r \, \frac{\partial G_{\rm E}({\bm r}_{\rm I}, {\bm r})}{\partial {\bm n}_s} \Phi_{\rm p}({\bm r},t).
    \end{equation}
Here $G_{\rm E}({\bm r}_{\rm I}, {\bm r})$ is the electrostatic Green's function satisfying the Dirichlet boundary conditions, $G_{\rm E}({\bm r}_{\rm I}, {\bm r})=0$ for ${\bm r}\in S$, and ${\bm n}_s$ is a unit vector normal to the surface. The electric field projected along the trap axis is then given by $ E_{\rm t}(t)=- {\bm e}_{\rm t}\cdot \nabla \Phi( {\bm r}_{\rm I},t)$ and the resulting electric-field spectrum can be written as
    \begin{equation}
    \label{eq:SE_General}
    S^\mathrm{(PP)}_\mathrm{E} = \int_S d^2 r_1 \int_S d^2 r_2 \, \mathcal{G}( {\bm r}_{\rm I}, {\bm r}_1) \mathcal{G}( {\bm r}_{\rm I}, {\bm r}_2) C_{\rm V}( {\bm r}_1, {\bm r}_2,\omega).
    \end{equation}
In this expression
    \begin{equation}
    \label{eq:mathcalG}
    \mathcal{G}({\bm r}_{\rm I}, {\bm r}) = {\bm e}_{\rm t}\cdot \vec \nabla \left[\frac{\partial G_{\rm E}({\bm r}_{\rm I}, {\bm r})}{\partial {\bm n}_s}\right]_{{\bm r} \in S}
    \end{equation}
is a geometric factor describing the electric field produced at the position of the ion from a small patch located at ${\bm r}$ on the electrode. The correlation function,
    \begin{equation}
    \begin{split}
    &C_{\rm V}({\bm r}_1,{\bm r}_2,\omega) = 2 \int_{-\infty}^\infty d\tau \, \langle \delta \Phi_{\rm p}({\bm r}_1,\tau)\delta \Phi_{\rm p}({\bm r}_2,0)\rangle e^{-i\omega \tau},
    \end{split}
    \end{equation}
where $\delta \Phi_{\rm p}({\bm r},\tau)= \Phi_{\rm p}({\bm r},\tau)-\langle\Phi_{\rm p}({\bm r},\tau)\rangle$,
contains all of the information about temporal and spatial correlations of the fluctuating patch fields at positions ${\bm r}_1$ and ${\bm r}_2$.

While the exact dependence of $C_{\rm V}({\bm r}_1,{\bm r}_2,\omega)$ on frequency and distance requires a detailed knowledge about the microscopic origin of the patch potentials, a reasonable approximation to a real surface potential can be obtained by assuming that the electrode is covered by $N_{\rm p}$ separate patches \citep{Low:2011}:
    \begin{equation}
    \label{eq:puzzle_chi}
    \Phi_{\rm p}({\bm r},t)= \sum_{i=1}^{N_{\rm p}} V_i(t) \chi_i({\bm r}).
    \end{equation}
Here the step function $\chi_i({\bm r})=0,1$ has a non-vanishing support only for values of ${\bm r}$ within the area, $A_i$, of the $i$-th patch, and $V_i(t)$ denotes the fluctuating potential of this patch. By assuming that the voltage fluctuations between different patches are uncorrelated and described by the spectral density $S_{\rm V}^i$, this model leads to a correlation function
    \begin{equation}
    \label{eq:Corr_patch_approx}
    C_{\rm V}({\bm r}_1,{\bm r}_2,\omega) = \sum_{i=1}^{N_{\rm p}} S_{\rm V}^i \chi_i({\bm r}_1)\chi_i({\bm r}_2).
    \end{equation}
This can be further simplified by assuming $S_{\rm V}^{i}\approx S_{\rm V}$. The voltage correlations described by Eq.~\eqref{eq:Corr_patch_approx} are non-zero only over the extent of individual patches which, for a sufficiently homogeneous distribution, implies a correlation length $r_{\rm c} \approx \sqrt{A_{\rm p}}$, where $A_{\rm p}$ is the average patch area.

\subsubsection{Distance scaling for a planar trap}

The influence of a finite patch size on the distance scaling of electric-field noise experienced by the ion is most apparent by considering the idealized case, where the ion is trapped above an infinitely extended planar electrode. In this case the electrode dimension $R_{\rm el}\rightarrow \infty$ is eliminated from the problem and the remaining length scales are the correlation length, $r_{\rm c}$, and the ion-surface distance, $d$. For a single, perfectly conducting plane specified by ${\bm r}=(x,y,z=0)$ the Green's function is simply that of a charge plus its image charge inside the metal:
    \begin{equation}
    G_{\rm E}({\bm r}_{\rm I}, {\bm r}) = \frac{1}{4\pi} \left( \frac{1}{ |{\bm r}_{\rm I}-{\bm r} |} -\frac{1}{ |{\bm r}_{\rm I}-{\bm r}- 2z {\bm e}_z| } \right).
    \end{equation}
Therefore, the resulting potential defined in Eq.~\eqref{eq:Phi_Patch} is equivalent to a potential being produced by a layer of surface dipoles covering the electrode, with a local dipole moment $d\mu({\bm r},t) = 2 \epsilon_0 \Phi_{\rm p}({\bm r},t) dS$ per surface element $dS$. Note that since this dipole layer is located symmetrically around $z=0$, there is a factor of two compared to a layer of dipoles on top of the surface.
For an ion located at ${\bm r}_{\rm I}=(0,0,z=d)$ the geometric factor given in Eq.~\eqref{eq:mathcalG} is explicitly given by $\mathcal{G}({\bm r}_{\rm I}, {\bm r}) = g_D(x,y)/(2\pi)$, where
    \begin{eqnarray}
    g_D (x,y)&=& \frac{2 d^2 - x^2-y^2}{|d^2+x^2+y^2|^{5/2}},\qquad {\bm e}_{\rm t} ={\bm e}_z, \label{eq:gDz}\\
    g_D (x,y)&=& \frac{3dx}{|d^2+x^2+y^2|^{5/2}},\qquad {\bm e}_{\rm t} ={\bm e}_x, \label{eq:gDx}
    \end{eqnarray}
describes electric-field fluctuations perpendicular and parallel to the electrode surface, respectively. In the limit of small patches, $r_{\rm c} \ll d$ and a surface coverage fraction of $\sigma_{\rm p}$ the integrals in Eq.~\eqref{eq:SE_General} can be evaluated in straightforward manner. As a result one finds
    \begin{equation}
    \label{eq:SE_SmallPatches}
    S^{({\rm PP})}_{\mathrm{E},\eta=\perp,\parallel} \simeq s_\eta \frac{3 \sigma_{\rm p} A_{\rm p}}{16 \pi d^4} S_{\rm V}
    \end{equation}
for the field fluctuations perpendicular $(s_\perp=1)$ and the parallel $(s_\parallel=1/2)$ to the surface. Eq.~\eqref{eq:SE_SmallPatches}
shows that the microscopic structure of the electrode surface can substantially modify the distance scaling of the electromagnetic noise, and would lead, at least for this simple geometry, to the often-cited $d^{-4}$ scaling of ion-trap heating rates.

On gold surfaces static patch potentials with sizes ranging from 10\,nm to 10\,$\mu$m have been experimentally measured using Kelvin probes \citep{Camp:1992,Rossi:1992}. The sizes of patches which could fluctuate at radio frequencies are unknown. From the theory presented here, in the limit $r_{\rm c}\ll d$ the predicted scaling of the electric-field noise given in Eq.~\eqref{eq:SE_SmallPatches} is insensitive to the exact details of the patch correlation function. This is not the case for moderate or large patch sizes $d \lesssim r_{\rm c}$. For example, an exponential correlation $C_{\rm V}({\bm r}_1,{\bm r}_2,\omega) \sim e^{-|{\bm r}_1-{\bm r}_2|/r_{\rm c}}$ leads to a divergent spectrum $S_\mathrm{E}\sim d^{-1}$ for $r_{\rm c} \gg d$ \citep{Dubessy:2009}. In contrast, a sharp cutoff, $C_{\rm V}({\bm r}_1,{\bm r}_2,\omega) \sim \Theta(r_{\rm c}- |{\bm r}_1-{\bm r}_2|)$, where $\Theta(r)$ is the unit step function, results in $S_\mathrm{E}\sim d^{+1}$, and thus a vanishing spectrum for large patch sizes \citep{Low:2011}.

\begin{figure}[t]
\begin{center}
\includegraphics[width=\columnwidth]{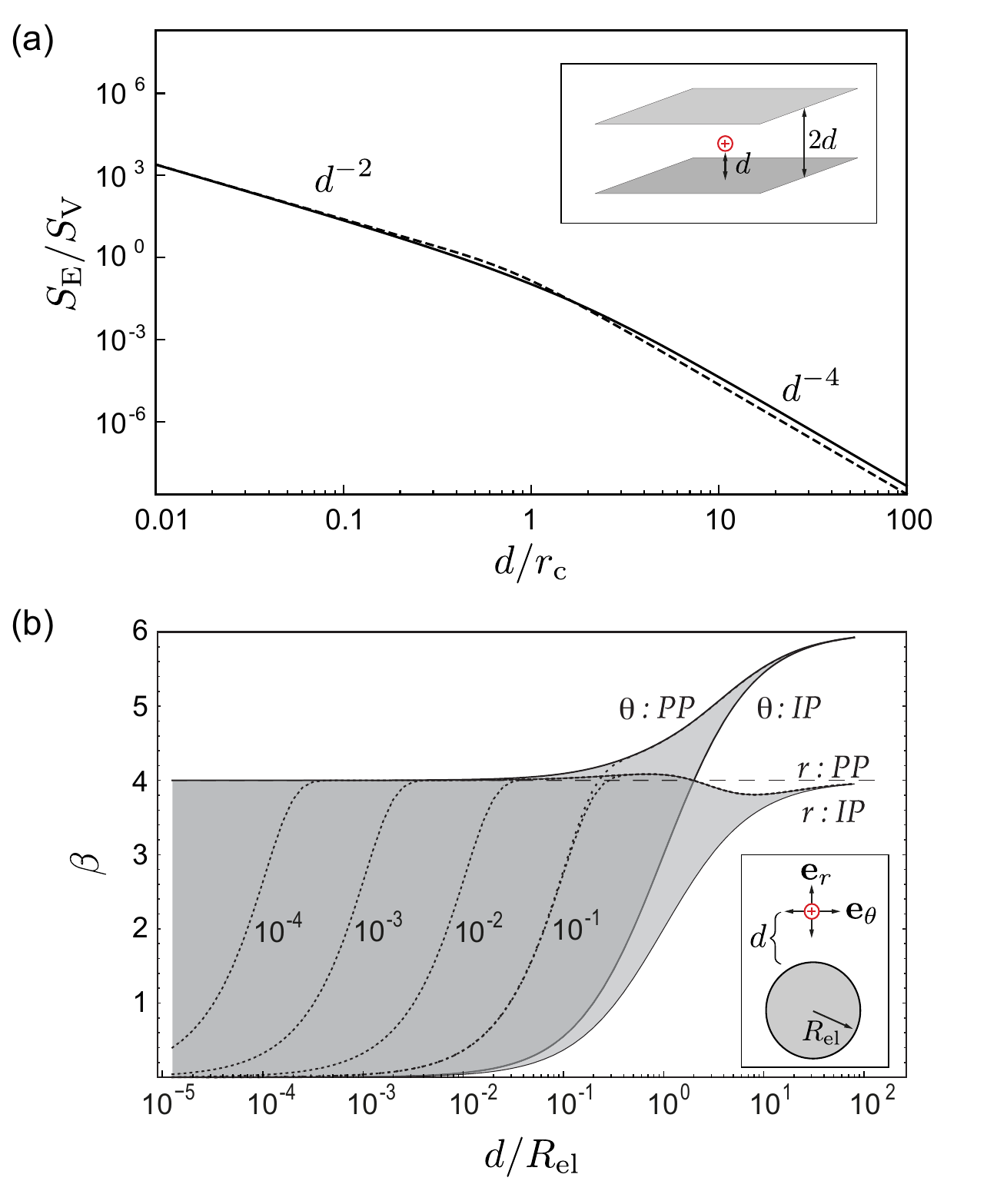}
\caption{Electric-field noise from patch potentials. (a) Distance dependence of the electric-field noise, $S_\mathrm{E}$, for an ion trapped between two infinite parallel plates (inset) assuming an exponential (solid line) and step like (dashed line) cutoff for the spatial voltage correlation function $C_{\rm V}({\bm r}_1,{\bm r}_2,\omega)$. (b) Plot of the local distance-scaling coefficient, $\beta(d)$, defined in Eq.~\eqref{eq:LocalBeta} for an ion trapped above a sphere of radius $R_{\rm el}$ (inset). The solid lines represent the limits for point-like patches (PP) for which $r_{\rm c}\rightarrow 0$, and infinite patches (IP) for which $r_{\rm c}\rightarrow \infty$. The dotted lines show the results for intermediate values of $r_{\rm c}/R_{\rm el}$ as indicated in the plot. [Figure adapted from \citet{Low:2011} and used with permission.]}
\label{fig:PatchPotentialDistanceScaling}
\end{center}
\end{figure}

The prediction of such qualitatively and unexpectedly different behavior for essentially minor changes to $C_{\rm V}$  is a consequence of the rather unphysical configuration of an infinite plane without additional ground electrodes. As a more appropriate approximation to a real trap, an ion confined between two planar electrodes separated by $2d$ can be considered. The Green's function for this configuration can be evaluated analytically, by taking into account the additional image dipoles at the planes $z=\pm 4jd$, where $j=1,2,\dots$. Therefore, the analysis of a single plane can be repeated by replacing
    \begin{equation}
    g_D(x,y)\rightarrow \sum_{j=0}^\infty \frac{2 d^2(2j+1)^2 - x^2-y^2}{|d^2(2j+1)^2+x^2+y^2|^{5/2}},
    \end{equation}
and the summation can be carried out after changing to a Fourier representation \citep{Dubessy:2009}. The resulting electric-field spectrum is plotted in Fig.~\ref{fig:PatchPotentialDistanceScaling}(a), for the two types of correlation function, $C_{\rm V}$,  given above. Consistent with the discussion of Johnson-like noise in a parallel-plate configuration, $S_\mathrm{E}$ exhibits a $d^{-2}$ scaling at larger patch sizes or small electrode separations, where the ion essentially sees the constant homogeneous field of a plate capacitor. At $r_c\approx d$, this then crosses over to the $d^{-4}$ scaling derived in Eq.~\eqref{eq:SE_SmallPatches}, with effective patch areas $A_{\rm p}=2\pi r_{\rm c}^2$ and $A_{\rm p}=\pi r_{\rm c}^2$ for the exponential and step-like cutoff, respectively. Therefore, the configuration of two parallel plates can be used as a minimal toy model to describe the crossover from Johnson noise to patch-potential noise in ion traps.

\subsubsection{Influence of the electrode geometry}

As already discussed in Sec.~\ref{subsubsec:Johnson_DistanceScaling} for the case of Johnson noise, the expected distance scaling of the electric-field noise in realistic traps can considerably differ from that of an infinite planar geometry when the shape and finite size of the electrodes is taken into account. For patch-potential noise, this dependence is further complicated by the existence of a second length scale given by the patch size, $r_{\rm c}$. \citet{Low:2011} have carried out a detailed analysis predicting the electric-field noise scaling for various planar, spherical and spheroidal electrode geometries. The authors introduced a `local' scaling coefficient,
    \begin{equation}
    \label{eq:LocalBeta}
    \beta(d)= - \frac{\partial \ln ( S_\mathrm{E})}{\partial \ln d},
    \end{equation}
and studied the dependence of $\beta(d)$ on the electrode distance, $d$, the electrode geometry, and the characteristic patch size, $r_{\rm c}\approx R_{\rm p}$. An example of this analysis is shown in Fig.~\ref{fig:PatchPotentialDistanceScaling}(b), where $\beta(d)$ is plotted for an ion at a distance $d$ above a spherical electrode of radius $R_{\rm el}$. A similar plot for a needle-like electrode (Fig.~\ref{fig:NeedleAnalysis}) is discussed in more detail in Sec.~\ref{subsec:NeedleTrap}. Fig.~\ref{fig:PatchPotentialDistanceScaling}(b) clearly shows that, for $d\gg R_{\rm el}$, Johnson noise and patch-potential models lead to a similar scaling. In the case of heating in a direction normal to the sphere's surface this gives $\beta=4$. Only for distances $d\lesssim R_{\rm el}$, do the predictions from the two types of noise start to deviate significantly.

A note of caution must be added to the interpretation of this plot for distances $d< r_{\rm c}$. As shown above for the example of a planar geometry, the scaling in this regime of an ion above a single electrode can significantly differ from that of an ion located symmetrically between two electrodes. Similarly, it is expected that the scaling of a more realistic trap configuration modeled by two spheres will differ from the values shown in Fig.~\ref{fig:PatchPotentialDistanceScaling}(b). However, this does not significantly affect the scaling for $d\geq r_{\rm c}$.

\subsection{Two-level fluctuator models}
\label{subsec:TwoLevelFluctuators}

\begin{figure}[t]
\begin{center}
\includegraphics [width=\columnwidth]{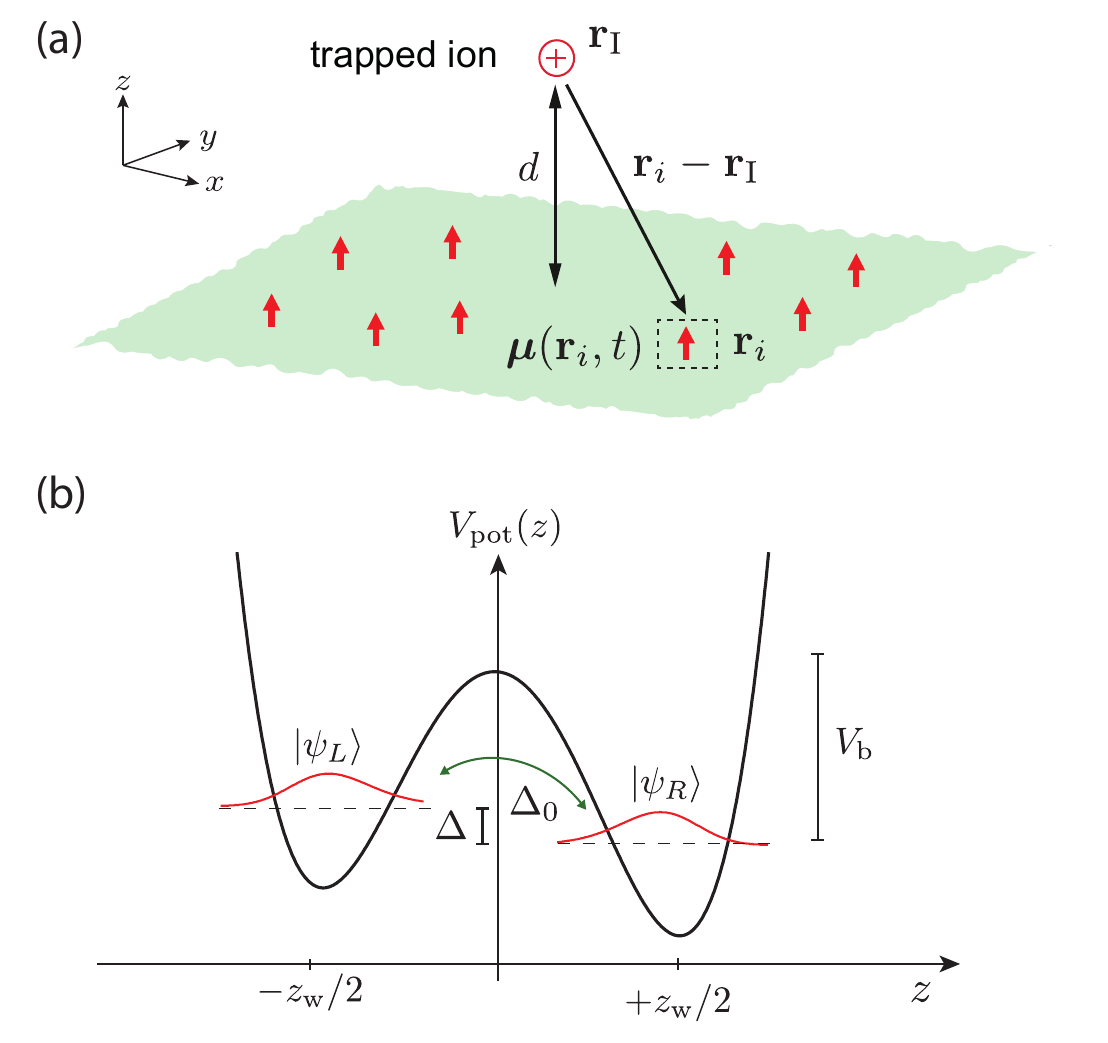}
\caption{(a) General setup considered in Secs.~\ref{subsec:TwoLevelFluctuators}--\ref{subsec:AdatomDiffusion} for the analysis of the electric-field noise generated from microscopic processes above a planar electrode. The noise processes are modeled by a distribution of point-like dipoles, ${\boldsymbol \mu}({\bm r}_i,t)$, located within a small layer above the metal surface. (b) Double-well potential representing the energy of an atomic or electronic two-state system.}
\label{fig:TLF_Setup}
\end{center}
\end{figure}

The assumption of homogeneous lattices and surfaces is only a crude approximation for real solids where surface corrugations or lattice dislocations create local minima in the otherwise periodic potential landscape. In highly disordered systems or amorphous solids, this can lead to the formation of so-called two-level fluctuators (TLFs) \citep{Phillips:1987}, where electrons, atoms or groups of atoms become localized in one of two nearby potential minima. The scenario considered here is depicted in Fig.~\ref{fig:TLF_Setup}. The localization of charge means that TLFs on a surface can be considered as a specific class of patch potentials.

Quantum tunneling through the barrier or thermally activated transitions over the barrier induce random jumps between the minima, causing fluctuations of the local dipole moment associated with the two distinct configurations. Phenomenological models based on the existence of a large ensemble of TLFs have been successfully used to explain the unusual low-temperature properties of glasses \citep{Anderson:1972,Phillips:1972} or the appearance of current and voltage fluctuations with a 1/$f$ frequency scaling \citep{Dutta:1981, Paladino:2014}. More direct measurements on two-level systems have recently been performed with superconducting qubits, where TLFs in the insulating layer of a Josephson junction have been spectroscopically resolved \citep{Martinis:2005} and the coherent manipulation of individual two-level defects has been demonstrated \citep{Neeley:2008,Lisenfeld:2010}.

In ion traps TLFs can potentially form in a disordered insulating layer on top of the metal electrode and lead to heating. Even though the precise microscopic origin or the number of defects is unknown, general predictions for the temperature and frequency scaling of this noise process already follow from a few reasonable assumptions regarding the distribution of energy scales and relaxation times of the TLFs. The following section first discusses the electric-field noise spectra expected from a random distribution fluctuating dipoles, and more specifically that from two-state systems, located in the vicinity of the electrode surface. Sec.~\ref{subsubsec:ThermallyActivatedTLF} and~\ref{subsubsec:TunnellingStates} then review two basic physical processes: thermal activation and quantum tunneling. These are most commonly considered in the solid-state literature for the modeling of noise with a $1/f$ spectrum. A less-frequently discussed fact is that non-uniform distributions of barrier heights in these models can lead to non-trivial temperature dependencies and deviations from the strict $1/f$ scaling of the noise. These effects could play a role in observed temperature variations of $\alpha$~\citep{Labaziewicz:2008_10} and are briefly described in Sec.~\ref{subsubsec:NonUniformTLFDistribution}.

\subsubsection{Electric-field noise from fluctuating dipoles}

For an ion which is trapped close above a metallic electrode, charge fluctuations within the electrode are efficiently shielded and net charges on the surface are compensated by their respective image charges in the metal. It can therefore be assumed that the relevant microscopic noise processes on the electrode are well described by a distribution of point-like dipoles ${\boldsymbol \mu}({\bm r}_i,t)$ located at position ${\bm r}_i$ within a thin surface layer of thickness $h\ll d$ [see Fig.~\ref{fig:TLF_Setup}(a)]. The total field at the position of the ion, $E_{\rm t}(t)={\bm e}_\mathrm{t}\cdot {\bm E}({\bm r}_\mathrm{I},t)$, is then given by
    \begin{equation}
    E_{\rm t}(t)= \frac{1 }{4\pi \epsilon_0}\sum_i \frac{\left(3 ({\bm e}_\mathrm{t} \cdot {\bm n}_i) {\bm n}_i- {\bm e}_\mathrm{t}\right) \cdot {\boldsymbol \mu}({\bm r}_i,t)}{|{\bm r}_i-{\bm r}_\mathrm{I}|^3} ,
    \end{equation}
where ${\bm n}_i=({\bm r}_i-{\bm r}_\mathrm{I})/|{\bm r}_i-{\bm r}_\mathrm{I}|$ is the unit vector pointing from the ion to dipole $i$. For a planar electrode, and assuming a sufficiently homogeneous distribution of dipoles with surface density $ \sigma_{\rm d}$, the electric-field fluctuations perpendicular and parallel to the surface are given by
$ S^\mathrm{({\rm TLF})}_{\mathrm{E},\parallel}=$\,$S^\mathrm{({\rm TLF})}_{\mathrm{E},\perp}/2$ and
    \begin{equation}\
    \label{eq:SE_Dipoles}
    S^{\rm (TLF)}_{\mathrm{E},\perp} = \frac{3\pi}{2} \frac{\sigma_{\rm d} }{(4\pi \epsilon_0)^2 d^4} \times \bar S_\mu.
    \end{equation}
Here $\bar S_\mu = \frac{1}{N_A} \sum_{i=1}^{N_A} S_\mu^i$ denotes the averaged dipole-fluctuation spectrum of a large number, $N_A\gg 1$, of fluctuators located within an area $A\ll \pi d^2$, where $S_\mu^i = \int_{-\infty}^\infty d\tau \, \langle \delta \mu_i (\tau) \delta \mu_i (0)\rangle e^{-i\omega \tau}$ is the spectrum of a single dipole with fluctuating dipole moment $\delta \mu_i(t)$ perpendicular to the surface. Note that Eq.~\eqref{eq:SE_Dipoles} corresponds to the zero-correlation-length limit of the patch-potential model and therefore leads to the same $d^{-4}$ distance scaling in the considered limit of a flat trap geometry.

\subsubsection{Two-level fluctuators}

A single TLF can be formed either by a particle trapped in two proximal potential minima or by an extended atomic complex or lattice dislocation, which can switch between two energetically favorable configurations. Both cases can be modeled by an (effective) particle moving in a double-well potential, as shown in Fig.~\ref{fig:TLF_Setup}(b). For the temperature regime of interest the dynamics of the TLF can be restricted to the two states $|\psi_L\rangle$ and $|\psi_R\rangle$, with the corresponding wave functions being localized in the left or right minimum, respectively. If the two stable configurations are associated with different dipole moments, the dipole operator for a single TLF is $\hat \mu(t) = (\mu/2) \sigma_z(t)$, where $\sigma_z=|\psi_R\rangle\langle \psi_R| -|\psi_L\rangle \langle \psi_L|$ is the Pauli operator for the population difference between the two states. For a charged particle the dipole moment is approximately $\mu=|e| z_{\rm w}$, where $z_{\rm w}\sim \mathrm{\AA}$ is the separation between the wells. In general, $\mu$ must be evaluated from the charge density of the full atomic complex in the two stable configurations. If the TLF is in contact with a thermal reservoir, the mean value of the population imbalance relaxes to a stationary value according to
    \begin{equation}
    \frac{d}{dt} \langle \sigma_z (t)\rangle = - \frac{( \langle \sigma_z(t)\rangle - \langle \sigma_z\rangle_{\rm eq} )}{T_1}.
    \end{equation}
Here $\langle \sigma_z\rangle_{\rm eq}=- \tanh\left( E_{\rm TLF}/2k_\mathrm{B}T\right)$ is the equilibrium value for a two-level system with energy difference $E_{\rm TLF}$, and $T_1$ is the characteristic relaxation time. With reference to Fig.~\ref{fig:TLF_Setup}, $E_{\rm TLF}=\sqrt{\Delta^2+\Delta_0^2}$ and so is equal to $\Delta$ in the classical limit where tunneling can be neglected. According to the quantum regression theorem \citep{Gardiner:2004}, the population fluctuations $\langle \delta \sigma_z(t)\delta \sigma_z(0)\rangle$ obey an exponential decay which, for a single TLF, results in a simple Lorentzian shape for the dipole-fluctuation spectrum:
    \begin{equation}
    \label{eq:TLFSpectrum}
    S_\mu= \frac{\mu^2}{2\cosh^2(E_{\rm TLF}/(2k_\mathrm{B}T))}\frac{T_1 }{1+\omega^2T_1^2}.
    \end{equation}
For a large ensemble of fluctuators this spectrum must be averaged over a distribution of parameters and
    \begin{equation}
    \label{eq:Sbar}
    \bar S_\mu= \frac{\mu^2}{2} \int \, dE dT_1 \frac{ P(E,T_1)}{\cosh^2(E/(2k_\mathrm{B}T))}\frac{T_1 }{1+\omega^2T_1^2} .
    \end{equation}
Here $P(E_{\rm TLF},T_1)$ is the probability density (per energy and time interval) of finding a TLF with energy difference $E_{\rm TLF}$ and relaxation time $T_1$, assuming approximately the same dipole moment, $\mu$, for all TLFs.

\subsubsection{Thermally activated fluctuators}
\label{subsubsec:ThermallyActivatedTLF}

For high temperatures or large barrier widths, quantum tunneling events can be neglected and the switching of the TLF is mainly induced by thermally activated transitions over the potential barrier of height $V_\mathrm{b}$. For $E_{\rm TLF}\ll k_\mathrm{B} T < V_\mathrm{b}$ the switching rates between the two minima are approximately the same and the relaxation time is given by the Arrhenius law \citep{Arrhenius:1889,Haenggi:1990}:
    \begin{equation}
    \label{eq:T1Thermal}
    T_1(V_\mathrm{b})\simeq \tau_0 e^{\frac{V_\mathrm{b}}{k_\mathrm{B} T}}.
    \end{equation}
The timescale, $\tau_0$, is approximately given by the oscillation period of the particle in a single well and typically assumes values in the range of $\tau_0\approx 10^{-12}-10^{-14}$\,s \citep{Gomer:1990,Ovesson:2001}. The height of the potential barrier is determined by electronic energy scales in the range of $V_\mathrm{b}\sim 0.1-1$ eV \citep{Gomer:1990}. The exponential dependence of the relaxation rate on $V_\mathrm{b}$ results in a large range of possible switching times which, particularly at room temperature or below, lie within the relevant range of the trapped-ion oscillation time. With no further knowledge regarding the physical origin of the TLF it is reasonable to consider, as a first approximation, TLFs with a uniform distribution of energies, $E_{\rm TLF}$, up to a maximum energy, $E_{\rm max} < k_\mathrm{B} T$. It is also reasonable to assume (at least initially) that they take a uniform distribution of activation energies, $P(V_\mathrm{b})=const.$, within an interval $\Delta V$ between $V_{\rm min}$ and $V_{\rm max}$. The corresponding distribution of relaxation times is \citep{Phillips:1987}
    \begin{equation}
    P(E_{\rm TLF}, T_1)= \frac{1}{E_{\rm max}} \frac{k_\mathrm{B}T}{\Delta V} \frac{1}{T_1}.
    \end{equation}
Within the frequency range $\exp(-V_{\rm max}/k_\mathrm{B}T) < \omega \tau_0 < \exp(-V_{\rm min}/k_\mathrm{B}T)$ this distribution results in an averaged spectrum,
    \begin{equation}
    \label{eq:SwThermal}
    \bar S_\mu \simeq \frac{1}{\Delta V}\frac{\pi k_{\rm B}T}{4 \omega}\mu^2,
    \end{equation}
which exhibits an $\omega^{-1}$ dependence on the frequency and a linear scaling with temperature. Note that Eq.~\eqref{eq:SwThermal} has been derived under the assumption that $E_{\rm TLF}\ll k_\mathrm{B}T$ and that the number of TLFs is independent of temperature. Deviations from these assumptions, as well as non-uniform distributions $P(V_\mathrm{b})$ for the activation energies, can lead to a different temperature-scaling behavior, as discussed below.

To estimate some typical values, in amorphous solids the density of TLFs is found to be around $\sim 10^{46}$ J$^{-1}$ m$^{-3}$ \citep{Phillips:1987}. We consider a system at $T=300$\,K with a contamination layer of height $h=10$ nm \citep{Daniilidis:2014}, which corresponds to a few tens of monolayers of adsorbates on top of the metal electrode. Assuming only TLFs of energy $E_{\rm max}\lesssim k_\mathrm{B} T$ are active one expects an areal TLF density of $\sigma_{\rm d}=4\times 10^{17}$ m$^{-2}$. By assuming potential barriers of about $V_\mathrm{b}\sim \Delta V\sim 1$ eV and a characteristic dipole moment of $\mu\sim 5$\,D \citep{Martinis:2005} the resulting electric-field spectral density at $\omt/2\pi=1 $\,MHz and at a distance $d=100\, \mu$m above the trap is about $S_\mathrm{E}\approx5 \times 10^{-12} $\,\B. This value is around the level seen experimentally in good (low-noise) traps (cf. Fig.~\ref{fig:S_E_d}).

\subsubsection{Tunneling states}
\label{subsubsec:TunnellingStates}

At lower temperatures, thermally activated switching events are strongly suppressed and, particularly for a small barrier width or low effective particle mass, $m_{\rm TLF}$, quantum-mechanical tunneling through the barrier becomes important. In this case the TLF must be described by a quantum-mechanical two-level system with Hamiltonian \citep{Phillips:1987}
    \begin{equation}
    \label{eq:HTLF}
    \hat H_{\rm TLF}=\frac{\Delta}{2} \sigma_z - \frac{\Delta_0}{2} \sigma_x,
    \end{equation}
where $\sigma_z$ and $\sigma_x$ are the usual Pauli operators in the subspace of the localized states $|\psi_L\rangle$ and $|\psi_R\rangle$.
Here $\Delta$ is the `classical' energy difference between the two wells (cf. Fig.~\ref{fig:TLF_Setup})
and $\Delta_0$ is the tunnel coupling.
For a separation $z_{\rm w}$ between the wells, the tunnel amplitude depends exponentially on the barrier height, $V_\mathrm{b}$, and can be estimated by \citep{Phillips:1987}
    \begin{equation}
    \label{eq:TunnelingAmplitude}
    \Delta_0\approx \frac{\hbar}{\tau_0} e^{-\lambda},\qquad \lambda \approx \sqrt{2m_{\rm TLF} z_{\rm w}^2 V_\mathrm{b}/\hbar^2},
    \end{equation}
where $\tau_0^{-1}$ is the attempt frequency, $\lambda$ is the tunneling parameter. The Hamiltonian given in Eq.~\eqref{eq:HTLF} can be diagonalized and written as
    \begin{equation}
    \hat H_{\rm TLF}= \frac{E_{\rm TLF}}{2} \tilde \sigma_z,\qquad E_{\rm TLF}=\sqrt{\Delta^2+\Delta_0^2},
    \end{equation}
where $\tilde \sigma_z$ denotes the Pauli operator in the rotated eigenbasis $|\tilde 0\rangle = \cos(\phi/2) |\psi_L\rangle + \sin(\phi/2)|\psi_R\rangle$ and $ |\tilde 1\rangle = \cos(\phi/2) |\psi_R\rangle - \sin(\phi/2)|\psi_L\rangle$ and $\tan(\phi)=\Delta_0/\Delta$.

The TLF is coupled to phonons via deformation-potential interactions, which induce transitions between the energy eigenstates $|\tilde 0\rangle$ and $|\tilde 1\rangle$.
The resulting relaxation rate of the TLF can be written as
    \begin{equation}
    T_1^{-1}= \left(\frac{\Delta_0}{E_{\rm TLF}}\right)^2 T^{-1}_{\rm min}(E_{\rm TLF}),
    \end{equation}
where $T_{\rm min}(E_{\rm TLF})$ is the minimum relaxation time in the case of a symmetric double well, $\Delta=0$. This rate is approximately given by \citep{Phillips:1987}
    \begin{equation}
    \label{eq:T1Tunneling}
    T^{-1}_{\rm min}(E_{\rm TLF})= \frac{E_{\rm TLF}^3 }{2 \pi \rho\hbar^4 } \left( \frac{\Xi_{\rm l}^2}{v_{\rm l}^5}+ \frac{2\Xi_{\rm t}^2}{v_{\rm t}^5}\right) \coth\left( \frac{E_{\rm TLF}}{2k_\mathrm{B}T}\right),
    \end{equation}
where $\rho$ is the density of the electrode material,
$\Xi_{\rm l}$ ($\Xi_{\rm t}$) and $v_{\rm l}$ ($v_{\rm t}$) are the deformation potential constants ($\sim$ 1 eV)
and the sound velocity for longitudinal (transverse) phonon modes, respectively.
For $E_{\rm TLF}=\Delta_0$ Eq.~\eqref{eq:T1Tunneling} shows a quadratic dependence of the relaxation rate on the tunneling coupling, $\Delta_0$, which in turn depends exponentially on the tunneling parameter, $\lambda$, i.e. on the barrier height, $V_\mathrm{b}$, and the well separation, $z_{\rm w}$. Consequently, the phonon-assisted tunneling mechanism leads to a broad distribution of relaxation times, $T_1$, similar to the case of thermally activated TLFs. By assuming, as above, a flat distribution for the parameter $\lambda$ and the double-well asymmetry $\Delta\in\{-\Delta_{\rm max},\Delta_{\rm max}\}$,
the resulting distribution of energies and relaxation rates for tunneling states is \citep{Phillips:1987}
    \begin{equation}
    \label{eq:RelaxationDistribution}
    P(E_{\rm TLF},T_1)= \frac{P_0 }{2 T_1 \sqrt{1-T_{\rm min}(E_{\rm TLF})/T_1}}.
    \end{equation}
Here $P_0= 1/(\Delta_{\rm max} \log(\Delta_{0,{\rm max}}/\Delta_{0,{\rm min}}))$ is a normalization constant, where $\Delta_{0,{\rm max}}$ and $\Delta_{0,{\rm max}}$ are the minimum and maximum values of $\Delta_{0}$.
Note that in the rotated eigenbasis the dipole operator is $\hat \mu(t)= (\mu/2)\cos(\phi)\tilde \sigma_z(t)$ and, when integrating over the distribution $P(E_{\rm TLF},T_1)$ in Eq.~\eqref{eq:Sbar}, an additional factor $\cos^2(\phi)=(\Delta/E_{\rm TLF})^2=1-T_{\rm min}(E_{\rm TLF})/T_1$ appears. For low temperatures it can be assumed that the maximum TLF energy $E_{\rm max} \gg k_\mathrm{B}T$. The evaluation of the resulting integrals over $E_{\rm TLF}$ and $T_1$ then leads to an averaged dipole-fluctuation spectrum of the form
    \begin{equation}\label{eq:SwTunnel}
    \bar S_\mu \approx P_0\frac{\pi k_{\rm B}T}{4 \omega}\mu^2.
    \end{equation}

By assuming a similar range for the distributions of energy offsets, $\Delta_{\rm max}$, and barrier heights, $\Delta V$, this result is very similar to that in Eq.~\eqref{eq:SwThermal}, which was derived for thermally activated TLFs. It may be noted that for the same temperature, similar noise levels for tunneling states and thermally activated TLFs are expected. However, it should be emphasized that the common linear scaling with $T$ in Eqs.~\eqref{eq:SwThermal} and \eqref{eq:SwTunnel} is coincidental. In the low-temperature regime the factor $T$ accounts for the number of TLFs that are thermally occupied and contribute to the spectrum. In the high-temperature limit the scaling with $T$ is related to the thermally activated switching rate.

\subsubsection{Non-uniform distributions of activation energies}
\label{subsubsec:NonUniformTLFDistribution}

In Eq.~\eqref{eq:SwThermal} the $\omega^{-1}$ scaling of the noise spectrum with frequency and its linear scaling with temperature follow from the assumption of a uniform distribution of activation energies, $V_\mathrm{b}$. To explain the deviations from this behavior which are observed in current- and voltage-noise spectra in metals,
TLF models with non-uniform distributions $P(V_\mathrm{b})$ have been proposed \citep{Dutta:1981}.
By rewriting Eq.~\eqref{eq:Sbar} as an integral over $P(V_\mathrm{b})$ it can be shown that the dominant contribution to the averaged spectrum comes from TLFs with a barrier height $V_\mathrm{b}\approx -k_\mathrm{B}T \log(\omega \tau_0)$ and for a non-uniform $P(V_\mathrm{b})$ the modified spectrum is of the form
    \begin{equation}
    \label{eq:SwPV_b}
    \bar S_\mu \approx \mu^2 \left(\frac{\pi k_\mathrm{B}T}{4\omega}\right) P\left(V_\mathrm{b} = -k_\mathrm{B}T \log(\omega \tau_0)\right).
    \end{equation}

This result shows that any $P(V_\mathrm{b})\neq const.$ leads to a non-linear temperature scaling of the TLF noise combined with corrections to the strict $\omega^{-1}$ frequency dependence. By defining a local frequency-scaling exponent $\alpha(\omega,T)= -\partial \ln \bar{S}_\mu/\partial \ln \omega$, Eq.~\eqref{eq:SwPV_b} predicts a general relation between frequency and temperature dependence of the noise \citep{Dutta:1981}:
    \begin{equation}
    \alpha(\omega,T)= 1-\frac{1}{\ln(\omega \tau_0)} \left[ \frac{\partial \ln \bar S_\mu}{\partial \ln T} -1\right].
    \end{equation}

For example, a power-law distribution $P(V_\mathrm{b})\sim V_\mathrm{b}^{\gamma-1}$ of activation energies results in $T^\gamma$ scaling of the noise \citep{Labaziewicz:2008_10}, and
    \begin{equation}
    \alpha(\omega)= 1-\frac{(\gamma-1)}{\ln(\omega \tau_0)}.
    \end{equation}
For the frequency range of interest $ \log(\omega \tau_0)\approx -10$ and, for $\gamma >1$, a small increase in the scaling exponent $\alpha$ is expected.
A more natural scenario would involve a peaked distribution of barrier heights.
Considering, for example, a Lorentzian distribution of barrier heights centered at $V_0$ and with a width $\Delta V$ \citep{Dutta:1981},
    \begin{equation}
    P(V_\mathrm{b})= \frac{1}{\pi} \frac{\Delta V}{\Delta V^2+(V_\mathrm{b}-V_0)^2)}.
    \end{equation}

\begin{figure}[t]
\begin{center}
\includegraphics [width=\columnwidth]{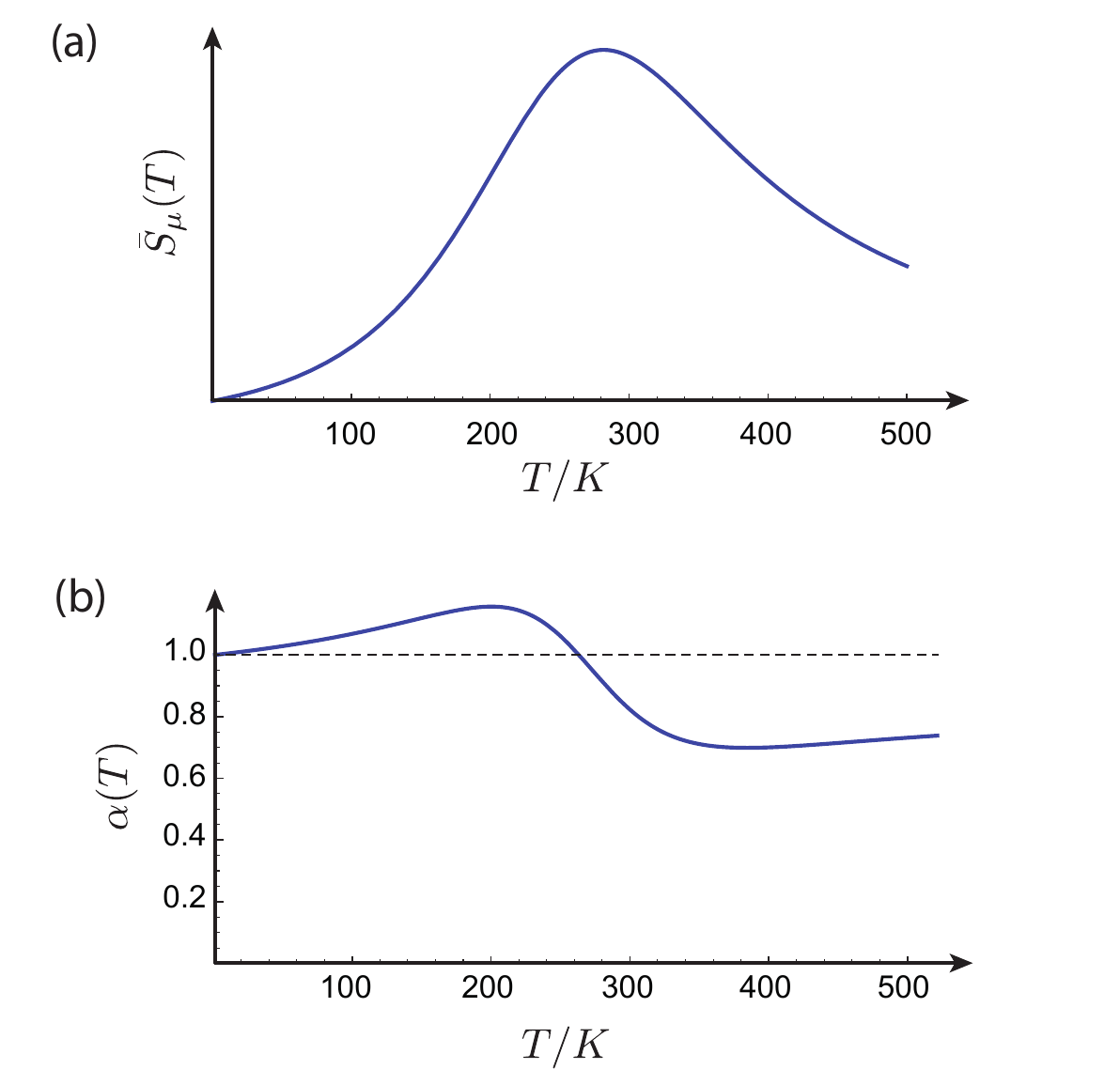}
\caption{Temperature-dependent behavior of non-uniformly distributed TLFs. (a) Temperature dependence of the averaged dipole-fluctuation spectrum $\bar S_\mu$ (in arbitrary units) produced by an ensemble of TLFs with a Lorentzian distribution of activation energies centered around $V_0=0.3$ eV and with a width $\Delta V=0.15$ eV. For the same parameters (b) shows the temperature dependence of the local scaling exponent $\alpha(\omega,T)$.}
\label{fig:TLF_Temp}
\end{center}
\end{figure}

For this distribution, and a particular choice of some physically reasonable parameters ($V_0=0.3$ eV, $\Delta V=0.15$ eV and $\omega \tau_0\approx 10^{-6}$), the temperature dependence of $\bar S_\mu$ is plotted in Fig.~\ref{fig:TLF_Temp} together with the local scaling exponent $\alpha(\omega,T)$. This example illustrates that an ensemble of TLFs with a peaked distribution of activation energies can exhibit a maximum as a function of temperature with a corresponding change of the scaling exponent from $\alpha >1$ to $\alpha< 1$. For the present example the maximum value of $\bar{S}_\mu$ occurs at $T\approx 300$\,K. Noise measurements in other systems at much lower frequencies are consistent with $V_0\approx 1$ eV~\citep{Dutta:1981}. For ion-trap experiments where $\omega\sim$~MHz, the corresponding maximum would in this case occur at much higher temperatures of $T \approx 850$ K.\\

In the case of tunneling states, a non-uniform distribution $P(\lambda)$ for the tunneling parameter can lead to modifications of the spectrum given in Eq.~\eqref{eq:SwTunnel}. In analogy to Eq.~\eqref{eq:SwPV_b} one obtains a scaling $\bar S_\mu\sim P(\lambda=\lambda_0)$ where $\lambda_0$ determines the dominant contribution in the average in Eq.~\eqref{eq:Sbar} and depends logarithmically on $\omega$. However, since the tunneling amplitude, $\Delta_0$, does not depend on temperature, the modifications are less pronounced than in the case of thermally activated TLFs and do not lead to a significant change in the temperature scaling.

\subsection{Adatom dipoles}
\label{subsec:AdatomDipoles}

Even on a very regular surface and at low background pressures, atoms or molecules in the atmosphere will stick to the metal electrodes and create extended layers, small patches or individual defects. The induced dipole moments of adsorbed atoms locally lower the work function of the metal and provide one of the mechanisms that can lead to the formation of patch potentials described in Sec.~\ref{subsec:PatchPotentials}. For alkali atoms adsorbed on a gold surface the induced dipole moments are typically a few debye and individual adatomic dipoles can be measured and visualized using field-emission microscopy \citep{Gomer:1961}. This method also allows the observation of adatom diffusion, a process which is further analyzed in Sec.~\ref{subsec:AdatomDiffusion}.

Surface contamination has been observed to have an influence on trapped-ion heating rates in several experiments \citep{DeVoe:2002, Turchette:2000, Letchumanan:2007, Hite:2012,Daniilidis:2014}. This suggests that adatoms may play an important role in the electric-field noise generated at small distances. The direct effect that a large, artificially created surface dipole has on trapped ions been observed by \citet{Harlander:2010}. Recently \citet{SafaviNaini:2011,SafaviNaini:2013} have analyzed the electric-field noise spectrum which is expected from a random distribution of adatomic dipoles on a planar electrode. This model, which is briefly summarized here, is predicated on the idea that random fluctuations of surface dipoles can arise from phonon-induced transitions between different bound vibrational states of the adatom-surface potential.

\subsubsection{Adatoms}

\begin{figure}[t]
\begin{center}
\includegraphics [width=\columnwidth]{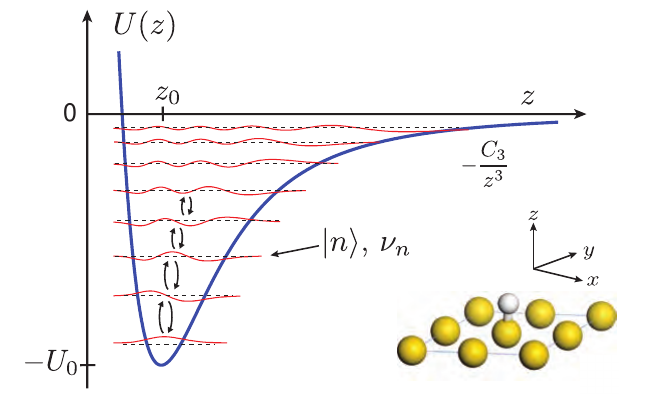}
\caption{Sketch of the typical shape of the adatom-surface potential $U(z)$ approximated by the model potential given in Eq.~\eqref{Uz} for $w=5.8$. The dashed lines indicate the energies $\hbar \nu_n$ of the bound vibrational states with wave functions indicated by the red solid lines. [Reprinted figure from \citet{SafaviNaini:2013} and used with permission. \copyright 2000 by the American Physical Society.]}
\label{fig:AdatomPotential}
\end{center}
\end{figure}

Atoms or molecules are attracted towards a nearby surface by the van der Waals potential, $U_{\rm vdW}(z) \sim -C_3/z^3$, where $z$ is the distance to the surface and $C_3$ is a function of the dynamic polarizability, $\alpha_{\rm p}(\omega)$, of the particle. At short distances the electronic wave functions of the adatom and the surface constituents start to overlap and lead to a sharply repulsive potential wall. The resulting shape of the full adatom-surface potential, $U(z)$, is sketched in Fig.~\ref{fig:AdatomPotential}, and can be calculated for specific adatoms using \emph{ab initio} numerical methods. For many studies of adatom-surface interactions it is sufficient to replace the actual potential by an approximate analytic model, for example of the form \citep{Hoinkes:1982}
    \begin{equation}
    \label{Uz}
    U(z)= \frac{w}{w - 3}
    U_0 \left[\frac{3}{w} \,e^{w (1-z/z_0)} -\left( \frac{z_0}{z}\right)^3\right],
    \end{equation}
which is parameterized by the equilibrium position, $z_0$, the depth of the potential, $U_0$, and the dimensionless parameter $w\sim 5-10$ characterizing the width of the potential well. The potential depth, $U_0$, depends strongly on the surface and adatom species and can range from a few eV for strongly bound (chemisorbed) atoms like H, to values of $\sim 10$ meV for weakly bound (physisorbed) atoms like He or Ne \citep{SafaviNaini:2013}. The potential minimum at $z_0$ occurs at atomic distances of a few \AA\, from the surface.
The characteristic frequency is set by the frequency difference between the two lowest levels: $\nu_{10}=\nu_1-\nu_0$. Using a harmonic approximation of the potential well this can be estimated to be
    \begin{equation}
    \label{eq:AdatomPotentialEstimate}
    \nu_{10}\approx \zeta\sqrt{\frac{U_0}{m_{\rm ad} z_0^2} },
    \end{equation}
where $\zeta=\sqrt{3(w^2-4w)/( w -3)}$ is a numerical factor and $m_{\rm ad}$ is the mass of the adatom.

The attractive adatom-surface potential arises from a rearrangement of the electronic wave functions of the adatom and the electrons in the metal. This creates an induced dipole moment, $\mu(z)$, perpendicular to the surface, which depends on the adatom-surface distance, $z$. At large distances a scaling $\mu(z) ~\sim 1/z^4$ is expected \citep{Antoniewicz:1974} and $\mu(z\approx z_0)$ can reach several debye when the adatom touches the surface.

\subsubsection{Phonon-induced dipole fluctuations}

The potential, $U(z)$, supports several bound vibrational states, $|n\rangle$, with vibrational frequencies $\nu_n$ as indicated in Fig.~\ref{fig:AdatomPotential}.
Due to the dependence of $\mu(z)$ on the adatom-surface distance, each vibrational state acquires a different average dipole moment $\mu_n=\langle n|\mu(z)|n\rangle$. \citet{SafaviNaini:2011} suggested that at non-zero temperatures, phonon-induced transitions between different vibrational states, $|n\rangle$, lead to a fluctuating adatom dipole moment, $\mu(t)$.

Phonon-induced transitions between two vibrational states $|n\rangle$ and $|m\rangle$ arise from fluctuations of the closest surface atoms, which thereby modulate the potential $U(z)$. For $n>m$ and a transition frequency $\nu_{nm}=\nu_n-\nu_m>0$ the resulting transition rates are approximately given by \citep{SafaviNaini:2011}
    \begin{eqnarray}
    \label{eq:rates1}
    \Gamma_{n\to m}&=&\frac{\pi \bar g(\nu_{nm}) }{3 \hbar M \nu_{nm}} \vert \langle n \vert U^\prime(z)\vert m \rangle \vert ^2 \left(n_{\rm B}(\nu_{nm})+1\right),\\
    \label{eq:rates2}
    \Gamma_{m\to n}&=&\frac{\pi \bar g(\nu_{nm}) }{3 \hbar M \nu_{nm} } \vert \langle n \vert U^\prime(z)\vert m \rangle \vert ^2 n_{\rm B}(\nu_{nm}),
    \end{eqnarray}
where $M$ is the surface-atom mass and $n_{\rm B}(\omega)=1/(e^{\hbar \omega/(k_\mathrm{B}T)}-1)$ is the Bose-Einstein distribution for a particular temperature, $T$. In the above expressions $\bar g(\omega)$ denotes the partial (projected) phonon density of states (PDOS) of a surface atom (which takes into account the fact that different phonon modes can couple more efficiently or less efficiently to surface atoms than to bulk atoms). The decay rate from the first excited vibrational state to the ground state at zero temperature $\Gamma_0\equiv \Gamma_{1\to0}(T=0)$ defines a characteristic transition rate which, by using the bulk PDOS $g(\omega)= 3 a_{\rm L}^3\omega^2/(2\pi^2 v^3)$ as a reference, can be estimated as
    \begin{equation}\label{eq:Gamma0}
    \Gamma_0\approx \frac{1}{4\pi} \frac{\nu_{10}^4 m_{\rm ad}}{v^3\rho}.
    \end{equation}
Here $\rho$ is the density of the bulk material, $v$ is the velocity of sound in the material and  $a_{\rm L}$ is the lattice constant.

Considering typical numbers, for a gold surface ($\rho=19.3$\,g/cm$^3$ and $v=3240$\,m/s) with an adsorbate covering of atomic mass $m_{\rm ad}=100$\,amu, and given a typical vibrational frequency $\nu_{10}/2\pi=1$\,THz, Eq.~\eqref{eq:Gamma0} predicts values around $\Gamma_0\approx 3\times 10^{10}$ s$^{-1}$. Much lower values in the MHz regime are expected for very heavy or weakly bound adatoms, where vibrational frequencies are reduced. Furthermore, it has been shown by \citet{SafaviNaini:2013} that, due to a mass mismatch, the presence of an additional monolayer of contaminants on the metallic electrode can considerably modify the surface PDOS. This also significantly reduces $\Gamma_0$ compared to the estimate in Eq.~\eqref{eq:Gamma0} and in general values of $\Gamma_0$ ranging from a few $10^7$ s$^{-1}$ to a few $10^{10}$ s$^{-1}$ can be expected.

\subsubsection{Noise spectrum}
\label{subsubsec:ADF_NoiseSpectrum}

For a planar trap geometry the spectral density of electric-field noise produced by a homogeneous distribution of independent adatoms is given by Eq.~\eqref{eq:SE_Dipoles}, with $\bar S_\mu$ being replaced by the dipole-fluctuation spectrum $S_\mu$ of a single adatom. The fluctuating dipole moment of a single adatom is given by $\mu(t)= \sum_n \mu_n P_n(t)$, where the occupancies of the vibrational levels, $P_n(t)$, evolve according to the rate equation
    \begin{equation}
    \dot P_n(t) = - \sum_{m\neq n}\Gamma_{n\rightarrow m} P_n(t) + \sum_{m\neq n} \Gamma_{m\rightarrow n} P_m(t).
    \end{equation}
From this equation the steady-state populations $\langle P_n(t)\rangle$ and correlation functions $\langle P_n(t)P_m(0)\rangle$ can be obtained numerically and used to calculate $\langle \delta \mu (t)\delta \mu(0)\rangle$.

\begin{figure}[t]
\begin{center}
\includegraphics[width=\columnwidth]{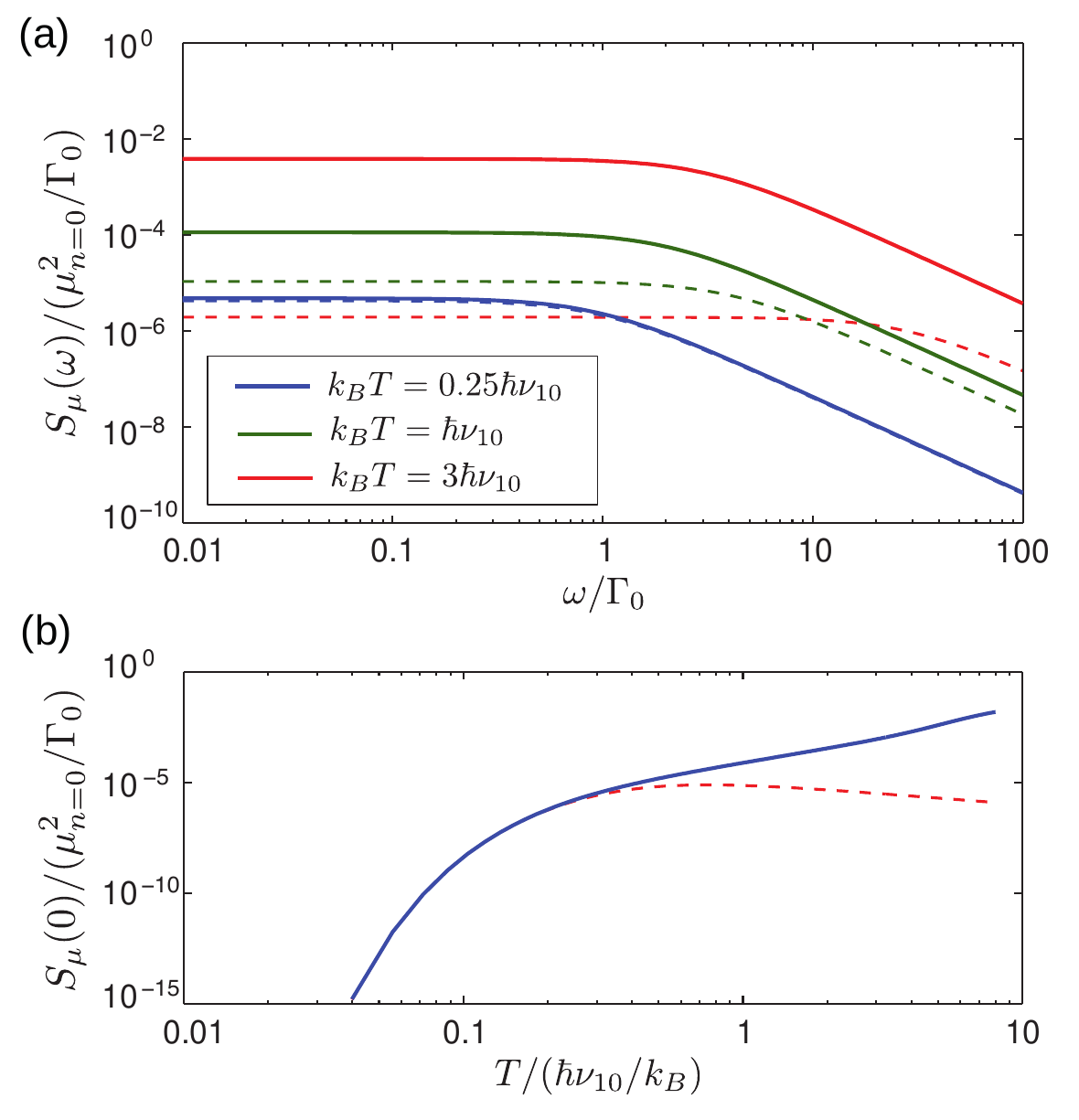}
\caption{Dipole-fluctuation spectrum $S_\mu$ of a single adatom. (a) Dependence of the full spectrum on the rescaled frequency $\omega/\Gamma_0$ for three different temperatures. The dashed lines show the corresponding spectrum under a two-level approximation.
(b) Temperature dependence of $S_\mu(\omega\rightarrow 0)$ for the full multilevel model (solid line) and a two-level system (dashed line).}
\label{fig:AdatomSpectrum}
\end{center}
\end{figure}

Figure~\ref{fig:AdatomSpectrum} shows the resulting noise spectrum as a function of frequency for different temperatures.
For this specific example an intermediate binding energy of $U_0=0.25$ eV ($\approx 3000$ K) and a mass of $m_{\rm ad}=100$\,amu has been assumed,
where $\nu_{10}/2\pi\approx 1.3$\,THz. Note that hydrocarbon chains of similar mass have been identified as electrode contaminates in a recent experiment by \citet{Daniilidis:2014}.
At low temperatures only the lowest two vibrational levels are occupied and $S_\mu$ can be approximated by a Lorentzian TLF spectrum [see Eq.~\eqref{eq:TLFSpectrum}]:
    \begin{equation}
    S_\mu\simeq (\mu_0-\mu_1)^2 \frac{2\Gamma_0}{\omega^2+\Gamma^2_0} \,e^{-\frac{\hbar \nu_{10}}{k_\mathrm{B}T}}.
    \end{equation}
The spectrum is flat for $\omega \ll \Gamma_0$ and decays as $\omega^{-2}$ in the opposite limit of large frequencies. The fluctuations are thermally activated with a characteristic temperature $T^*=\hbar \nu_{10}/k_\mathrm{B}$ defined by vibrational frequencies. In the example shown in Fig.~\ref{fig:AdatomSpectrum} this corresponds to a temperature of $T^*\approx 60$ K. Due to the multi-level structure of the bound adatom the full spectrum does not saturate at higher temperatures, but instead increases as $\sim T^{\gamma}$, where $\gamma\approx 2.5$.  For a surface density $\sigma_{\rm d}=10^{18}$ m$^{-2}$ which, for an Au(111) surface, corresponds to a coverage fraction of approximately 10\%, an induced dipole moment of $\mu_{n=0}=5$\,D and assuming $\Gamma_0/2\pi=10^7$\,s$^{-1}$ the predicted electric-field noise level at a distance $d=100\,\mu$m above the trap is around $S_\mathrm{E}\approx 10^{-13}$\,\B. Taking into account the distance scaling of $\beta=4$ [see Eq.~\eqref{eq:SE_Dipoles}] this is comparable to the absolute level of heating seen in some small traps (see Fig.~\ref{fig:S_E_d}).

The mechanism described by \citet{SafaviNaini:2011} shows that phonon-induced fluctuations of individual adatomic dipoles can contribute to the electric-field noise observed in good ion traps. It should be emphasized, however, that the predicted noise levels depend sensitively on the microscopic parameters of the adatom-surface interaction. For many atoms or more complicated molecules these are not known. Eq.~\eqref{eq:TLFSpectrum} shows that to realize a noise level which can be seen in experiments and exhibits a finite frequency variation at $\omega\sim$ MHz requires low values for $\nu_{10}$ and $\Gamma_0$. This is expected for very weakly bound or very heavy adatoms and molecules. Most numerical studies in the literature have focused on extreme cases of alkaline atoms or noble gases absorbed on metals, which either lead to strongly bound atoms or, in the latter case, lead to very shallow potentials which do not support binding at room temperatures. Therefore, the relevance of this noise mechanism for specific surface conditions found in ion-trap experiments remains subject of future experimental and theoretical investigations.

\subsection{Adatom diffusion}
\label{subsec:AdatomDiffusion}

Apart from fluctuations in the magnitude of their induced dipole moment, ${\boldsymbol \mu}(t)$, adsorbed atoms and molecules can contribute to the electric-field noise by diffusing on the surface, thereby changing the spatial distribution of dipoles over time. Adatom-diffusion-induced noise processes have been studied, for example, in the context of field-emission microscopy \citep{Gomer:1961,Kleint:1963E,Timm:1966}, where the field-emission current depends sensitively on the total number of adatoms within a small area at the end of a sharp tip. Adatoms which diffuse in and out of that area change the average work function and lead to fluctuations of the emission current. The observed plateau of this noise at low frequencies and the scaling $\sim\omega^{-3/2}$ \citep{Timm:1966} for large frequencies agree well with predictions from simple diffusion models \citep{Burgess:1953,VanVliet:1965,Gesley:1985}. The diffusion of adatomic dipoles has been suggested as a potential mechanism for ion-trap heating \citep{Wineland:1998_05}. This hypothesis is supported by the $\sim\omega^{-3/2}$ scaling of heating rates observed in experiments at NIST \citep{Turchette:2000,Hite:2012}.

\subsubsection{Adatom diffusion on surfaces}

The motion of individual adatoms on a planar surface is well described by a 2D diffusion process with diffusion constant $\mathcal{D}= a_{\rm L}^2 \Gamma_{\rm hop}/\ell$ \citep{Gomer:1961}. Here $a_{\rm L}$ is the lattice constant, $\Gamma_{\rm hop}$ is the hopping rate between neighboring adsorption sites, which are separated by an energy barrier $V_{\rm b}$, and $\ell$ is the coordination number which depends on the lattice geometry, e.g. $\ell=4$ for a square lattice. The hopping of adatoms between adjacent minima of the surface potential can be thermally activated $(\Gamma_{\rm hop}\approx \tau_0^{-1} e^{-V_{\rm b}/k_\mathrm{B}T}$) or be due to quantum tunneling. The resulting diffusion constant can be approximated by
    \begin{equation}\label{eq:D}
    \mathcal{D} \simeq \mathcal{D}_{\rm t} + \mathcal{D}_{0} e^{ -\frac{V_{\rm b}}{k_{\rm B}T} },
    \end{equation}
where the first term describes the diffusion due to quantum tunneling and the second term describes the thermally activated diffusion. For most surface diffusion processes one finds $\mathcal{D}_0\approx 10^{-7}$\,m$^{2}$s$^{-1}$ \citep{Gomer:1961,Zhdanov:1991},
which is consistent with an attempt frequency of $\tau_0^{-1}\approx10^{12}-10^{13}$\,s$^{-1}$
and a lattice spacing of a few \AA. In the temperature range $T=100-400$\,K and assuming a typical energy barrier of
$V_{\rm b}\sim 150$\,meV ($\sim 1750$\,K) \citep{Gomer:1961,Zhdanov:1991} the resulting diffusion constants are
$\mathcal{D}\approx 10^{-15}-10^{-9}$\,m$^2$s$^{-1}$. At very low temperatures thermally activated processes are strongly suppressed
and the diffusion constant saturates at a value $\mathcal{D}_{\rm t}$, which is set by a finite probability to tunnel through the energy barrier.
While a precise evaluation of $\mathcal{D}_{\rm t}$ is rather involved and beyond the scope of this review, an estimate for the saturation temperature can be obtained by comparing the thermally activated rate, $\Gamma_{\rm hop}$, with the coherent tunneling amplitude, $\Delta_0/\hbar$, given in Eq.~\eqref{eq:TunnelingAmplitude}. Apart from hydrogen adatoms, for which quantum effects are important even at room temperature, the saturation of the hopping rate typically occurs at few tens of kelvins.

\subsubsection{Adatom-diffusion-induced noise}

\begin{figure}[t]
\begin{center}
\includegraphics [width=\columnwidth]{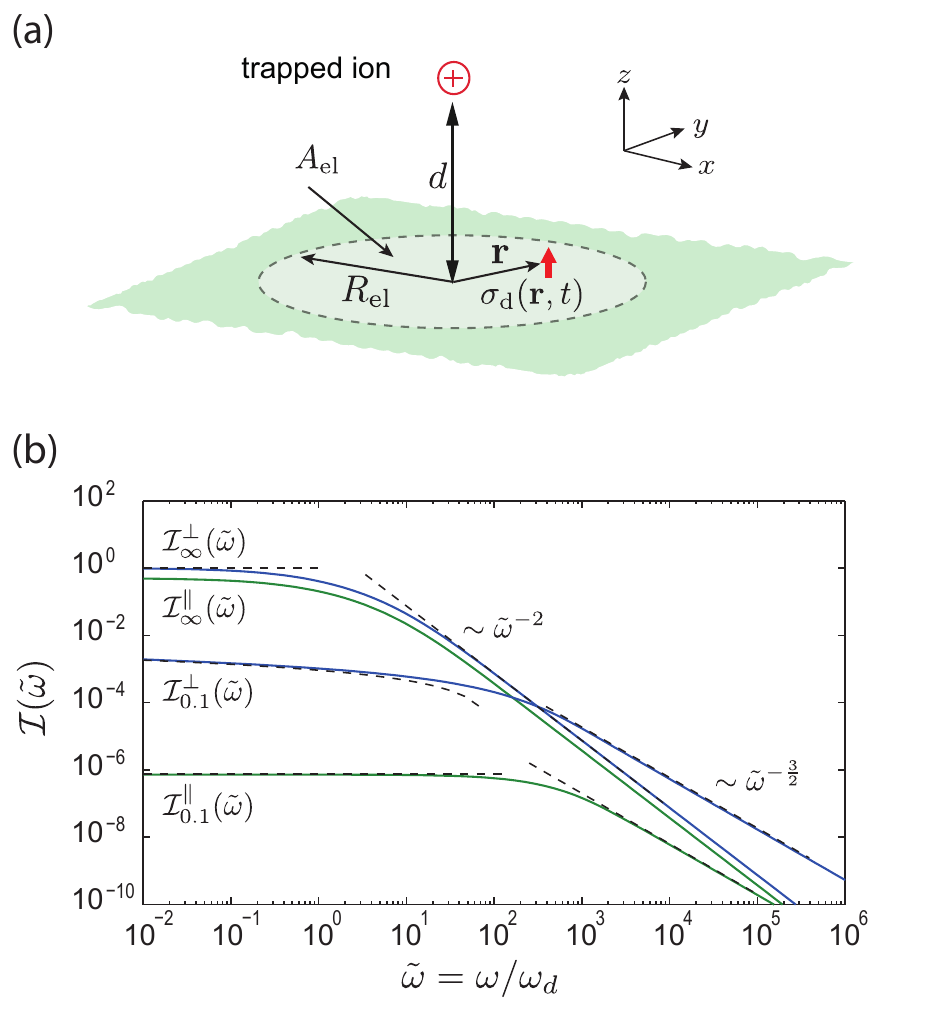}
\caption{Adatom diffusion model.
(a) Setup considered for the evaluation of electric-field noise generated by diffusing dipoles on a planar electrode. Within this model a needle-like electrode can be approximately described by only considering the electric field from dipoles within a small electrode area, $A_{\rm el}$.
(b) Dependence of the normalized spectral function $\mathcal{I}^\eta_{\tilde R}(\tilde \omega)$ defined in Eq.~\eqref{eq:Jtildeomega} on the scaled frequency
$\tilde \omega=\omega/\omega_{\rm d}$, where $\omega_{\rm d}=\mathcal{D}/d^2$. The dashed lines indicate the analytic limits given in Eqs.~\eqref{eq:Jinfty}, \eqref{eq:Jinfty_HighFreq}, \eqref{eq:Jtip_perp_low}, \eqref{eq:Jtip_perp}, \eqref{eq:Jtip_para_low} and \eqref{eq:Jtip_para}.}
\label{fig:DiffusionFigure}
\end{center}
\end{figure}

The diffusing adatoms can be modeled by a surface polarization density $\mathcal{P}({\bm r},t)= \mu \sigma_{\rm d}({\bm r},t)$, where $\sigma_{\rm d}({\bm r},t)$ is the areal density of adatoms with a fixed dipole moment $\mu$. For an ion located at a distance $d$ above a planar electrode [see Fig.~\ref{fig:DiffusionFigure}(a)], the resulting electric-field noise spectrum is
    \begin{equation}
    \begin{split}
    S^{\rm (AD)}_\mathrm{E}= &\frac{\mu^2}{8\pi^2\epsilon_0^2} \int_{\mathcal{S}}d^2r_1 \int_{\mathcal{S}} d^2r_2 \\
    & g_D ({\bm r}_1) g_D({\bm r}_2) C_\sigma({\bm r}_1, {\bm r}_2,\omega).
    \end{split}
    \end{equation}
Here the geometrical factors $g_D(\bm r)$ describe the dipole pattern and are given in Eqs.~\eqref{eq:gDz} and~\eqref{eq:gDx} for trap axes perpendicular and parallel to the electrode surface, respectively, and
    \begin{equation}
    \label{eq:Cn}
    C_\sigma({\bm r}_1, {\bm r}_2,\omega) = 2{\rm Re}\int_{0}^{\infty} d\tau \, \langle \delta \sigma_{\rm d}({\bm r}_1,\tau) \delta \sigma_{\rm d}({\bm r}_2,0)\rangle e^{-i\omega \tau}
    \end{equation}
is the correlation spectrum of the density fluctuations, $\delta \sigma_{\rm d}({\bm r},t) = \sigma_{\rm d}({\bm r},t) -\langle \sigma_{\rm d}({\bm r},t)\rangle$. At low adsorbate densities the diffusion of the individual dipoles is independent and the mean value of the density, $\sigma_{\rm d}({\bm r},t)$, obeys the 2D diffusion equation
    \begin{equation}
    \left(\partial_{\rm t} - \mathcal{D} \vec \nabla ^2 \right) \langle \sigma_{\rm d}({\bm r},t)\rangle =0.
    \end{equation}
For a sufficiently homogeneous surface the adatom distribution relaxes to the stationary value $\bar{\sigma}_{\rm d}$. For $\tau>0$ fluctuations around this value are given by \citep{Gesley:1985}
    \begin{equation}
    \langle \delta \sigma_{\rm d}({\bm r}_1,\tau) \delta \sigma_{\rm d}({\bm r}_2,0)\rangle =
    \frac{\bar{\sigma}_{\rm d}}{4\pi \mathcal{D} \tau} \, e^{-\frac{|{\bm r}_1-{\bm r}_2|^2}{4\mathcal{D}\tau}}.
    \end{equation}
Under these idealized conditions the correlation function [Eq.~\eqref{eq:Cn}] can be evaluated analytically and the general expression for the electric-field fluctuation spectrum is
    \begin{equation}
    \label{eq:SE_Diffusion}
    \begin{split}
    S^{\rm (AD)}_\mathrm{E}= &\frac{\mu^2 \bar{\sigma}_{\rm d}}{8\pi^3 \epsilon_0^2 \mathcal{D} } \int_{\mathcal{S}} d^2 r _1 \int_{\mathcal{S}} d^2 r _2 \, \\
    & g_D ({\bm r}_1) g_D ({\bm r}_2) {\rm Ker}_0\left(\sqrt{|{\bm r}_1-{\bm r}_2|^2 \omega/\mathcal{D}}\right).
    \end{split}
    \end{equation}
Here $ {\rm Ker}_0(x)$ is the zeroth-order Kelvin function \citep{Abramowitz:1972}. In contrast to the patch-potential model or the fixed dipoles considered above, the correlation function $C_\sigma({\bm r}_1,{\bm r}_2,\omega)$ is not separable into spatial and temporal correlations. The mean diffusion length of an adatom depends on the available time, which is inversely proportional to the frequency of the noise component considered. Consequently the distance and frequency scaling of $S^{\rm (AD)}_\mathrm{E}$ are tightly connected.

\subsubsection{Diffusion on smooth surfaces}
\label{subsubsec:PlaneAndNeedleDiffusion}

In the limit where adatoms diffuse freely over an infinite planar electrode the only relevant length scale is set by the ion-surface distance, $d$, which defines a characteristic frequency scale $\omega_{\rm d}= \mathcal{D}/d^2$. For a finite-sized trap, the electrode size, $R_{\rm el}$, introduces a second length scale with a corresponding frequency scale, $\omega_{\rm R}=\mathcal{D}/R_{\rm el}^2$. To obtain analytic estimates for diffusion-induced noise for both large and small trapping electrodes the full problem can be approximated by a scenario shown in Fig.~\ref{fig:DiffusionFigure}(a). Here adatoms diffuse on an infinite planar surface but, to account for a finite electrode size, only adatoms within an area
$A_{\rm el}=\pi R_{\rm el}^2$ contribute significantly to the noise seen by the ion.
Using the dimensionless parameters $\tilde{\omega}=\omega/\omega_{\rm d}$ and $\tilde{R}=R_{\rm el}/d$, Eq.~\eqref{eq:SE_Diffusion} can be rewritten as
    \begin{equation}
    S^{\rm (AD)}_{{\rm E},\eta=\perp,\parallel}= \frac{\mu^2 \bar{\sigma}_{\rm d}}{8\pi \epsilon_0^2 d^2 \mathcal{D} }\times \mathcal{I}^\eta_{\tilde R}\left(\tilde \omega\right),
    \end{equation}
where
$\mathcal{I}^\eta_{\tilde R} (\tilde \omega)$ is a dimensionless integral,
    \begin{equation}
    \label{eq:Jtildeomega}
    \begin{split}
    \mathcal{I}^\eta_{\tilde R}(\tilde \omega)=& \frac{1}{\pi^2}\int_{\leq \tilde R} d^2r_1 d^2r_2 \\
    &\tilde g_D({\bm r}_1) \tilde g_D({\bm r}_2) {\rm Ker}_0\left(\sqrt{|{\bm r}_1\!-\!{\bm r}_2|^2 \tilde \omega}\right),
    \end{split}
    \end{equation}
for which $\tilde g_D({\bm r}) = g_D({\bm r} d)$. The dependence of $\mathcal{I}^\eta_{\tilde R} (\tilde \omega)$ on $\tilde \omega$ is plotted in Fig.~\ref{fig:DiffusionFigure}(b) for the limiting cases $\tilde R\rightarrow \infty$ and $\tilde R\ll 1$.\\

We now consider two extremes of geometry: planar and needle. In each instance the absolute value of $S_\mathrm{E}$ is estimated, as are the expected scalings, $\alpha$, $\beta$, and $\gamma$, for reasonable trapping conditions.\\

\noindent $\bullet$ Infinite planar electrode\\
In the limit of the electrode dimension, $R_{\rm el}$, being large compared to the ion-surface distance, $d$, Eq.~\eqref{eq:Jtildeomega} can be evaluated for $\tilde R\rightarrow \infty$. In this case the parallel and perpendicular electric-field noise spectra only differ by a factor of two: $\mathcal{I}^\parallel_{\infty} (\tilde \omega)= \mathcal{I}^\perp_{\infty} (\tilde \omega)/2$.
The respective low- and high-frequency limits are given by
    \begin{eqnarray}
    \mathcal{I}^\perp_{\infty} (\tilde \omega\rightarrow 0) &\simeq& 1, \label{eq:Jinfty}\\
    \mathcal{I}^\perp_{\infty } (\tilde \omega\gg 1) &\simeq& \frac{15}{2\tilde \omega^2}. \label{eq:Jinfty_HighFreq}
    \end{eqnarray}
The crossover occurs at a frequency scale $\omega\approx \omega_{\rm d}=\mathcal{D}/d^2$, which is set by the ion-electrode separation, $d$. For an ion-electrode distance $d=100\,\mu$m and a diffusion constant of $\mathcal{D}=10^{-10}$\,m$^2$s$^{-1}$, the trapping frequencies of $\omt/2\pi \sim$\,MHz
are much larger than the frequency scale $\omega_{\rm d}/2\pi\approx 10$\,Hz. In this regime the resulting diffusion-induced-noise spectrum is
    \begin{equation}
    S^{({\rm AD})}_{\mathrm{E},\perp}\simeq \frac{15 \mu^2 \bar{\sigma}_{\rm d} \mathcal{D} }{16\pi \epsilon_0^2 d^6 \omega^2 },
    \end{equation}
and therefore $\beta=6$ and $\alpha=2$.
For adatom density of $\bar{\sigma}_{\rm d}=10^{18}$\,m$^{-2}$ and an induced dipole moment of $\mu=5$\,D, the overall noise level is around $S_\mathrm{E}\approx 10^{-18}$\,\B. Since $S_\mathrm{E}\sim \mathcal{D}$, diffusion-induced noise is exponentially suppressed at lower temperatures.\\

\noindent $\bullet$ Needle trap\\
In the opposite limit of a sharp needle trap the adatoms can still diffuse across the whole electrode, but the relevant contributions to the noise arise from adatoms within a small area $A_{\rm el}=\pi R_{\rm el}^2$ at the end of the tip. This situation is analogous to the diffusion processes studied in the context of field-emission-current noise \citep{VanVliet:1965,Gesley:1985} and is approximately captured by restricting the integration area in Eq.~\eqref{eq:SE_Diffusion} to a small disc of radius $R_{\rm el}\ll d$. In this limit the normalized electric-field fluctuations perpendicular to the surface scale as
    \begin{eqnarray}
    \mathcal{I}^\perp_{\tilde R} (\tilde \omega\rightarrow 0) &=& -4 \tilde R^4 \log(\sqrt{\tilde \omega }\tilde R),\label{eq:Jtip_perp_low} \\
    \mathcal{I}^\perp_{\tilde R} (\tilde \omega\gg 1) &=& \frac{\sqrt{32}\tilde R}{\tilde \omega^\frac{3}{2}}. \label{eq:Jtip_perp}
    \end{eqnarray}
The crossover occurs at a frequency scale $\omega\approx \omega_{\rm R}=\mathcal{D}/R_{\rm el}^2$,
which is set by the tip size, $R_{\rm el}$, rather than the ion-electrode separation, $d$. In the regime, $\omega\gg\omega_{\rm R}$, the overall noise spectrum for the small-tip geometry is
    \begin{equation}
    S^{\rm (AD)}_{\mathrm{E},\perp}\simeq \frac{\mu^2 \bar{\sigma}_{\rm d} R_{\rm el} \sqrt{\mathcal{D}} }{\sqrt{2}\pi \epsilon_0^2 d^6 \omega^\frac{3}{2} }.
    \end{equation}
For the field fluctuations parallel to the surface (which is to say, perpendicular to the axis of the needle) the respective limits are
    \begin{eqnarray}
    \mathcal{I}^\parallel_{\tilde R} (\tilde \omega\rightarrow 0) &=& \frac{3\tilde R^6}{4}, \label{eq:Jtip_para_low}\\
    \mathcal{I}^\parallel_{\tilde R} (\tilde \omega\gg 1) &=& \frac{9\tilde R^3}{\sqrt{2} \tilde \omega^\frac{3}{2}}.\label{eq:Jtip_para}
    \end{eqnarray}

Eq.~\eqref{eq:Jtip_perp} and \eqref{eq:Jtip_para} show that for different trap geometries the frequency scaling of the diffusion-induced noise can change from $\omega^{-2}$ to $\omega^{-3/2}$. The distance dependence in this case is still $S_\mathrm{E}\sim d^{-6}$ for perpendicular fluctuations and $S_\mathrm{E}\sim d^{-8}$ for parallel field fluctuations. For the same parameters as above and $R_{\rm el}/d=0.1$ the resulting field noise is $S_\mathrm{E}\approx 5\times 10^{-17}$\,\B.\\

In summary, the above analytic estimates shows that adatom diffusion on a small electrode can lead to a $\omega^{-3/2}$ scaling of the field noise, but in these simple scenarios the expected overall noise level is rather small and decays quickly with increasing $d$. This is related to the fact that compared to a spatially fixed, but fluctuating dipole, a dipole which is just displaced on the surface by a small distance $\Delta r$ changes the field at the position of the ion only by a fraction $\Delta r/d\ll 1$.

\subsubsection{Diffusion on corrugated surfaces}
\label{subsubsec:CorrugatedDiffusion}

The different analytic scalings for $S^{\rm (AD)}_\mathrm{E}$ presented above have been derived under the idealized assumptions that adatoms diffuse freely and independently from each other. Under more realistic conditions, surface corrugations, potential steps between atomic layers, and also the mutual influence of adatoms at high densities can modify these results. In general, potential barriers as well as high densities of adatoms tend to suppress diffusion and further reduce the associated noise. However, diffusion-induced noise can be considerably enhanced when taking into account other microscopic mechanisms that change the magnitude of the induced dipole moment, $\mu$.

Consider, for example, an adatom diffusing over a surface covered by patches which have a different work function than the rest of the electrode, such that the induced dipole moment changes by $\Delta \mu$ when the adatom diffuses onto one of the patches. The noise originating from diffusion over a single patch is then similar to the small electrode limit discussed above, with $R_{\rm el}$ replaced by the patch radius $R_{\rm p}$ and $\mu$ replaced by $\Delta \mu$. The absolute level of noise, however, would be enhanced by the number of patches, $N_{\rm p}\sim d^2/R_{\rm p}^2$. Overall, this generalized diffusion model predicts a scaling
    \begin{equation}
    S^{\rm (AD)}_{\mathrm{E},\perp}\sim \frac{(\Delta \mu)^2 \sqrt{\mathcal{D}}}{ d^4 R_{\rm p} \omega^{3/2}}.
    \end{equation}
For $R_{\rm p}\sim1-10\,\mu$m, and assuming otherwise the same parameters for the estimate made in Sec.~\ref{subsubsec:PlaneAndNeedleDiffusion}, noise levels of about $S_\mathrm{E}\sim 10^{-12}-10^{-14}$\,\B are expected. This is similar to the level of noise expected from other microscopic processes. Unless non-equilibrium processes are considered, the diffusion-induced noise is still exponentially suppressed with temperature and a strong reduction of this noise contribution below a few tens of kelvin is in general expected.

\section{Known Mechanisms}
\label{sec:KnownMechanisms}

Section~\ref{sec:ExperimentalOverview} considered the observed spectral density of electric-field noise in different trapped-ion experiments. It suggested that the different experiments may well be limited by different sources of noise. It also underlined the problems of attempting to draw general conclusions either from specific experiments or from the full data set. Sections~\ref{sec:Theory} and \ref{sec:Microscopic} then considered the theoretical expectations arising from models of different heating mechanisms. This section now considers a number of experiments in which it is possible to identify a specific (probable) heating mechanism. Each section heading gives a possible or likely cause of heating which has been identified for specific experiments. The sources of heating in those cases where it can be identified are seen to be diverse. Consequently, the electric-field noise observed in ion traps should not be viewed as a single homogeneous phenomenon, but as many separate phenomena which can be different in different experiments. Despite such diversity, the case studies in this section shed light on a number of possible solutions to the heating seen in some ion traps, which might be implemented in other experiments.

\begin{figure*}[t]
\begin{center}
\includegraphics[width=\textwidth]{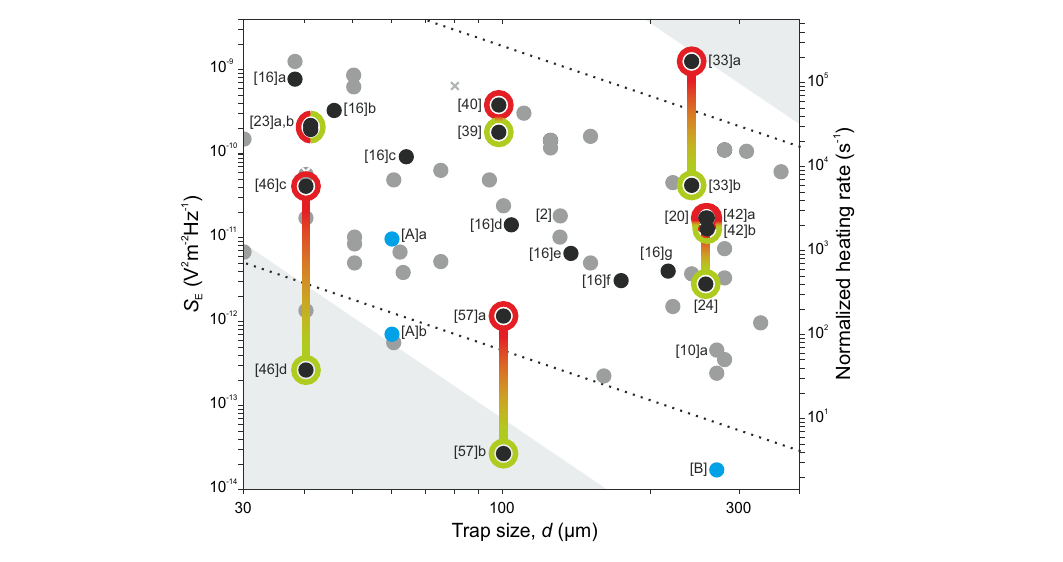}
\caption{Overview of traps for which the heating mechanisms are known. For comparison, the heating rate for the first use of ``anomalous heating" in the literature is shown as point [$\Monroe95$] \citep{Monroe:1995_11}. Experimental data points are taken from the relevant references in table~\ref{tab:References}. Points [A] and [B] are calculated values from \citet{Leibrandt:2007_01}. The gray points (noise sources unknown) are the same as those in Fig.~\ref{fig:S_E_d}. For orientation, the gray shaded regions and dotted lines are the same as those shown in Fig.~\ref{fig:S_E_d}, and indicate slopes of -4 and -2 respectively. Despite the experiments being limited by many different sources of noise, the reported values of $S_\mathrm{E}$ in the various experiments are broadly similar. Consequently, it is difficult to directly infer a particular noise source simply from the absolute level of noise.}
\label{fig:KnownMechanisms}
\end{center}
\end{figure*}

\subsection{Patch potentials}
\label{subsec:NeedleTrap}

\citet{Deslauriers:2006_09} used a trap consisting of a pair of tungsten needle electrodes for which the tip-to-tip separation, 2$d$, could be controllably varied. Single $^{111}$Cd$^+$ ions were ground-state cooled, and the axial heating rate was measured. Heating-rate measurements were made for seven different ion-electrode spacings in the range $38\,\mu\mathrm{m}<d<216\,\mu$m. The results are reproduced here in Fig.~\ref{fig:Deslauriers06_09Fig4}. Taking a weighted fit to the data, they report a distance-scaling exponent of $\beta=3.5(1)$. An analysis of the evidence for patch-potential heating mechanisms, over against Johnson noise, is provided here.  For comparison with the other results discussed in this section, the heating rates measured by \citeauthor{Deslauriers:2006_09} are plotted as points [\Deslauriers06]a-g in Fig.~\ref{fig:KnownMechanisms}.

\begin{figure}[t]
\begin{center}
\includegraphics[width=0.45\textwidth]{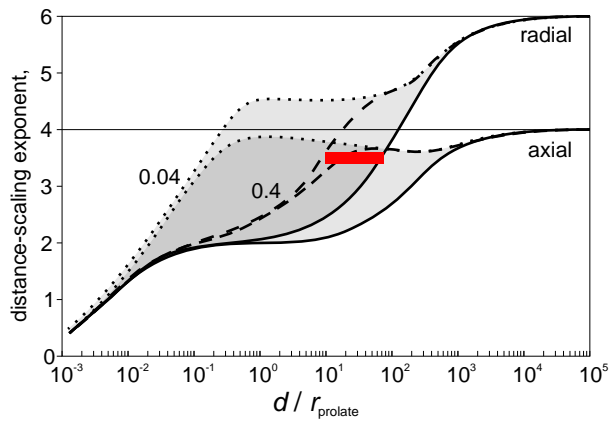}
\caption{Distance-scaling exponent, $\beta$, as a function of dimensionless distance, $d/r_\mathrm{prolate}$, for both axial and radial modes above a spheroidal needle of radius $r_\mathrm{prolate}$ and half-length 100~$r_\mathrm{prolate}$ [adapted from \citep{Low:2011}. Used with permission.] The solid lines are for the limit of a single patch covering the entire electrode. The dotted lines correspond to intermediate patches sizes or angular extent $\theta_\zeta$ = 0.4, 0.04. The red bar shows the approximate range of ion-electrode distances and the possible values of $\beta$ (one sigma) measured by the experiment of \citet{Deslauriers:2006_09}.}
\label{fig:NeedleAnalysis}
\end{center}
\end{figure}

As first calculated by \citet{Turchette:2000}, and discussed here in Sec.~\ref{subsec:PatchPotentials}, a patch-potential model which considers the ion to be trapped inside a sphere of radius $d$, with patches of radius $R_\mathrm{p} \ll d$ predicts a distance-scaling exponent for the spectral density of electric-field noise of $\beta=4$. \citet{Deslauriers:2006_09} pointed out that the exact value of $\beta$ could vary depending on the details of the trap geometry. \citet{Low:2011} approximated the needle trap by a single prolate spheroidal needle, and showed the way in which, for this geometry, $\beta$ varied as a function of both the trap geometry and the patch size. The approximation of a single prolate spheroid was chosen as it is analytically solvable. They showed that the observed distance scaling of $\beta=3.5$ might be explained by a patch-potential model, as depicted here in Fig.~\ref{fig:NeedleAnalysis}.

When \citet{Deslauriers:2006_09} originally considered the possibility of Johnson noise, they assumed that Johnson noise would exhibit a distance-scaling exponent of $\beta=2$. Taking the geometry into account, the distance scaling for Johnson noise in their trap would be expected to exhibit $\beta=2.5(2)$ (see Sec.~\ref{subsubsec:Johnson_DistanceScaling}). This scaling prediction does not only hold specifically for Johnson noise, but generally for any source of noise which generates a spatially homogeneous field at the position of the ion.

The above analysis provides a strong indication that, in the experiment by \citet{Deslauriers:2006_09}, the noise arose from localized potential fluctuations on the electrode. There remain numerous physical mechanisms -- some of which are detailed in Sec.~\ref{sec:Microscopic} -- that could underlie such localised fluctuations and lead to the observed behavior.  Investigations to identify the physical mechanism responsible could include a simultaneous measurement of distance and frequency dependence of the noise in both the axial and radial directions.

\subsection{Electromagnetic pickup}
\label{subsec:DualTrap}

\citet{Britton:2008PhD} fabricated a segmented surface trap based on a silicon-on-insulator (SOI) substrate, with the ion 41\,$\mu$m above the surface. The design had two separate trapping zones: one had bare doped-silicon electrodes, 100\,$\mu$m thick; the other had doped-silicon electrodes on which a 1\,$\mu$m layer of gold was evaporated. The two different electrode surfaces were explicitly designed to demonstrate and characterize the effect of surface material on the ``anomalous" heating rate.

Heating-rate measurements were made in the gold and bare-silicon experimental zones. In both zones the spectral density of electric-field noise was measured to vary in the range $50 \times 10^{-12} < S_\mathrm{E}/$\,\B$ < 500 \times 10^{-12}$. The result for the gold and silicon traps are shown as points [$\Britton08$]a,b in Fig.~\ref{fig:KnownMechanisms}. This level of noise is relatively unremarkable, though it is around an order of magnitude higher than that seen in NIST's best traps \citep{Epstein:2007,Ospelkaus:2011}. The significance of this difference in absolute value is unclear as trap to trap variations at this level are not uncommon, even for nearly identical traps being investigated at NIST \citep{Britton:2008PhD}. What is significant, however, is that the heating rate varied across the stated range from day to day. While such fluctuations have been observed in other systems \citep[e.g.][]{Wang:2012PhD} it is uncommon for experiments at NIST.

To identify the source of the heating, extensive investigations into a variety of possible sources were undertaken. The results of these investigations are summarized in table~\ref{tab:BrittonInvestigation}. The level of heating seen was consistent with that expected if ambient field noise in the lab were coupled to the apparatus and happened to be on resonance with the trap motional frequency. (The theory of coupling to ambient field noise is discussed here in Sec.~\ref{subsec:EMPickup}.) This mechanism could also explain why the observed heating rate fluctuated from day to day, if the injected noise had a resonance which only sometimes overlapped with the trap frequency.

Electromagnetic pickup has been positively identified as the dominant source of heating in one other trap. The work of  \citet{Poulsen:2012} identified pickup as being a significant source of heating at a very specific frequency, corresponding to  a switched mode power supply. This is shown as point [$\Poulsen12$]b in Fig.~\ref{fig:S_E_d}. In this instance, having identified the source of noise, it was a relatively simple source to avoid, either by changing the power supply or by operating the trap at a different secular frequency. It is, however, interesting to note that minimizing pickup is not a concern restricted to traps in one particular size regime. Trap [$\Poulsen12$]b is the largest Paul trap for which heating rates have been measured, while traps [$\Britton08$]a,b are some of the smallest.

\begin{table*}
\begin{center}
\begin{tabular}{l r}
\hline
Source & Magnitude \\
\hline \\
Imperfect micromotion compensation & 10 times too low \\
Noise from the DAC cards (after filtering) & 100 times too low \\
Noise from an improved battery source (after filtering) & 1000 times too low \\
Johnson noise from filters & 100 times too low \\
Ambient laboratory fields coupled to trap apparatus & $\sim$ required magnitude \\
\hline
\end{tabular}
\caption[Heating sources considered by \citet{Britton:2008PhD}.]
{Heating sources considered by \citet{Britton:2008PhD}. A comparison is given of the calculated or measured level of noise from different sources and the level of noise that would be required to account for the ion-heating rates observed. Britton ultimately concluded that pick up of environmental noise could cause considerable heating if it were to overlap with the ion's motional frequencies. The relative magnitudes of the effects given here are specific to  Britton's experiment, and could be different for other apparatuses.}
\label{tab:BrittonInvestigation}
\end{center}
\end{table*}

\subsection{Technical noise}
\label{subsec:PowerSupply}

\citet{Schulz:2008} fabricated a three-dimensional, gold-on-alumina, segmented trap with the ion 257\,$\mu$m from the nearest electrode.
The electrodes were made from 500\,nm of evaporated gold with an RMS roughness of less than 10\,nm. The heating rate was measured for successive iterations of improving the drive electronics.

The initial setup was limited by noise from analog-output cards, which were used to provide DC voltages to the segmented electrodes. The noise from these cards caused heating rates high enough to preclude sideband cooling. By way of improvement, the electronics was changed such that the DC electrodes were controlled using an op-amp circuit powered by lead-gel batteries.\footnote{Personal communication, U. Poschinger, Mainz, 2012.} In this configuration, heating rates were measured which can be used to infer a spectral density of electric-field noise of $S_\mathrm{E}= 17(3) \times 10^{-12}$\,\B for $\omega=2\pi \times 1.18 $\,MHz. This is a rather unremarkable heating rate, given the size of the trap, as shown by point [$\Schulz08$] in Fig.~\ref{fig:KnownMechanisms}. Subsequent to this measurement, the DC voltage supply was further improved \citep{Poschinger:2009} by adding a transistor push-pull stage at the output of the DC voltage supply.\footnote{Personal communication, U. Poschinger, Mainz, 2012.} This lowered the output impedance of the circuit, which could make the system less prone to noise pickup in the wiring to the trap. The additional stage may also have provided some degree of low-pass filtering. Following this improvement, the heating rate was reduced, with the corresponding noise level inferred changing from $S_\mathrm{E}/\mathrm{V}^2/\mathrm{m}^2\mathrm{Hz}=18(3) \times 10^{-12}\rightarrow 3(1) \times 10^{-12}$. The improved result is plotted as point [$\Poschinger09$] in Fig.~\ref{fig:KnownMechanisms}.

It may initially seem surprising that such technical noise could cause problems. Very early in the discussions about heating in ion traps, \citet{Wineland:1998_06} highlighted the importance of static trap voltages being heavily filtered at the trap frequency, saying that heating due to power-supply noise and electrical pickup should be negligible for filter factors of $F < 10^{-4}$. Since these early considerations, the introduction of ion shuttling in segmented traps has given cause to revisit the practicability of 80\,dB filtering at the trap frequency. First proposed by \citet{Wineland:1998_05}, and implemented by \citet{Rowe:2002}, ions are shuttled by applying time-varying (quasi-static) voltages instead of static voltages. In the limit of fast shuttling, these voltages may need to be varied at frequencies close to the trap frequency \citep{Walther:2012, Bowler:2012}. While non-trivial, this is possible using more-involved filters \citep{Blakestad:2010PhD}, or novel switching methods \citep{Alonso:2013}. Nonetheless, even in instances not requiring fast shuttling, heating from technical sources can only be neglected \textit{after} great care has been taken to make it negligible.

\subsection{Technical and Johnson noise}
\label{subsec:Filters}

\citet{Harlander:2012PhD} used a three-dimensional, gold-on-alumina, segmented trap with the ion 257\,$\mu$m from the nearest electrode; the trap had the same essential design as that used by \citet{Schulz:2008}, and described here in Sec.~\ref{subsec:PowerSupply}. The initial 500\,nm layer of evaporated gold was gold-electroplated to a thickness of 10-15\,$\mu$m \citep{Splatt:2009PhD}. Ion-heating rates were measured
as a function of frequency, initially for weak low-pass filtering with a high cutoff frequency, and then for more aggressive filtering. In the first configuration, the scaling of the heating rate was that expected from RC filtered technical noise. After adding more aggressive filtering electronics, both the scaling and the absolute value of the heating rate were in line with what would be expected from Johnson noise in the filters' resistors.

\begin{figure}[t]
\begin{center}
\includegraphics[width=0.45\textwidth]{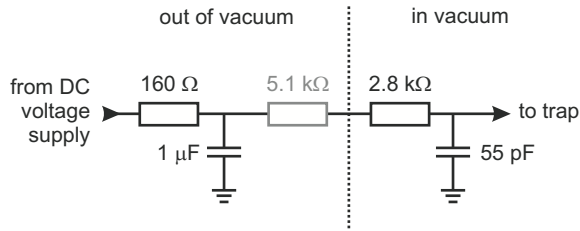}
\caption{
Filters used by \citet{Harlander:2012PhD}. The network initially consisted of an RC filter outside vacuum (cutoff at 1\,kHz) and a second RC filter inside the vacuum (cutoff at 1\,MHz). A resistor was added (shown in gray) to reduce the cutoff  of the second stage to 370\,kHz.}
\label{fig:MicreonFilters}
\end{center}
\end{figure}

In the first configuration the trapping voltages were generated by lead batteries, referenced with a 10\,V precision reference. This was filtered by out-of-vacuum RC filters with a cutoff frequency of 1\,kHz and in-vacuum RC filters with a cutoff frequency of 1\,MHz (shown in Fig.~\ref{fig:MicreonFilters}). As in the work of \citet{Schulz:2008}, the high cutoff was chosen to allow fast shuttling. The absolute value of the spectral density of electric-field noise was $15 \times 10^{-12}$\,\B at 1\,MHz (plotted as point [$\Harlander12$]a in Fig.~\ref{fig:KnownMechanisms}) and scaled with $\alpha=2.3(2)$. This is the scaling that would be expected from essentially white technical noise being filtered by the filter network close to the trap.

A 5.1\,k$\Omega$ resistor was added to the second filter stage, but outside the vacuum (shown in Fig.~\ref{fig:MicreonFilters}). The absolute heating rates were reduced by a factor of 1.6(8) over the frequency range measured, with $S_\mathrm{E} = 10 \times 10^{-12}$\,\B at 1\,MHz  (plotted as point [$\Harlander12$]b in Fig.~\ref{fig:KnownMechanisms}). By simulating the trap geometry and considering the nearest four segments, the absolute level of electric-field noise expected at the position of the ion due to the Johnson noise of the filters is calculated to be $S_\mathrm{E} = 10 \times 10^{-12}$\,\B at 1\,MHz. The observed heating rate can therefore be fully accounted for by Johnson noise in the filters. It may be additionally noted that the frequency scaling of the noise would be expected to become steeper due to the additional external filtering. This is consistent with the observed scaling in this configuration of $\alpha=2.9(7)$. This strongly indicates that, following the suppression of technical noise, the system was limited by Johnson noise in the filters.

\subsection{Surface effects}
\label{subsec:SurfaceEffects}

\citet{Daniilidis:2011} fabricated a gold-on-sapphire, segmented surface trap with the ion 240\,$\mu$m above the surface. The electrodes were made of 5\,$\mu$m of electroplated gold with 17\,nm RMS roughness. Initially heating rates were measured
and used to infer a spectral density of electric-field noise of $S_\mathrm{E}\approx 10 \times 10^{-12}$\,\B for $\omega=2\pi \times 1$\,MHz at the trapping position where the ions were loaded. This noise level remained constant for several months, until the background pressure was briefly increased to 10$^{-7}$\,mbar. After this the heating rates were measured along the entire length of the trap. They were found to have increased by a factor of $\sim 10$ in the region where the ions had been loaded.  The results before and after the change are shown as points [$\Daniilidis11$]a,b in Fig.~\ref{fig:KnownMechanisms}. By contrast, the heating rates away from the loading zone were comparable to that originally measured in the `pristine' loading zone. The uncooled ion lifetimes also decreased by roughly one order of magnitude in the parts of the trap where increased heating rates were observed. The new heating behavior of the trap then remained stable over more than six months.

Given the behavior changed only in the loading region, it seems that the effects of noise sources such as technical noise and RF pickup are excluded in this experiment at the noise levels observed. It is also not considered likely that the change was due to coating from the Ca oven, as the atomic beam was directed along the axis of the trap, and would be expected to coat the trap evenly along its length. \citeauthor{Daniilidis:2011} conjecture several possible mechanisms, including bombardment of the trap electrodes by electrons created during trap loading or by $^{40}$Ca$^+$ ions, which are created by photoionization outside the trapping volume. These can impinge on the gold surface with energies of up to 100 eV when accelerated by the RF field. $^{40}$Ca$^+$ ions with energies of a few eV could get physisorbed onto the trap electrodes and later form chemical compounds, while ions at higher energies could sputter material from the trap surface. In addition, they note that it is possible that the laser light used for ion creation and detection locally alters the chemical composition.

\subsection{Contamination removed by laser cleaning}
\label{subsec:LaserCleaning}

The first systematic investigation of in-situ cleaning of a trap \citep{Allcock:2011} used a SiO$_2$-on-Si, segmented surface trap with aluminium electrodes. The ion was 84\,$\mu$m above the trap surface and 98\,$\mu$m from the nearest electrode. The aluminium electrodes were 2.4\,$\mu$m thick, with a 2-3\,nm native oxide and 8\,nm RMS roughness. Any exposed surfaces on the silicon substrate were coated with 114\,nm of gold. The heating of the trap was first characterized
by measuring the absolute value of the heating rate and the frequency scaling, $\alpha$ \citep{Allcock:2012}. Part of the trap was then cleaned using nanosecond pulses of laser light at 355\,nm. The heating rate was then characterized again, measuring the absolute value and the frequency scaling in both the cleaned region and in an uncleaned region \citep{Allcock:2011}.

The heating rate was reduced when laser energy densities over 50\,mJ/cm$^2$
were used to clean the gold-covered side walls. The inferred spectral density of electric-field noise was reduced by around a factor of two
from 380 $\times$ 10$^{-12}$\,\B to 180 $\times$ 10$^{-12}$\,\B at 1\,MHz. The frequency scaling, $\alpha$, before and after cleaning, also changed from 0.93(5) to 0.57(3). The heating rate and the frequency scaling in the uncleaned region did not significantly change over the measurement period. The results before and after cleaning are shown as points [$\Allcock12$] and [$\AGH11$] respectively in Fig.~\ref{fig:KnownMechanisms}.

Given the behavior changed only in the region being cleaned, the effects of noise sources such as technical noise and RF pickup are excluded in this experiment at the noise levels observed. It seems instead that the reduction in heating is directly linked to removal of material from the electrode surfaces. The frequency scaling of heating due to fluctuating dipoles adsorbed on the trap surface varies as a function of the adsorbed atomic or molecular species \citep{SafaviNaini:2011, SafaviNaini:2013} (see Sec.~\ref{subsec:AdatomDipoles} here). Selective cleaning of a particular subset of adsorbates could therefore account for the change in $\alpha$. Alternatively the heating rate may have been reduced due to changes in the surface topography of the electrode \citep{Dubessy:2009,Low:2011}. \citet{Allcock:2011} raise both of these possibilities, and at present the data do not provide a basis to favor one option over the other.

The heating reduction described above seemed to saturate at the levels stated and further cleaning did not lead to continued improvements. \citeauthor{Allcock:2011} suggest that this may implicate additional heating mechanisms which are limiting at this level.

Subsequent to these measurements, higher energy densities of the cleaning laser were used. Energy densities of 180\,mJ/cm$^2$ removed parts of the gold coating over the silicon substrate and densities of 255\,mJ/cm$^2$
caused damage to the surface of the aluminium electrodes. This increased the absolute heating rate (though did not return it to as high as the pre-cleaning values) and increased the frequency scaling to $\alpha=0.88(3)$.

\subsection{Contamination removed by ion-beam cleaning}
\label{subsec:PlasmaCleaning}

A different method of in-situ trap cleaning was pioneered by \citet{Hite:2012} using a gold-on-crystalline-quartz trap with the ion 40\,$\mu$m above the trap surface. The gold of the electrodes was 10\,$\mu$m thick. The trap surface was cleaned by bombardment with a beam of argon ions, which is a well-established cleaning technique in surface science. The rate at which a trapped ion was heated was characterized before and after the trap surface was cleaned with argon-ion-beam bombardment.

Initially, Auger electron spectroscopy (AES) of a surface trap revealed it to have 2-3 monolayers of carbon on the surface. These were thought to be most likely from hydrocarbon deposition from the gas phase, where the presence of hydrogen is undetectable by AES. Following cleaning with an argon-ion beam, AES of the trap showed the surface to be contaminant-free gold. (The method is sensitive to a coverage fraction of $\sim$1\% , and no impurities were seen at this level.) A duplicate trap was used to  measure axial heating rates
which, before cleaning, gave an inferred spectral density of electric-field noise of $S_\mathrm{E} = 40(1) \times 10^{-12}$\,\B for $\omega=2\pi \times 3.6$\,MHz. The result is plotted as point [$\Hite12$]c in Fig.~\ref{fig:KnownMechanisms}. The frequency scaling was measured to be $\alpha=1.53(7)$. The trap was then cleaned using an argon-ion beam and the heating rate was remeasured. This gave an inferred spectral density of electric-field noise of $S_\mathrm{E}=0.25(1) \times 10^{-12}$\,\B, a factor of 160 lower than before cleaning. The point is plotted as [$\Hite12$]d in Fig.~\ref{fig:KnownMechanisms}. (Points [$\Hite12$]c,d in Fig.~\ref{fig:KnownMechanisms} are for improved cleaning parameters and at a higher trap frequency than points [$\Hite12$]a,b in Fig.~\ref{fig:S_E_d}). The frequency dependence after cleaning remained essentially unchanged at $\alpha=$1.57(4). It seems that the source of noise before cleaning was a few monolayers of (hydro)carbon on the surface of the gold, which could be cleaned off by argon-ion-beam cleaning. This situation seems to match well to the the adatom diffusion considered theoretically in Sec.~\ref{subsec:AdatomDiffusion}.

The process was shown to be repeatable by re-exposing the trap to air, and then vacuum-baking again. The field noise before cleaning was then $90(10)\times 10^{-12}$\,\B, which returned to $0.75(5) \times 10^{-12}$\,\B after argon-ion-beam cleaning. With no further processing this rose slightly to $1.1 \times 10^{-12}$\,\B but then remained constant. Subsequent experiments showed that the  improvements obtained by argon-ion-beam cleaning persisted for at least ten weeks \citep{Hite:2013}.

The frequency scaling of this noise source (both before and after cleaning) is different to that observed by \citet{Allcock:2011}. The absolute level before cleaning was also lower than the level seen by \citet{Allcock:2011} after cleaning. Despite the efficacy of surface cleaning in both instances, the difference of the absolute heating rates and frequency scalings suggest that \citet{Hite:2012} may not be limited by the same heating mechanism as \citet{Allcock:2011}.

Following the work of \citeauthor{Hite:2012}, \citet{Daniilidis:2014} performed similar investigations of in-situ ion-beam cleaning. Using a copper-aluminium alloy trap on a fused-quartz substrate, with $d=100\,\mu$m they saw a similar decrease in the level of noise following argon-ion-beam cleaning: it was reduced from $1.2 \times 10^{-12}$\,\B to $0.02 \times 10^{-12}$\,\B at frequencies just below 1\,MHz. This suggests that their experiment was also limited by surface contaminants. For comparison, these are plotted as [$\DGB13$]a,b in Fig.~\ref{fig:KnownMechanisms}. The value of $\alpha$ was measured to be 1.27(23) before cleaning, and 0.95(28) after cleaning. In contrast to \citet{Hite:2012}, they observed that -- even after cleaning -- the trap surface was not  atomically clean and oxide free. Nonetheless, they found that after cleaning they were no longer limited by these surface contaminants, but rather by technical noise, observing a resonance in the frequency spectrum around 800\,kHz consistent with noise from their filter board.

\subsection{Conclusions}

In conclusion it can be seen that, in certain instances, it is possible to directly link the ion heating to specific aspects of the experiment, or to specific changes made. While the absolute level of heating observed in many of these experiments is rather similar, the sources of heating in those cases where it can be identified are seen to be diverse.  This conclusion, which was first hinted at by the wide-ranging frequency-scaling behavior reported in Sec.~\ref{subsec:FrequencyScaling}, has now been explicitly borne out. As such one cannot investigate the electric-field noise observed in ion traps as some homogeneous phenomenon, but rather as many  different phenomena which must be carefully teased apart.

\section{Open Questions}
\label{sec:OpenQuestions}

Having looked at the mechanisms which can be more-or-less established, there exist a number of experiments where the noise is well characterized, but which do not lend themselves to an unambiguous identification of the noise source. These are discussed here in turn.

\subsection{Cryogenic noise floor}
\label{subsec:CryoNoiseFloor}

It has been shown (see Sec.~\ref{subsec:TemperatureScaling}) that cooling an ion trap down to cryogenic temperatures reduces the observed ion-heating rate -- and by implication reduces the electric-field noise. At MIT temperature scalings in the range $2 < \gamma < 4$ have been observed until a noise floor is reached around 30\,K. \citet{Labaziewicz:2008_10} suggest that the effect can be accounted for by a continuous spectrum of thermally activated random processes such as charge traps or adsorbate diffusion. (This is discussed further in Secs.~\ref{subsec:TwoLevelFluctuators} - \ref{subsec:AdatomDiffusion} of this review). While such models may be able to account for the observed behavior, they are not the only mechanisms which could,  in principle, explain the observed features. While the mechanisms suggested below are speculative, they indicate that the observed effects do not conclusively or unambiguously point to a particular source, and that one or more of a number of mechanisms may be at work in cryogenic systems.

\subsubsection{Johnson and technical noise}
\label{subsubsec:JohnsonNoiseAtMIT}

While every effort is usually made to keep cabling short, cryostats generally require longer wires between the trap-drive electronics and the trap. If the driving and control electronics are held at room temperature [which is often, though not always the case, cf. the work of \citet{Poitzsch:1996} and \citet{Gandolfi:2012}] short wires would provide little thermal resistance, and lead to a significant heat load at the cold stage \citep{Ekin:2007}. Instead, longer wires are used, allowing the heat to be exchanged between the wires and the cold-finger. It is also advantageous --- from the perspective of the thermal load --- for these wires to have a low thermal conductivity, and consequently a higher ohmic resistance. Using longer wires of higher resistance has the advantage of a much lower heat load at the cold stage. However, it also means that, in addition to the basic resistance of the trap electrodes (which at cryogenic temperatures may be $\sim$10\,m$\Omega$) there is also Johnson noise from the cabling, which may have a resistance of $\sim$10\,$\Omega$.

As the temperature is reduced the resistance of most cables scales approximately linearly with $T$. This would lead to an expected value of $\gamma=2$ (see Sec.~\ref{subsec:JohnsonNoise}), as observed by \citet{Labaziewicz:2008_10} in experiments with good (i.e. low heating-rate) traps. The experiment uses phosphor-bronze cables \citep{Labaziewicz:2008PhD}, for which the resistance only changes by around 20\% between 300 and 4\,K \citep{Tuttle:2010}. If the experiment were limited by this resistance then  one would expect a still smaller value of $\gamma$. However, if the experiment were limited by other resistances in the system, a value of $\gamma=2$ is unsurprising. At some temperature -- which can vary significantly depending on the purity of the material  \citep{Ekin:2007} -- the resistivity would plateau, becoming independent of temperature, and leading to an exponent of $\gamma=1$. Even if the resistance of the cabling were to go to zero, at some point the noise would be limited by the technical noise of the drive electronics. In this case, where the driving electronics is outside the cryostat, and where the RF and DC sources, leads and filters remain at a constant temperature throughout the measurement \citep{Labaziewicz:2008_10} this noise level is independent of the temperature at the trap, and would appear as a leveling off of $S_\mathrm{E}$.

\subsubsection{Magnetic pickup}
\label{subsubsec:MagneticShieldingAtMIT}

In order to adequately exchange heat between the electrical cabling and the cold finger, the cabling in a cryostat is wrapped several times around each heat stage in the cryostat \citep{Ekin:2007}. This means that the level of EM pickup could be considerably higher than in typical room-temperature systems, where lengths of wire are generally kept to an absolute minimum (see Sec.~\ref{subsec:EMPickup} for a discussion of EM pickup).

The cryostat shrouds (at room temperature) provide a good level of shielding for electric-field noise, though, viewports, imperfect seams and other defects can allow non-negligible amounts of EMI through \citep{Miller:1966, Bridges:1988, Rajawat:1995}. As the temperature is reduced, the resistance of the copper heat shields decreases, and the magnetic-shielding effect increases. This effect has already been noted in connection with the improved coherence times of hyperfine qubits when shielded from variations in magnetic fields at hertz frequencies \citep{Brown:2011_09}. At radio frequencies, the improvement of shielding with reduced temperature might be expected to lead to a reduction in the ion-heating rate. Such an effect would also reach a plateau when the heating became limited by some effect other than EM pickup.

\subsubsection{Fluctuating adatom dipoles}
\label{subsubsec:AdatomDipoleNoiseAtMIT}

As was discussed in detail in Sec.~\ref{subsubsec:ADF_NoiseSpectrum}, adatoms whose dipole moment undergoes thermally induced fluctuations can exhibit the noise characteristics of the behavior observed by \citeauthor{Labaziewicz:2008_10} Considering a planar geometry, and choosing some reasonable parameters for the adsorbate properties, this mechanism might be expected to display $\gamma \approx 2.5$ above an activation temperature of a few tens of kelvins. with an activation temperature of $T\sim 65\,$K. This is in broad agreement with the experimental observations. In trap  [\LGL08]a [for which $\gamma=3.0(2)$] the frequency scaling was measured at cryogenic temperatures, and found to be $\alpha \approx 1$. The scaling predicted in Sec.~\ref{subsubsec:ADF_NoiseSpectrum} for fluctuating adatom dipoles was $\alpha=0$ at low frequencies and $\alpha=2$ at high frequencies. The simple model here contains many unknown parameters, and it may be that -- with the particular parameters of the contamination in this instance --  the trap was being tested near the characteristic frequency.\\

In summary, behavior observed by \citet{Labaziewicz:2008_10} does not immediately rule out the involvement of Johnson noise or EM pickup.
Whether the appearance of a noise floor at $T\approx 30$ K is due to an exponential suppression of thermally activated fluctuators or set by a combination of different sources of electrical noise in the circuit is still an open question.

\subsection{Noise above superconductors}
\label{subsec:SuperconductorsAtMIT}

The ion-heating rates above superconducting surfaces have been measured in niobium traps and niobium nitride traps \citep{Wang:2010_12}. Nb (or NbN)-on-sapphire surface traps, with an ion height of 100\,$\mu$m, were operated at around 6\,K ($T_\mathrm{c}^\mathrm{Nb}=9.2$\,K; $T_\mathrm{c}^\mathrm{NbN}=16$\,K). The inferred electric-field noise above both traps was comparable to the lowest value of electric-field noise measured in Au, Ag and Al traps of the same design in the same apparatus. Furthermore, the temperature of a single  niobium trap was varied across the superconducting transition. The heating rates above and below $T_\mathrm{c}$ were comparable.

\citeauthor{Wang:2010_12} therefore conclude that the heating observed in this experiment was not due to the buried defects in, or the resistance of, the trap electrodes. They further surmise that the only remaining option is for the observed heating is predominantly a surface effect. However, from the wide variety of possible mechanisms which may play a role in the heating observed in ion traps, there may be other possible explanations yet to be excluded, as discussed above in Sec.~\ref{subsec:CryoNoiseFloor}: Johnson noise in cabling, technical noise from drive electronics, and EM pickup would all be expected to be independent of electrode material and remain essentially unchanged when the trap itself becomes superconducting.

\subsection{Frequency-independent heating rates}
\label{subsec:NoiseAtAarhus}

One of the largest traps in which heating rates have been measured  -- Trap [$\Poulsen12$] \citep{Poulsen:2012} -- which also had the lowest heating rate of any room temperature ion trap, exhibited a heating rate which was independent of the motional frequency over the range  280-585\,kHz, except for a narrow resonance at 295\,kHz. The trap was made of stainless-steel rods at a distance of 3500\,$\mu$m from the ion and plated with 5\,$\mu$m gold. The resonance was traced to a nearby switched-mode power supply, though the cause of the underlying heating rate (indicating a noise scaling with $\alpha=-1$) is unclear. Understanding the situation in this trap may be of significance  to quantum information-processing experiments, particularly in the low-frequency regime: certain gates which use the ions' motional dipole to mediate the interaction \citep{Cirac:2000} work faster at lower trap frequencies, as the ions are less tightly localized.  This, however, can cause technical challenges \citep{Kumph:2011} as the heating rate (normally) increases at lower trap frequencies which cancels or even outweighs any benefit of the increased gate speed. Should a regime be found in which this were not the case, there may be new options available for scalable quantum computing.

Bare black-body radiation noise, at the temperatures and frequencies of interest, would be expected to increase with frequency and scale with $\alpha=-2$,  as discussed in Sec.~\ref{subsec:Blackbody}. By additionally considering (conjectural) external sources of radio-frequency noise,  or frequency-dependent shielding effects, the observed scaling may be consistent with technical noise. This verdict is not, however, definitive, and further investigation would be required in order to be conclusive. A few possible avenues are mentioned here.

Firstly, and most obviously, the trap is large. It is almost a factor of three bigger than the next largest Paul trap in which a heating rate has been measured: trap [$\Home06$] \citep{Home:2006PhD}. Moreover, it is almost a factor of ten bigger than the next largest trap in which $\alpha$ has been measured: trap [$\Turchette00$]g \citep{Turchette:2000}]. It may be interesting to know what happens between these sizes.

A second interesting feature of the trap is that it is fabricated of stainless steel.  This is not a common trap material, having only previously been used in a few large traps \citep{Rohde:2001, Home:2006PhD, Benhelm:2008_06}, none of which have been used to measure $\alpha$.  Stainless steel has the unusual property of having better magnetic field shielding at lower frequencies  (over the range 0.1-10\,MHz) \citep{Kaye:1995}.  While vacuum chambers are standardly made of stainless steel,  it is not clear what shielding effect -- if any -- might be gained from traps made of stainless steel.

A final unusual aspect to the Aarhus trap's operation may be noted. Almost all traps in the literature, and all traps in which $\alpha$ have been measured,  create the trapping potential by driving some electrodes with an alternating voltage,  $V_\mathrm{rf}  \cos(\Omega_\mathrm{rf}t)$, while keeping the remaining electrodes at RF ground. In the trap used by \citeauthor{Poulsen:2012} some electrodes were driven with an alternating voltage, ($V_\mathrm{rf}/2)  \cos(\Omega_\mathrm{rf}t)$, while the remaining electrodes were driven 180$^\circ$ out of phase, with $-(V_\mathrm{rf}/2) \cos(\Omega_\mathrm{rf}t)$. Provided the far-off ground (such as the vacuum chamber) is a long way away and well shielded from the ions by the trap electrodes, these two methods are identical but for the former having a quasi-uniform field oscillating at the trap-drive frequency. In most traps, however, the endcaps -- being close to the ions and held at RF ground -- break the symmetry between the two situations. By applying RF voltages to the endcaps \citeauthor{Poulsen:2012} preserved the idealized behavior even against such symmetry breaking. It is not immediately obvious why this would affect heating rates, but is mentioned for completeness.

\subsection{Very low absolute heating rates}
\label{subsec:TurchettesWeirdTraps}

Trapping $^9$Be$^+$ in a molybdenum ring trap with an ion electrode distance of $d=125$~$\mu$m, \citet{Turchette:2000} observed a spectral density of electric-field noise of $140\times 10^{-12}$~\B at a frequency of $\omega/2\pi=7.9$~MHz, with a frequency scaling of $\alpha\sim 0$. The trap was removed from vacuum and cleaned with HCl to remove the Be coating deposited by the atomic source. It was then electropolished in phosphoric acid, and rinsed in distilled water followed by methanol. Following this treatment the noise was re-measured and found to be $0.7\times 10^{-12}$~\B at around the same frequency. The trap also exhibited a frequency scaling of $\alpha=6.0(2)$ (over the range measured: 3.2~MHz~$ < \omega < 8.2$~MHz). The trap before and after cleaning is referred to in this review as [$\Turchette00$]c,h respectively.

In a second, larger trap ($d=280$~$\mu$m), made in the same piece of metal as trap [$\Turchette00$]c, the spectral density of electric-field noise was measured to be $7.4\times 10^{-12}$~\B at a frequency of $\omega/2\pi=3.5$~MHz. Following the same cleaning treatment as described above, the noise was measured to be $0.014\times 10^{-12}$~\B at $\omega/2\pi=3.3$~MHz; a factor of 500 reduction from the initial measurement. The trap also exhibited a frequency scaling of $\alpha=4.0(8)$ (over the range measured: 1.3~MHz~$ < \omega < 3.5$~MHz), suggesting that even lower heating rates might have been attainable at higher trap frequencies. The trap before and after and after cleaning is referred to in this review as [$\Turchette00$]d,i respectively.

It is not clear why the ultimate heating behavior in these traps was different from the initial heating behavior, and also different from other traps. It was conjectured by \citeauthor{Turchette:2000} that the modified behavior for traps [$\Turchette00$]h,i was due to a less-than-usual deposition of beryllium on the electrodes though, even with the advent of photoionization loading, and the very low fluxes this requires \citep{Kjaergaard:2000, Deslauriers:2006_12, Brownnutt:2007}, the effect has never been reproduced. Should the result be reproduced, it would, obviously, be of significant interest to the ongoing search for low heating rates.

\section{Outlook}
\label{sec:Outlook}

Given the present state of understanding in the literature, as summarized in this review, there are several ways in which ion-trap work, and more generally any work relating to fluctuating electric fields near surfaces, might proceed. The noise may be viewed as a problem to be removed, and steps can be taken to reduce it without further effort to understand the exact causes. Alternatively, either for its own sake, or in an attempt to be better able to minimize it, the noise may be further investigated. This could take the form of continued heating-rate measurements in ion traps, or of characterizing -- experimentally or theoretically -- particular parts of the system separately. In practice, any given project is likely to draw on a combination of these options, to differing degrees, depending on its specific aims. They are discussed here in turn.

\subsection{Reduction of noise}
\label{subsec:Outlook_Pragmatic}

Inasmuch as ion-trapping experiments wish to study the basic nature of light-matter interactions, probe the foundations of quantum mechanics, or build better clocks, electric-field noise -- whatever its causes -- is almost always a distraction from the physics under investigation. In this regard a pragmatic approach, which does not interest itself with what causes the noise beyond knowing how to suppress it,  and which allows experiments to move forward, may be attractive. To this end, a number of things should be considered.

There are multiple sources of electric-field noise to be reckoned with, a number of which can contribute to heating at or near a level which can limit trapped-ion experiments. If great effort is taken to suppress one (and only one) effect it is likely that the experiment will be limited by a combination of other effects. Consequently one should take every precaution to suppress all effects.

Ideally, the laboratory environment should be kept as electronically quiet as possible. However, because it is impossible to know all the sources of RF radiation a priori, shielding becomes the only possible way to maintain a repeatable experiment. Ambient electric fields can be blocked by a Faraday cage, as used in the experiment of \citet{Daniilidis:2014}. Mu-metal shielding, as used by \citet{Monz:2011PhD}, attenuates both electric and magnetic field noise, though has drawbacks due to its greater cost, weight and bulk. At low temperatures the shielding effect  of copper shrouds in a cryostat can offer significant advantages \citep{Brown:2011_09}.

Any electronic devices attached to the experiment, such as voltage sources for DC or RF electrodes, should be as low-noise as possible. Noisy power supplies \citep{Poschinger:2009} and DC control \citep{Harlander:2012PhD, Britton:2008PhD} can contribute noise at a level which can limit experiments. Starting from initially-quiet sources, these should be filtered as heavily as possible, noting also that non-optimal filter design can introduce noise at a level which can limit experiments \citep{Harlander:2012PhD}. The filters should be situated as close to the trap as possible to limit pickup of noise from fluctuating electric or magnetic fields after the filters. It has been known for a long time \citep{Wineland:1998_05}, but bears repeating, that technical noise can be neglected in ion-trap experiments if -- and only if -- great pains are taken to reduce it as far as possible.

Any dielectrics in the system should be as low-loss as possible. This includes dielectrics in electronic components such as capacitors as well as any other material which is exposed to RF radiation near the trap. These might include the trap substrate, mountings for the trap and any in-vacuum optics.

The trap-electrode surfaces should be as clean as possible. In at least some cases, exposure of the electrodes to flux from the atomic oven has been linked to increased heating \citep{Turchette:2000, DeVoe:2002, Brownnutt:2007}. This can be minimized by using more efficient ionization and loading methods \citep{Kjaergaard:2000, Cetina:2007}, by cleaner loading systems \citep{DeVoe:2002, Brownnutt:2007, Sage:2012}, or by shielding the trap electrodes from the flux \citep{Britton:2009, Allcock:2011, Doret:2012}. This may not ultimately prove sufficient as the act of ionizing atoms may degrade the trap surface \citep{Daniilidis:2011}. This problem can be mitigated in a segmented trap by using separate zones for loading and for operations \citep{Blakestad:2009, Doret:2012}.

A variety of noise mechanisms can be suppressed by operating traps at cryogenic temperatures.  This could be expected to improve `typical' traps by two orders of magnitude, while traps with initially high heating rates may be improved even more. Cryostats have a number of advantages besides suppression of heating rates, including fast installation of new traps, ultra-low background pressures, and the possibility of using novel materials. Weighed against this is their expense (both the initial costs of the cryostat and running costs related to cryogens) and possible increased vibration.

To remove the causes of noise created by surface contaminants the traps may be cleaned in-situ. Modest improvements have been demonstrated using laser ablation cleaning \citep{Allcock:2011}. Improvements of around two orders of magnitude have been reported using cleaning by argon-ion bombardment \citep{Hite:2012, Daniilidis:2014}. Both methods should be used with care as they aggressively remove material from the trap surface and entail the possibility of damaging the trap.

Once every effort has been made to reduce the sources of noise, there are still further options for reducing the heating rates. Almost all noise sources which are likely to limit ion traps have lower spectral densities at higher frequencies. Consequently, heating rates can be reduced by operating experiments so that ions have higher motional frequencies.

An often-overlooked conclusion of \citet{Turchette:2000} regards another rather obvious possibility for reducing the heating rate:  bigger traps have smaller heating rates. They noted that with little sacrifice in the trap secular frequency (which determines the fastest gate speed for many gate implementations) a significant decrease in the heating rate vs logic gate speed appears possible by using larger traps. The greater amount of data now available suggests that increasing the trap size will not necessarily bring the level of improvement initially anticipated by \citeauthor{Turchette:2000}, though it will bring some. Moreover, given the possibility of ultra-fast gates \citep{GarciaRipoll:2003, Bentley:2013} it may be that in future the limit to gate speeds set by the motional frequency is relaxed. Thus, while it often seems to be anathema, unless there is some other specific reason for requiring small traps, larger ion-electrode separations could be advantageous.

Finally, it has become common to plot $S_\mathrm{E}(d)$, rather than $\dot{\bar{n}}(d)$, as this removes the influence of the ion mass, and therefore gives a fairer inter-trap comparison of the noise. However, it should not be forgotten that it is often heating rate, $\dot{\bar{n}}$, and not the noise per se, which is of interest for many trapped-ion applications. There are, obviously, many considerations which go into selection of ion species to be used \citep{Hughes:1998, Wineland:2003, Lucas:2003}. One aspect which may be added to the list of considerations is that the heating rate is inversely proportional to the ion mass.  Were a heavier-mass ion to be used then, to maintain the same trap frequency and stability parameter, for a given size of trap, the RF voltage must be increased in proportion to the ion mass. In larger traps where very high voltages are used already, further increases in voltage may not be feasible. However, for smaller traps, higher voltages may not be prohibitive. By way of illustration for the same level of electric-field noise, and assuming the motional frequency is unchanged, an $^{171}$Yb$^+$ ion will be heated 19 times more slowly than a $^9$Be$^+$ ion. All other things being equal, this is a similar equivalent benefit as might otherwise be expected by increasing the trap size by a factor of about ten, or by operating the trap at liquid-nitrogen temperatures.

It seems at present that there is not one magic bullet which makes all of the problems go away, but rather steps must be taken to ensure the noise is reduced simultaneously on multiple fronts. In opting to implement multiple of these solutions, however, it should be noted that they may not all be independent. For example, if cryogenic operation reduces noise by freezing out the motion of surface contaminants, and if ion-beam cleaning reduced noise by removing surface contaminants, then an ion-beam-cleaned trap might not be expected to be improved further by operating it at cryogenic temperatures. Unfortunately, how different effects behave in combination is largely conjectural at this stage; suggestions for future experiments in this direction are given in Sec.~\ref{subsubsec:NoiseCombinations}.

\subsection{Investigation of noise in traps}
\label{subsec:outlook_InvestigateI}

Appreciation of the effects underlying electric-field noise in ion traps can aid in finding ways to minimize or mitigate the influence it has on experiments. As electric-field noise above surfaces plays a role in a wide variety of experiments beyond ion traps, an increased understanding in one system may serve to shed light on others. It is also possible that the noise may prove useful; it may, for example, offer new ways to analyze surfaces to complement or extend the existing surface-science toolbox. Whatever the reason, further systematic investigation of the noise may be of interest.

As this review has pointed out, there are a number of qualifications which must be made concerning any use of trapped ions to characterize noise. The ion only provides information regarding the total noise: how this relates to the several mechanisms potentially causing it requires careful interpretation. Furthermore, characterizing the noise in one experiment does not necessarily say anything about the noise in an experiment across the hall.  As such, while the experiments suggested here will fill in certain gaps in the literature, one should be cautious of claiming too general a result. Given the results may only apply to a particular experiment, all the more care should be taken in generalizing results beyond ion traps to, for example, nanocantilever systems or atom-trap systems. Given these caveats, we turn to a number of parameters with respect to which electric-field noise in ion traps may be investigated.

\subsubsection{Frequency}
\label{susubsec:FrequencyOutlook}

It may be expected that $\alpha$ varies between experiments, and even takes different values for different parameter ranges in a single experiment. Many noise sources (such as EMI, EM pickup, Johnson noise from filter networks, technical noise, and TLFs) can give rise to resonances at particular frequencies. Other noise sources (such as filtered Johnson noise, fluctuating adatom dipoles and adatom diffusion) exhibit a roll-off at higher frequencies. The characteristic frequencies at which such resonances or roll-offs occur can provide significant information about the underlying mechanism and go a long way to constraining theory.

Most experimental results to date for frequency scaling infer that, within some range of uncertainty, over the frequency range measured, and at the resolution taken, $\alpha$ is constant for that experiment. Other experiments clearly show the presence of resonances in the spectrum \citep{Poulsen:2012, Daniilidis:2014}. Future work should be considered to search for changes in $\alpha$. This may necessitate higher-resolution scans of the trap frequency and require  measurements over a wider range of frequencies. Obviously, this is more time consuming than taking a smaller number of points, but the information provided by any features in the spectrum would provide much greater insight into possible mechanisms.

Consider noise sources such as EMI, EM pickup, Johnson and technical noise: the absolute level of noise, as well as the frequency scaling expected, can vary greatly depending on the exact details of the apparatus. These include the details of the driving electronics, the wiring, and the filter design. Future investigations of heating which may either implicate or exclude such noise sources would benefit from including comprehensive details of the associated electronics.

\subsubsection{Distance}
\label{susubsec:DistanceOutlook}

As this review has gone to pains to highlight, Figs.~\ref{fig:S_E_d} and \ref{fig:CryoS_E_d} are of limited use in elucidating a distance-scaling law for noise. To date, only one experiment has made a controlled measurement of the distance scaling \citep{Deslauriers:2006_09}. Given the complexity of the issues at hand, the case is clearly not closed for investigating distance-scaling behaviors. There are many different mechanisms at play and great care must be taken to tease these apart and understand them. Just as multiple experiments measuring frequency scalings have shed light on the multifaceted nature of the noise (see Sec.~\ref{subsec:FrequencyScaling}), so multiple distance-scaling measurements may provide a fuller picture.

\begin{figure}[t]
\begin{center}
\includegraphics[width=\columnwidth]{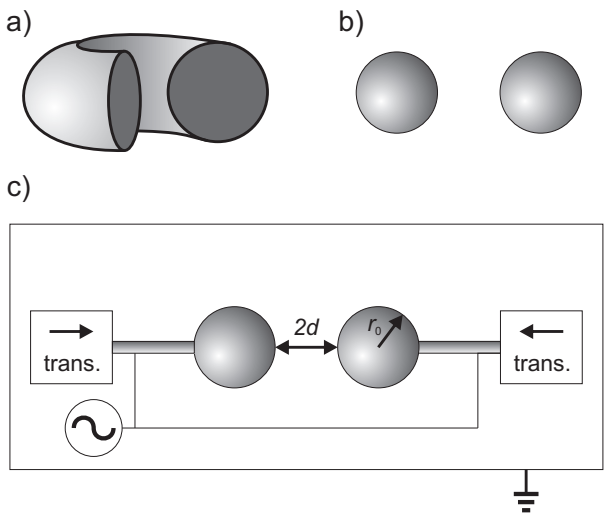}
\caption{Only two analytical solutions to the Laplace equation are capable of trapping: toroidal solutions (a) and  bispherical solutions (b). Toroidal solutions are already well known as ring traps, though bispherical solutions would be simpler to scale in a controllable way. This could be achieved in a manner similar to the work of \citet{Deslauriers:2006_09}, by mounting the spheres on a pair of translation stages (c). The vacuum chamber would then provide a far-off ground.}
\label{fig:LaplaceTraps}
\end{center}
\end{figure}

For ease of interpretation, such experiments could be carried out in traps with as simple a geometry as possible. Of the geometries which support an analytical solution to the Laplace equation \citep{Morse:1953} two form a trapping potential, shown in Fig.~\ref{fig:LaplaceTraps}. The toroidal solution is a ring trap, sometimes referred to as a Paul-Straubel trap \citep{Yu:1991}. The bispherical solution has never been used in practice, though it is approximated by a needle trap without sleeves.

Another possibility for scaling the ion-electrode separation uses electrical, rather than mechanical, means to move the ion closer to the trap. This can be achieved by varying the amplitude of the RF voltage applied to some -- but not all -- electrodes. By this method the position of the RF null, and thence the position of the ion, can be moved. This was first used by \citet{Herskind:2009} to move the ions small distances. It was later proposed as a way to measure heating rates by \citet{Kim:2010}. In this latter proposal a surface ring trap has multiple concentric ring electrodes to which different RF voltages can be applied, thereby changing the ion height, $d$.

As there is no mechanical motion involved, and with sufficient control of the variable RF voltages, this can be done in a highly controlled and repeatable way. Also, being a planar trap, the geometry is -- for certain noise mechanisms -- simple to model. For a mechanism such as fluctuating adatom dipoles the trap can be treated as a near-infinite plane: the dimensions of the individual electrodes are unimportant. By contrast, Johnson noise on different electrode segments would be essentially uncorrelated, and so the noise due to particular electrodes would need to be calculated and summed. Technical noise sources may provide a more complicated picture as the noise they inject into the system may vary as a function of the RF voltage being supplied. With careful design of the experiment it should be possible to distinguish between such sources.\\

As has been stressed throughout this paper, while the ion-electrode separation is known relatively easily, and while it is an important number for many practical purposes, it is not necessarily useful for elucidating heating mechanisms. Some papers provide characteristic distances (assuming a certain noise model), or sufficient detail that they may be calculated, though this is not always the case. It would be helpful for future literature to include traps' characteristic distances, for a selection of likely noise models. Calculating the characteristic distances for arbitrary geometries is not trivial, and the results can depend sensitively on how well the model reflects, for example, slight imperfections in the real trap geometry. Should the experimentally measured value of $\beta$ differ from the value from simulations, a decision must be made between believing the interpretation of the experiment or the simulation. It would therefore be useful to have a simulation-independent method of verifying how a given type of noise scales in a particular apparatus. For Johnson noise (or other sources which are correlated across the entire electrode) this can be achieved by deliberately applying excess white noise to one or more trap electrodes and measuring the heating rate of the ion as a function of ion-electrode separation \citep{Daniilidis:2011}. By this method it is possible to measure the characteristic length scale of various trap features. This can act as a benchmark against which the simulation results are tested, or used to obviate the need for simulations altogether.

Given the complexity of modeling trap geometries, and of changing the distance to the trapping electrodes, it may be advantageous to characterize the heating due to external objects which are independent of the trap operation. It has been proposed that this could be achieved by using an ion trap with an open geometry and varying the distance between it and some external object \citep{Maiwald:2009, Harlander:2010, Brownnutt:2012}.

At room temperature, heating-rate measurements have been performed in traps varying in size by over two orders of magnitude: $30\,\mu\mathrm{m} < d < 3500\,\mu\mathrm{m}$. At cryogenic temperatures the range is much smaller: $40\,\mu\mathrm{m} < d < 230\,\mu\mathrm{m}$. Given the ease and repeatability of fabricating surface traps, and the short times required to install new traps \citep{Wang:2012PhD}, it would be straightforward to measure traps larger than $d=230\,\mu\mathrm{m}$. This could provide a reasonably standardized geometry to determine the distance scaling and, given the current paucity of data in this range, even a few results in larger traps would provide significant information.

\subsubsection{Correlation length}
\label{subsubsec:CorrelationLengthOutlook}

Different mechanisms for electric-field fluctuations often results in drastically different spatial correlations of the noise and systematic measurements of noise-correlation lengths with single and multi-ion systems could shed further light into the origin of the underlying physical processes.

If the noise is correlated over distances of the order of the ion-electrode separation or larger, a single ion can be used to detect such correlations by measuring heating rates along different directions and at different positions in the trap.  For example, the Johnson noise considered by \citet{Leibrandt:2007_01} is correlated across each electrode, but uncorrelated between electrodes. Consequently,  the axial components of the noisy electric field from a single segment cancel near the center of the electrode and at this location the axial heating would be much lower than the heating observed near the edge or between two segments. Such a behavior could be relatively simple to measure experimentally and, if present, would indicate a correlated noise source.

As discussed in Sec.~\ref{subsec:NeedleTrap}, another method of inferring the correlation length of noise on surfaces can be performed with traps with movable electrodes. As can be seen from Fig.~\ref{fig:NeedleAnalysis}, in the limit of $d \gg R_\mathrm{el}$ (in this example $R_{\rm el}\sim r_{\rm prolate}$) the value of $\beta$ is the same for both Johnson noise (infinite patch size) and for models with finite correlation lengths. For $d\sim R_{\rm el}$ the ion-electrode separation at which $\beta$ departs from this value is related to the patch size. By measuring $\beta(d)$ it is in principle possible to determine the size of the patches, or more generally, the noise-correlation length, $r_{\rm c}$. This method requires detailed modeling of the electrode configuration and will be limited by the minimal achievable ion surface distance.

\begin{figure}[t]
\begin{center}
\includegraphics[width=\columnwidth]{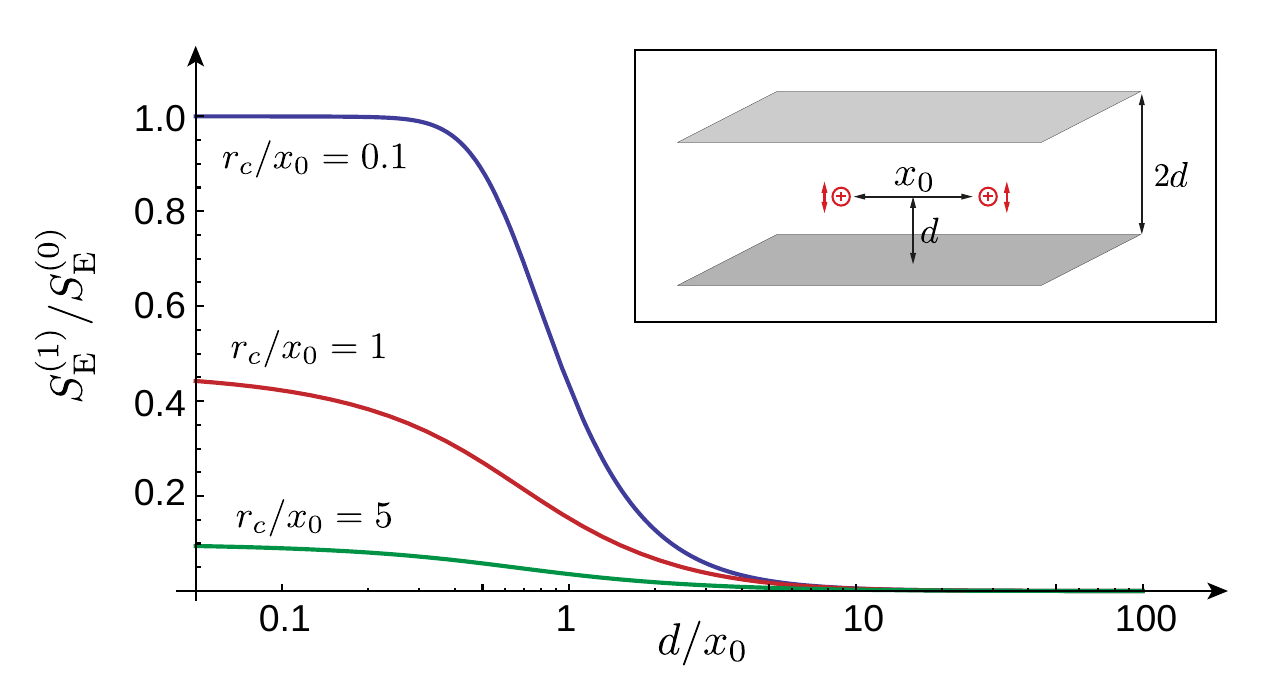}
\caption{Effect of noise correlations in multi-ion traps. The ratio between the spectral noise densities $S_{\rm E}^{(1)}$ and $S_{\rm E}^{(0)}$ evaluated at equal frequencies is plotted as a function of the ion surface distance and different correlations lengths, $r_{\rm c}$. For this plot an exponential cutoff for the noise correlations on the surface has been assumed.}
\label{fig:CorrelationDistanceScaling}
\end{center}
\end{figure}

More generally, noise-correlation effects appear in multi-ion systems, where the heating rates for different normal modes depend strongly on the correlations of the noise over the extent of the ion crystal (see Sec.~\ref{subsubsec:MultiIonHeating}). It should be noted that the field correlations at the positions of the ions are different from the correlations on the surface and depend again on the trap-surface distance and details of the trap geometry. This is illustrated in Fig.~\ref{fig:CorrelationDistanceScaling}, where, for a simple planar geometry, the heating of the common and the relative transverse mode for two ions separated by a distance $x_0$ is considered. The two modes couple to the symmetric $\delta E^{(0)}=(\delta {\bf E}_z(t,{\bf r_1})+\delta {\bf E}_z(t,{\bf r_2}))/\sqrt{2}$ and anti-symmetric $\delta E^{(1)}=(\delta {\bf E}_z(t,{\bf r_1})-\delta {\bf E}_z(t,{\bf r_2}))/\sqrt{2}$ field combinations, respectively. The corresponding spectral densities of the electric-field noise, $S_{\rm E}^{(0)}(\omega)$ and $S_{\rm E}^{(1)}(\omega)$, are compared at different ion-surface spacings. Noise sources with different correlations lengths become distinguishable for $d\lesssim x_0$, while for distances $d\gg x_0$ the noise always appears to be correlated. Traps where ions are confined in separate but proximate wells and $x_0\sim d$ \citep{Harlander:2011, Brown:2011_03, Wilson:2014} would be particularly suitable to search for correlation effects.

\subsubsection{Temperature}
\label{susubsec:TemperatureOutlook}

So far it has been shown, at least in the traps where it has been measured, that the noise usually scales less strongly at low temperature (6\,K$<T\lesssim50$\,K) than at higher temperatures (50\,K$\lesssim T<300$\,K). Despite this, the exact scaling behavior observed by \citet{Chiaverini:2014} is different to that  measured by \citet{Labaziewicz:2008_10}. This may not necessarily be surprising, given the wide variety of noise sources in different experiments. Such measurements therefore bear repetition by other groups to see what happens in different experiments, possibly limited by different noise sources.

If the noise floor is set by injected noise from sources outside the cryostat then, if the temperature were further reduced, these may remain the limiting factor. However, it may be that traps at millikelvin temperatures have new regimes of operation. In such a case it may also be possible to separate out the damping rate, $\Gamma$, from the occupation number of the thermal bath, $\bar{N}$ [see Eqs.~\eqref{eq:GammaNbar} - \eqref{eq:nbardot_t}].

No controlled measurements have been made regarding the effects of elevated temperature on heating rates.  It is feasible to operate ion traps at 400\,K or hotter \citep{Chwalla:2009PhD}. By increasing the temperature in a controlled manner, even a modest change of $\sim$100\,K above room temperature would be expected to increase the heating rate by a factor of 2 -- 3 (assuming a  range of possible exponents $2 < \gamma < 4$). This should be easily resolved, and would provide information on heating in a new regime.  Interestingly, for some mechanisms it is also predicted (see Fig.~\ref{fig:TLF_Temp}, Sec.~\ref{subsubsec:NonUniformTLFDistribution}) that the heating rate may \textit{decrease} at higher temperatures.

Combinations of heating or cooling different parts of the apparatus could also shed light on possible heating mechanisms \citep{Chiaverini:2014}. For example, if the vacuum chamber were heated while the trap was held at a constant temperature it may be expected that shielding of EMI will decrease, due to the increased resistance of the chamber. As a result the ion-heating rate may increase, despite the constant trap temperature.  Alternatively, heating (or cooling) the electronics while leaving other parts of the apparatus at ambient temperature would change the heating rate if Johnson or technical noise are limiting factors, while leaving the it unchanged if the experiment is limited by other heating mechanisms.

\subsubsection{Equilibrium dynamics}
\label{subsubsec:EquilibriumDynamics}

Prior to the advent of laser cooling, hot ions ($\sim$1\,eV) were routinely cooled in Penning traps using resistive cooling \citep{Wineland:1975_02, Itano:1995}. In principle, resistive cooling (where hot ions are cooled towards equilibrium with an electronic circuit) is the flip side of Johnson-noise or technical-noise heating (where cold ions are heated towards equilibrium with an electronic circuit).

Resistive cooling has been used and characterized in Penning traps for cooling clouds ($N\sim10^5$) of electrons \citep{Dehmelt:1968, Wineland:1975_02}. It is difficult to compare these results to the work at hand, however, as there exist thermalization mechanisms in a cloud which allow coupling between different modes of motion; these are not present in few-ion crystals or single trapped ions. Work on resistive cooling of single ions in Penning traps is much more scarce [with exceptions such as \citet{VanDyck:1995} and \citep{Djekic:2004}]. All such experiments (with either single ions or clouds) are performed in relatively large traps traps where $d > 2$\,mm. Separate experiments have implemented resistive cooling in Penning traps where the electronics was held at  room temperature \citep{Dehmelt:1968}, 80\,K \citep{Wineland:1975_02}, and 4.2\,K \citep{Gabrielse:1995, Djekic:2004}, though no systematic study has been made of the equilibrium behavior as a function of temperature.

In Paul traps, characterization of resistive heating has been made by \citet{Church:1969}. This experiment cooled a cloud of 10$^4$ protons in a spherical Paul trap with $d = 1.6$\,mm. The initially hot ions ($T=12,000$\,K) were cooled to 900\,K by coupling them to a resonant tank circuit at 330\,K. The discrepancy between the final ion temperature and the temperature of the circuit indicates the presence of some other heating mechanism. This was tentatively attributed to electron bombardment degassing producing sufficient distortion to cause heating of the ions by non-linearities in the RF field. However, such RF heating has very different outworking in a cloud compared to an ion crystal \citep{Chen:2013}.

Despite the relevance for ion-heating discussions, no experimental study has been made of resistive cooling -- nor of the equilibrium dynamics of trapped-ion heating -- for geometries differing significantly from hyperbolic surfaces (such as planar traps), or for small traps ($d < 1$\,mm). Nor have characterizations been made in Paul traps for single ions, or for few-ion crystals. Moreover, no consideration has been made regarding the adequacy of the resistive cooling model as a function of frequency, trap size or temperature. Given the dearth of experimental data, experiments to characterize the equilibrium behavior  of trapped ions -- either at 300\,K or 6\,K -- should be considered.

Beyond such regimes, it may be interesting to cool the external circuits further: to millikelvin temperatures. Assuming there are no other effects to interfere with the resistive cooling process, it may be possible to reach the quantum regime of the ion's motion with this method. From here one could measure $\Gamma$ and $\bar N$ independently (see Sec.~\ref{subsec:MasterEquation}). Knowledge about the stationary occupation number, $\bar n(t\rightarrow \infty)$, provides an additional analytic tool to distinguish between equilibrium and out-of-equilibrium noise processes, or to identify whether the noise arises from a low- or high-temperature source.

\subsubsection{Noise combinations}
\label{subsubsec:NoiseCombinations}

Characterization of the observed noise as a particular change is made -- such as varying the temperature, or cleaning the trap surface -- can provide information about possible noise sources, although it is not without ambiguity. By changing several parameters in combination, each of which has been characterized separately, one can infer more than was possible from varying parameters individually. The combinations and permutations are obviously many, but one possible experiment is outlined here by way of an example.

Cleaning the electrodes with an Ar-ion beam reduces the ion-heating rate observed; an effect attributed to the removal of  surface contaminants. Operating traps at cryogenic temperatures also reduces the ion-heating rates observed, possibly due to suppression of thermally activated processes involving contaminants on the electrode surface. If the conjectured mechanisms are correct it would be expected that, having cleaned the surface, operation at cryogenic temperatures would bring no further improvement. A temperature scaling of $\gamma=0$ might then be expected.

There are other possible causes for reduced noise at cryogenic temperatures. If the initial reduction in noise as the temperature is reduced from 300 -- 30\,K is due to improved magnetic-field shielding then it would be expected that $\gamma \neq 0$ when an ion-beam-cleaned trap is cooled. In this situation, however, a cryostat housed in mu-metal shielding might be expected to exhibit $\gamma=0$.

\subsection{Further avenues for investigation of noise}
\label{subsec:Outlook_InvestigateII}

There are a number of aspects of electric-field noise near surfaces which have direct implications for ion traps, but which are not best solved at the trap level. These require a combination of experimental and theoretical investigation, and are addressed here in turn.\\

\noindent $\bullet$ RF heating\\
Throughout most parts of this paper micromotion heating, or RF-heating, has been neglected. This was implicit from disregarding  the higher-order terms in Eq.~\eqref{eq:MathieuSolution}.  If $q_x$ is not sufficiently small \citep{Blatt:1986} or when the ion is displaced from the RF null \citep{Blakestad:2011} (see also App.~\ref{app:ExcessMM}) these higher-order terms become more important. Targeted work in this direction could provide further insight into the role of micromotion on the observed heating and how to use it for measuring spectral densities of electric-field noise at RF frequencies.  \\

\noindent $\bullet$ Instabilities in small traps\\
While all ion traps considered here are operated in a nominally stable region of the solution to the Mathieu equations, the effect of higher-order terms in the potential can cause the ions to become very energetic or unstable \citep{Wang:1993, Alheit:1996}.  The microscopic surface structure of miniature ion traps has been found to affect the performance of ion traps in mass-spectrometer applications for ion-electrode distances ranging from one to several hundred micrometres \citep{Xu:2009}. Also, if ion traps are made smaller so that the surface roughness is of a comparable scale to features of the trap geometry, then the quadrupole assumption made in Eq.~\eqref{eq:MathieuClassical} is not appropriate. Consequently the condition that $q_x\ll 1$ is not sufficient to ensure that the higher-order coefficients, $C_{2j}$ in Eq.~\eqref{eq:HeatingRateGoldenRule}, are small. A quantitative analysis of anharmonicities in small ion traps due to microscopic surface defects would shed light on another possible heating mechanism. These may become increasingly important as traps are miniaturized further.\\

\noindent $\bullet$ Spectrum of space-charge noise\\
Section~\ref{subsec:SpaceCharge} considered heating due to space charges from electron emission. Electron emission can usually be treated as white noise. However, in ion traps, field-emitted electrons would be correlated with the applied RF. It may be worthwhile to calculate what effect this pulsed white noise has on the electric-field spectrum and so on the ion heating. The secondary effects of small amounts of electron bombardment on the anode would likewise be useful to investigate, so that these sources of noise could be eliminated.\\

\noindent $\bullet$ Characterization of TLFs\\
When considering TLF models in Sec.~\ref{subsec:TwoLevelFluctuators}, a number of different cases were considered, both with regards to the fluctuators' spatial and energetic distribution. Different distributions lead to different predicted behavior. By experimentally measuring how, for example, the frequency scaling exponent, $\alpha$, varies as a function of temperature, it may be possible to constrain certain aspects of the model. This, in turn could constrain, or shed light on, what the physical mechanisms might be which underlie the TLF behavior.\\

\noindent $\bullet$ Characterization of adsorbates\\
When considering adsorbates in Secs.~\ref{subsec:AdatomDipoles} and \ref{subsec:AdatomDiffusion} a number of estimates had to be made regarding, for example, the types of adsorbate covering the surface and the way those adsorbates are bound. There are two respects in which these estimates could be greatly improved. Firstly, it is not known experimentally exactly what kind of surface coverage occurs: what materials, in what conformations, etc. Materials science experiments are required to characterize the surfaces of ion traps, as the ion sees them. Secondly, most theoretical studies of adsorbate behavior have focused on extreme cases of either alkaline atoms (which are strongly bound to the surface) or noble gases (which are very weakly bound to the surface). Theoretical work is needed to develop models in the intermediate regime for atoms such as carbon and oxygen which are most likely what contributes to noise in practice.

With this in mind, the investigation comes full circle in new experiments which integrate surface-analysis tools into ion-trap experiments to allow the microscopic structure and composition of surfaces to be measured, while using the ion to measure the electric-field noise \citep{Hite:2012, Daniilidis:2014}. Ion traps can then contribute to the toolbox of ways to investigate the surface contamination which may have been limiting them.

\subsection{Conclusion}
\label{subsec:FinalConclusions}

In conclusion, it is clear that the topic of heating in ion traps is complex and multifaceted. Consequently it necessarily draws on the insight of many diverse disciplines. At the same time, it can be expected to provide insight of interest to numerous disciplines.

The picture provided by ions is complicated by the fact that numerous physically distinct heating mechanisms can generate similar levels of noise, and may even manifest similar scaling behaviours over certain regimes. Equally, a single basic mechanism may manifest itself very differently depending on values of particular characteristic frequencies, distances, and temperatures for the situation at hand. These characteristic scales may vary with the experimental apparatus, the trap material, or the type and coverage fraction of surface impurities. Nonetheless, particularly when taken together, and in collaboration with methods from other fields, such a diversity of behaviour can potentially be turned to an advantage; ions offer a multitude of ways via which the noise might be investigated.

In laying out the current state of understanding for noise in ion traps -- both experimentally and theoretically -- it is hoped that the way forward for ion-trap research will become clearer. Firstly, in summarizing factors which may need to be considered, the review can provide a reference for those who have no interest in noise, other than to reduce it.  Secondly, it provides an overview of what  is known (and what is not known) to provide structure both to those who intend to study the noise further, or use the noise to study other physical questions.

\begin{acknowledgments}
This work was supported by
the European Research Council through the Advanced Research Project CRYTERION and the START grant Y 591-N16;
EU project SIQS;
the Austrian Science Fund (FWF) project SFB FOQUS;
Eranet NanoSci-E+ project NOIs;
and the Institute for Quantum Information GmbH.

We owe a debt of gratitude for the patient assistance of the ion-trapping community whose work we have reviewed, particularly from current and past members in the groups of
I.~Chuang (MIT, Cambridge, MA),
C.~Monroe (University of Maryland, MD), and
D.J.~Wineland (NIST, Boulder, CO).
We also thank H.~Sadeghpour for insightful discussions on the theoretical aspects.
\end{acknowledgments}

\appendix

\section{Derivation of the heating rate, $\Gamma_{\rm h}$}
\label{Appendix_Derivation}

This appendix summarizes the derivation of the heating rate, $\Gamma_{\rm h}$, given in Eq.~\eqref{eq:HeatingRateGoldenRule} using time-dependent perturbation theory. For the following derivation to be valid it must be assumed that the disturbance of the ion by the noisy electric field is weak, $|d_{\rm I}\delta E(t)|\ll \hbar \omt$, and that the correlation time of $\delta E(t)$ is short compared to the resulting heating time, $1/\Gamma_\mathrm{h}$. In the interaction picture with respect to the trap Hamiltonian, $\hat H_{\rm t}(t)$, the motional state of the ion can be written as $|\psi(t)\rangle =\sum_{n=0}^\infty c_n(t)|n\rangle$, where the amplitudes $c_n(t)$ evolve as
\begin{equation}
\dot c_n = i \frac{d_{\rm I}}{\hbar}\left( \sqrt{n} u(t) c_{n-1} + \sqrt{n+1} u^*(t) c_{n+1}\right)\delta E(t).
\end{equation}
The ion is initially prepared in the lowest vibrational state $|0\rangle$, and for short times, $c_0(t)\simeq 1$ and $c_{n>1}(t)\simeq 0$. Then, by integrating the equation for $\dot c_1(t)$ the probability, $P_1(\Delta t)=|c_1(\Delta t)|^2$, of finding the ion in the first vibrational state after a time $\Delta t$ is
\begin{equation}
\begin{split}
&P_1(\Delta t)= \frac{d_{\rm I}^2}{\hbar^2} \left\langle \Big|\int_0^{\Delta t}Ê dt^\prime \, u(t^\prime)\delta E(t^\prime)\Big|^2\right\rangle\\
=& \frac{2 d_{\rm I}^2}{\hbar^2} {\rm Re} \int_0^{\Delta t} dt_1 \int_0^{t_1} dt_2 \, u(t_1) u^*(t_2) \langle \delta E(t_1)\delta E(t_2)\rangle.
\end{split}
\end{equation}
For times, $\Delta t$, which are much longer than the correlation time of $\delta E(t)$ the lower integration bound for $t_2$ can be shifted to $-\infty$. The heating rate is defined as $\Gamma_{\rm h}=P_1(\Delta t)/\Delta t$ and using the expansion of $u(t)$ given in Eq.~\eqref{eq:u_expansion} it can be written as
\begin{equation}
\begin{split}
\Gamma_{\rm h}= &\frac{2d_{\rm I}^2}{\hbar^2}{\rm Re} \sum_{n,m} C_{2n} C_{2m}^* \left[\frac{1}{\Delta t} \int_0^{\Delta t} dt_1\, e^{i\OmRF(n-m)t_1}\right] \\
&\times \int_{0}^\infty d\tau \, e^{i\OmRF(b/2+m)\tau} \langle \delta E(t_1)\delta E(t_1-\tau)\rangle.
\end{split}
\end{equation}
For times $\Delta t\gg \OmRF^{-1}$ the integral over $t_1$ averages approximately to zero for $n\neq m$ and it is equal to $\Delta t$ for $n=m$. For a stationary noise process $\langle \delta E(t_1)\delta E(t_1-\tau)\rangle=\langle \delta E(\tau)\delta E(0)\rangle=\langle \delta E(-\tau)\delta E(0)\rangle^*$ and the expression above reduces to the result given in Eq.~\eqref{eq:HeatingRateGoldenRule}.

\section{Excess micromotion and heating}
\label{app:ExcessMM}

This appendix discusses modifications of the heating rate, $\Gamma_{\rm h}$, for an ion which is displaced from the RF null by a static electric field, $E_{\rm stat}$. The following analysis is restricted to a one-dimensional motion of the ion along the $x$-axis and to small stability parameters, $q_x$, $a_x$.
While the current analysis uses a fully quantum-mechanical description of the ion, an equivalent conclusion can be obtain from a classical pseudopotential analysis \citep{Blakestad:2009}.

In the presence of a static offset field, the quantized motion of the ion is described by the total trap Hamiltonian
\begin{equation}
\hat H_{\rm t}(t)=\frac{\hat p^2}{2\mi} +  \frac{|e|}{2} \left[ \Phi_{\rm dc}'' +  \Phi_{\rm rf}'' \cos(\OmRF t) \right] \hat x^2 - |e| E_{\rm stat} \hat x.
\end{equation}
The position operator, $\hat x(t)$, in the Heisenberg representation obeys the
inhomogeneous Mathieu equation,
\begin{equation}\label{eq:InhomoMathieu}
\frac{d^2}{dt^2} \hat x(t) +\frac{|e|}{\mi}\left[\Phi_{\rm dc}''+\Phi_{\rm rf}''\cos(\OmRF t) \right] \hat x(t)= \frac{|e|}{m_{\rm I}}  E_{\rm stat},
\end{equation}
with general solution
\begin{equation}
\hat x(t) = x_{\rm inh}(t) + \hat x_0(t).
\end{equation}
Here $\hat x_0= \sqrt{\hbar/(2\mi \omega_{\rm t})}[ u(t) \hat a^\dag + u^*(t) \hat a]$ is identical to the position operator defined in Eq.~\eqref{eq:PositionOperator} for $E_{\rm stat}=0$ and $x_{\rm inh}(t)$ is a particular solution of the inhomogeneous Mathieu equation. To lowest order in $q_x$,
\begin{equation}
x_{\rm inh} (t)\simeq  \Delta x \left[ 1 +\frac{q_x}{2} \cos(\OmRF t) \right] = \Delta x + x_{\rm emm}(t),
\end{equation}
where $\Delta x= |e| E_{\rm stat}/(\mi \omega_{\rm t}^2)$ is the mean displacement of the ion from the RF null and $x_{\rm emm}(t)\sim \Delta x$ is the excess micromotion. Again, the quantum state of the ion can be expressed in terms of the number states, $|n\rangle=(\hat a^\dag)^n/\sqrt{n!}|0\rangle$, of a reference oscillator with time-independent  annihilation and creation operators, $\hat a$ and $\hat a^\dag$. However, in the lab frame, the center of the wave function is modulated by $x_{\rm emm}(t)$.

The coupling of the ion to an additional noisy field is described by the Hamiltonian $\hat H_{\rm ion-field}(t)= |e| \delta \Phi(t, \hat x)$. In the interaction picture with respect to $\hat H_{\rm t}(t)$ and expanded around the mean position, $\Delta x$, the resulting ion-field coupling is given by
\begin{equation}
\begin{split}
\hat H_{\rm ion-field}(t)\simeq&  |e|  \left[x_{\rm emm}(t) + \hat x_0(t) \right] \delta \Phi^\prime(t,\Delta x)\\
+&\frac{|e|}{2} \left[x_{\rm emm}(t) + \hat x_0(t) \right]^2  \delta \Phi^{\prime\prime}(t,\Delta x).ÊÊ
\end{split}
\end{equation}
The first term represents the direct coupling of the noisy electric field, $\delta E_{\rm t}(t)= - \delta \Phi^\prime(t,\Delta x)$, to $\hat x_0(t)$. This contribution is the same as for an ion  trapped at the RF null and is  already included in the result for the heating rate, $\Gamma_{\rm h}$, in Eq.~\eqref{eq:HeatingRateGoldenRule}. The second term is proportional to the gradient of the noisy electric field, $\delta E^{\prime}_{\rm t}(t)= - \Delta x \delta \Phi^{\prime\prime}(t,\Delta x)$. It contributes to heating via a mixing between $x_{\rm emm}(t)$ and $\hat x_0$ and therefore it is only relevant for a finite $\Delta x$. Explicitly, by expanding $\hat H_{\rm ion-field}(t)$  up to first order in $q_x$,
\begin{equation}
\begin{split}
&\hat H_{\rm ion-field}(t) \simeq -  d_{\rm I} \left[\hat a^\dag e^{i\omega_{\rm t} t}  + \hat a e^{-i\omega_{\rm t} t}Ê\right]  \\
&\times \left[ \delta E_{\rm t}(t) + \frac{q_x}{2}Ê\cos(\OmRF t)\left(\delta E_{\rm t}(t) + \Delta x \delta E^{\prime }_{\rm t}(t)\right)\right].
\end{split}
\end{equation}
Restricted to low-frequency noise, $\omega\sim \omega_{\rm t}$, only the first term in brackets is important and the heating of the displaced ion is identical to the heating of the ion located at the RF null. For noise at RF frequencies, $\omega\sim \OmRF$, and $\Delta x\neq0$, both the direct coupling and the effect of a finite field gradient become important. Specifically, for noise applied via the trapping electrodes, $\Delta x \delta E^{\prime }_{\rm t}(t)\simeq \delta E_{\rm t}(t)$. The resulting RF heating rate is then four times as large as predicted by Eq.~\eqref{eq:HeatingRateGoldenRule}.

\section{Summary of noise characteristics}
\label{Appendix_Summary}

One of the central messages of this paper is that  the issues involved in trapped-ion heating are complex, and  simplifications are made at ones own peril.
Even minor changes in certain details of the experiment can have significant impacts on the noise behaviour.
Weighed against this is a need to maintain some kind of overview. Table~\ref{tab:TheorySummary} summarizes some of the major theoretical results of this paper.
The table is intended for orientation; as a guide to reading the main paper.
In the interests of summarizing material,  the many caveats which go with these results have been omitted,
and the reader is directed to the sections of the main paper for detailed discussion.
The absolute levels of the noise are to be treated with particular caution: they have been calculated
(as described in the main paper) to give indicative values of what \textit{may} be seen.
Depending on the details of the experiment the values actually observed, for a given source, might differ from these by many orders of magnitude.

\newpage

\renewcommand{\arraystretch}{1.5}
\LTcapwidth=\textwidth
\begin{longtable*}{p{2.5cm} l | c | c | c | p{2cm} | l p{6cm}}
\caption[Summary of theories]
{Summary of theoretical predictions for different noise mechanisms. Unless otherwise stated, the values given here are for $\omega=2\pi \times 1$\,MHz, $d=100$\,$\mu$m, $T=300$\,K. The values presented in this table are given for the purpose of orientation only.
Details and discussion of the circumstances under which they hold are given in the relevant portions of the paper, to which the reader is referred.}
\label{tab:TheorySummary}
\endfirsthead
\endhead
\endfoot
\endlastfoot

\hline
Source                              & Geometry  & $\alpha$  & $\beta$                       & $\gamma$  & Abs./ (\B)    & Section                                   & Notes                                         \\
\hline
Black-body radiation                &           & -2        & 0                             & 1         & 10$^{-22}$    & \ref{subsec:Blackbody}                    &                                               \\
Near-field \hspace{2mm} black-body  & Plane     & -2        & 2-3                           & 1         & $10^{-17}$    & \ref{subsubsec:SurfaceFields}             & $\rho_{\rm e}=2.44\times 10^{-8}\,\Omega$m    \\
\hline
EMI                                 &           & 1         & 0                             & -         & 10$^{-14}$    & \ref{subsec:EMI}                          & Outdoors, near buildings                      \\
                                    &           & 3         & 0                             & -         & 10$^{-10}$    & \ref{subsec:EMI}                          & Indoors                                       \\
\hline
EM pickup                           &           & 1         & JN                            & $>0$      & 10$^{-10}$    & \ref{subsec:EMPickup}                     & Unshielded loop of area $100$\,cm$^2$;        \\
                                    &           &           &                               &           &               &                                           &$D=100$\,$\mu$m; Indoors                       \\
\hline
Johnson noise                       & Plane     & $\sim$    & 2                             & 2         & $\sim$        & \ref{subsec:JohnsonNoise}                 &                                               \\
                                    & Spheres   & $\sim$    & 4                             & 2         & $\sim$        & \ref{subsec:JohnsonNoise}                 & $R_{\rm el} \ll d$                            \\
                                    & Needles   & $\sim$    & 2.5                           & 2         & $\sim$        & \ref{subsec:JohnsonNoise}                 & Geometry approximating \citet{Deslauriers:2006_09}, $30\,\mu\mathrm{m} < d <  200\,\mu$m\\
                                    & 2 Level   & $\sim$    & U                             & 2         & $10^{-14}$    & \ref{subsec:JohnsonNoise}                 & Geometry from \citep{Leibrandt:2007_01}       \\
                                    & Surface   & $\sim$    & U                             & 2         & $10^{-12}$    & \ref{subsec:JohnsonNoise}                 & Geometry from \citep{Leibrandt:2007_01}       \\
Technical noise                     &           & $\sim$    & JN                            & $\sim$    & $10^{-12}$    & \ref{subsubsec:TechnicalNoise}            & 80~dB of filtering, $D=100$\,$\mu$m           \\
\hline
Space charge                        & Point     & 0         & 4                             & $\sim$    & 10$^{-15}$    & \ref{subsec:SpaceCharge}                  & 1\,$\mu$A electron-emission; $d\,=\,40$\,$\mu$m\\
\hline
Patch potentials                    & Plane     & U         & 4                             & U         & U             & \ref{subsec:PatchPotentials}              & $r_{\rm c} \ll d$                             \\
                                    & Plane     & U         & 2                             & U         & U             & \ref{subsec:PatchPotentials}              & $r_{\rm c} \gg d$                             \\
                                    & Sphere    & U         & 4($\perp$), 6($\parallel$)    & U         & U             & \ref{subsec:PatchPotentials}              & $r_{\rm c} \ll d$                             \\
                                    & Needle    & U         & 4($\perp$), 6($\parallel$)    & U         & U             & \ref{subsec:NeedleTrap}                   & $r_{\rm c} \ll d$                             \\
\hline
TLFs                                & Plane     & 1         & 4                             & 1         & 10$^{-12}$    & \ref{subsec:TwoLevelFluctuators}          & Uniform TLF distribution                      \\
\hline
Adatom dipoles                      & Plane     & 0         & 4                             & 2.5       & 10$^{-13}$    & \ref{subsec:AdatomDipoles}                & $\gamma=0$ for $T \lesssim 65$\,K;            \\
                                    &           &           &                               &           &               &                                           &$\alpha=2~\mathrm{for}~\omega \gtrsim \Gamma_0 \sim 10$\,MHz \\
\hline
Adatom diffusion                    & Plane     & 2         & 6                             & E         & 10$^{-18}$    & \ref{subsubsec:PlaneAndNeedleDiffusion}   &                                               \\
                                    & Needle    & 1.5       & 6($\perp$), 8($\parallel$)    & E         & 10$^{-16}$    & \ref{subsubsec:PlaneAndNeedleDiffusion}   &                                               \\
                                    & Patches   & 1.5       & 4($\perp$), 6($\parallel$)    & E         & 10$^{-14}$    & \ref{subsubsec:CorrugatedDiffusion}       &                                               \\
\hline
\multicolumn{8}{l}{Key:}\\
\multicolumn{8}{l}{$\sim$ : can take such a wide range of values that it is not considered sensible to state a single number.}\\
\multicolumn{8}{l}{U : value not known.}\\
\multicolumn{8}{l}{JN : scales in the same way as Johnson noise.}\\
\multicolumn{8}{l}{E : follows an exponential scaling, rather than a power law.}\\
\newpage
\end{longtable*}

\section{Variable names}
\label{Appendix_Variables}

\LTcapwidth=\textwidth

\begin{longtable}{|l|l|}
\multicolumn{2}{c}%
{\tablename\ \thetable\ }\\
\hline \textbf{ } & \textbf{Quantity} \\ \hline
\endfirsthead

\multicolumn{2}{c}%
{\tablename\ \thetable\ -- \textit{Continued from previous column} }\\
\hline \textbf{ } & \textbf{Quantity} \\ \hline
\endhead

\hline \multicolumn{2}{r}{\textit{Continued on next column}} \\
\endfoot

\hline
\endlastfoot

$A$                     &   area of interest on trap surface                                \\
$A_{\rm el}$            &   electrode area contributing to noise                            \\
$A_\mathrm{L}$          &   area of loop contributing to EM pickup                          \\
$A_\mathrm{p}$          &   average patch area                                              \\
$\hat{a}$               &   annihilation operator                                           \\
$\hat{a}^\dagger$       &   creation operator                                               \\
$a_{\rm L}$             &   lattice constant                                                \\
$a_x$                   &   Mathieu stability parameter (in $x$ direction)                  \\
$a_0$                   &   extent of ion's ground-state wave function                      \\
\hline
${\bm B}$               &   magnetic field                                                  \\
$B$                     &   magnetic-field strength normal to loop                          \\
$b$                     &   Floquet exponent                                                \\
\hline
$C$                     &   capacitance                                                     \\
$C_{2j}$                &   Floquet expansion coefficients                                  \\
$C_\sigma$              &   adsorbate density correlation spectrum                          \\
$C_{3}$                 &   van der Waals dynamic polarizability prefactor                 \\
$C_{\rm V}$             &   voltage correlation function                                    \\
$c$                     &   speed of light                                                  \\
$c_k$                   &   mode function of $k$-th motional mode                           \\
$c_n$                   &   amplitude of motional state $|n\rangle$                         \\
\hline
$D$                     &   characteristic trap dimension                                   \\
$\mathcal{D}$           &   diffusion constant                                              \\
$\mathcal{D}_{\rm t}$   &   diffusion rate due to quantum tunneling                         \\
$\mathcal{D}_{\rm 0}$   &   prefactor for thermally activated diffusion                     \\
$d$                     &   ion-electrode separation                                        \\
$d_{\rm I}$             &   dipole moment of ion                                            \\
$d_{k}$                 &   dipole moment of $k$-th motional mode                           \\
\hline
${\bm E}$               &   electric field                                                  \\
$E$                     &   energy                                                          \\
$E_{\rm stat}$          &   static electric field                                           \\
$E_{\rm t}$             &   component of the \textbf{E}-field along the trap axis           \\
$\delta E_{\rm t}$      &   fluctuating contribution of  $E_{\rm t}$                        \\
$E_{\rm max}$           &   largest energy difference in TLF distribution                   \\
$E_{\rm TLF}$           &   energy difference of a two level system                         \\
$\delta {\bm E}$        &   fluctuating component of the ${\bm E}$-field                    \\
$\delta E_k$            &   projection of ${\bm E}$-field noise onto $k$-th eigenmode       \\
$e$                     &   elementary charge                                               \\
$|e\rangle$             &   ion's electronic excited state                                  \\
${\bm e}_{i}$           &   unit vector along the $i$ axis, $i \in x,y,z$                   \\
${\bm e}_{\rm t}$       &   unit vector along the trap axis                                 \\
\hline
$F_{\rm a}$             &   external noise factor                                           \\
$F$                     &   filtering factor                                                \\
\hline
$G_{\rm E}$             &   Green's function of the electric field                          \\
$g$                     &   phonon density of states                                        \\
$\bar{g}$               &   partial phonon density of states                                \\
$|g\rangle$             &   ion's electronic ground state                                   \\
$g_{\rm D}$             &   geometrical factor $\propto$ E-field of a dipole                \\
$\tilde{g}_{\rm D}$     &   dimensionless version of $g_{\rm D}$                            \\
\hline
$\hat H$                &   Hamiltonian                                                     \\
$\hat H_{\rm ion-field}$&   ion-field interaction Hamiltonian                               \\
$\hat H_{\rm t}$        &   trap Hamiltonian                                                \\
$\hat H_{\rm TLF}$      &   Hamiltonian of a TLF                                            \\
$h$                     &   thickness of a surface layer on the electrodes                  \\
$\hbar$                 &   Planck's constant / 2$\pi$                                      \\
\hline
$\mathcal{I}_{\tilde R}$&   a dimensionless integral defined in Eq.~\eqref{eq:Jtildeomega}  \\
\hline
$k$                     &   phonon mode number                                              \\
$k_{\rm B}$             &   Boltzmann constant                                              \\
$k_x$                   &   $x$ component of laser wave vector                              \\
\hline
$L$                     &   inductance                                                      \\
$\ell$                  &   lattice coordination number                                     \\
\hline
$M$                     &   mass of an atom in the electrode                                \\
$m$                     &   phonon number                                                   \\
$|m\rangle$             &   phonon number state of ion or adatom                            \\
$m_{\rm ad}$            &   mass of adatom                                                  \\
$m_{\rm I}$             &   ion mass                                                        \\
$m_{\rm TLF}$           &   effective mass of a TLF                                         \\
\hline
$\bar{N}$               &   mean phonon occupation of a heat bath                           \\
$N_{\rm A}$             &   number of fluctuators within an area $A$                        \\
$N_{\rm I}$             &   number of ions in a crystal                                     \\
$N_{\rm p}$             &   number of patches                                               \\
$N_{\rm th}$            &   phonon occupation number in thermal equilibrium                 \\
$n$                     &   phonon number                                                   \\
$|n\rangle$             &   phonon number state of ion or adatom                            \\
$\bar{n}$               &   average phonon excitations of ions or adatoms                   \\
$\dot{\bar{n}}$         &   heating rate                                                    \\
$n_{\rm B}$             &   Bose-Einstein distribution                                      \\
${\bm n}_i$             &   unit vector from the ion to surface dipole $i$                  \\
${\bm n}_s$             &   unit vector perpendicular to the surface                        \\
\hline
$P$                     &   used for various probability distributions                      \\
$\mathcal{P}$           &   surface polarization density                                    \\
$P_{\rm 0}$             &   normalization constant for $P$                                  \\
$\hat{p}$               &   quantized momentum of the ion                                   \\
$p_{\rm |e\rangle}$     &   excitation probability                                          \\
$P_{\rm n}$             &   occupation probability of vibrational level $n$                 \\
\hline
$Q$                     &   quality factor                                                  \\
$q_x$                   &   Mathieu stability parameter (in $x$ direction)                  \\
\hline
$R$                     &   resistance                                                      \\
$R_{\rm el}$            &   electrode size or radius of curvature                           \\
$\tilde{R}$             &   dimensionless electrode size for adatom diffusion               \\
$R_{\rm L}$             &   real part of inductor's impedance                               \\
$R_{\rm p}$             &   patch radius                                                    \\
$r_{\rm c}$             &   correlation length                                              \\
$r_{\rm cc}$            &   radius of a cylindrical cone on the surface                     \\
\emph{\textbf{r}}$_{\rm I}$ &   position of the ion (center of the trap)                    \\
\emph{\textbf{r}}$_i$   &   position of a dipole on (or in) the trap surface                \\
$r_{\rm prolate}$       &   radius of spheroidal `needle' electrode                         \\
\hline
$S$                     &   electrode surface                                               \\
$S_{\rm B}$             &   spectral density of magnetic-field noise                        \\
$S_{\rm E}$             &   spectral density of electric-field noise                        \\
$\tilde{S}_{\rm E}$     &   median spectral density of electric-field noise                 \\
$S^{\rm (AD)}_{\rm E}$  &   $S_{\rm E}$ due to adatom diffusion                             \\
$S^{\rm (ADF)}_{\rm E}$ &   $S_{\rm E}$ due to adatom dipole fluctuations                   \\
$S^{\rm (BB)}_{\rm E}$  &   $S_{\rm E}$ due to black-body radiation                         \\
$S^{\rm (BBS)}_{\rm E}$ &   $S_{\rm E}$ due to black-body above a surface                   \\
$S^{\rm (EMI)}_{\rm E}$ &   $S_{\rm E}$ due to EMI                                          \\
$S^{\rm (JN)}_{\rm E}$  &   $S_{\rm E}$ due to Johnson noise                                \\
$S^{\rm (PP)}_{\rm E}$  &   $S_{\rm E}$ due to patch potentials                             \\
$S^{\rm (PU)}_{\rm E}$  &   $S_{\rm E}$ due to EM pickup                                    \\
$S^{\rm (SC)}_{\rm E}$  &   $S_{\rm E}$ due to space charge                                 \\
$S^{\rm (TLF)}_{\rm E}$ &   $S_{\rm E}$ due to two-level fluctuators                        \\
$S_{\rm V}$             &   spectral density of voltage noise                               \\
$\bar{S}_{\mu}$         &   averaged dipole-fluctuation spectrum                            \\
$S_{\mu}^i$             &   spectrum of dipole $i$                                          \\
$s$                     &   electrode-electrode separation                                  \\
$s_{\eta}$              &   orientation-dependent constant in Eq.~\eqref{eq:SE_SmallPatches}\\
\hline
$T$                     &   temperature                                                     \\
$T^*$                   &   characteristic temperature of adatom system                     \\
$T_{\rm c}$             &   critical temperature of superconductor                          \\
$T_{\rm min}$           &   min. relaxation time for symmetric double well                  \\
$T_1$                   &   relaxation time of a TLF                                        \\
$t$                     &   time                                                            \\
$\tilde{t}$             &   dimensionless time                                              \\
$t_{\rm w}$             &   waiting time after cooling the ion                              \\
$\Delta t$              &   time interval                                                   \\
\hline
$U$                     &   potential                                                       \\
$U_{\rm vdW}$           &   van der Waals potential                                         \\
$U_0$                   &   depth of adatom-surface potential                               \\
$u_{\rm em}$            &   spectral energy density of black-body radiation                 \\
$u$                     &   dimensionless solution of Mathieu Eq.~\eqref{eq:MathieuClassical}\\
\hline
$V$                     &   voltage applied to trap electrodes                              \\
$V_{\rm b}$             &   potential-barrier height for a TLF                              \\
                        &       \hspace{1cm} or adsorbate hopping                           \\
$V_{i}$                 &   fluctuating voltage on the $i$-th patch                         \\
$V_{\rm rf}$            &   RF voltage applied to trap electrode                            \\
$V_\mathrm{L}$          &   EMI voltage induce in a loop                                    \\
$V_{\rm max}$           &   largest TLF barrier height                                      \\
$V_{\rm min}$           &   smallest TLF barrier height                                     \\
$V_{\rm pot}$           &   potential forming a TLF                                         \\
$V_{\rm 0}$             &   most-likely TLF barrier height                                  \\
$\Delta V$              &   $V_{\rm max} - V_{\rm min}$                                     \\
$v$                     &   sound velocity                                                  \\
$v_{\rm l}$             &   longitudinal sound velocity                                     \\
$v_{\rm t}$             &   transverse sound velocity                                       \\
\hline
$W$                     &   work function                                                   \\
$\Delta W$              &   change in work function                                         \\
$\tilde{w}$             &   dimensionless width of surface-adatom potential                 \\
\hline
$X_0$                   &   amplitude of the ion's motion                                   \\
$x$                     &   position of the ion                                             \\
$\hat{\bm x}$           &   quantized position of the ion, 3D                               \\
$\hat{x}$               &   quantized position of the ion, 1D                               \\
$x_{\rm emm}$           &   amplitude of excess micromotion                                 \\
$x_{\rm inh}$           &   a solution to inhomogeneous Mathieu equation                    \\
$\hat{x}_0$             &   quantized relative position of the ion                          \\
$x_0$                   &   equilibrium separation of two ions in a trap                    \\
$\Delta x$              &   mean displacement of ion from RF null                           \\
\hline
$z$                     &   distance from the trap surface                                  \\
$z_{\rm w}$             &   separation between wells                                        \\
$z_{0}$                 &   equilibrium position of adatom above a surface                  \\
\hline
$\alpha$                &             frequency-scaling exponent\\
$\alpha_{\rm p}$        &               dynamic polarizability of a particle\\
\hline
$\beta$                    &              distance-scaling exponent\\
\hline
$\Gamma$               &             damping rate\\
$\Gamma_{\rm h}$        &        rate of heating from $n=0$ to $n=1$\\
$\Gamma_{n\to m}$        &        transition rate from $|n\rangle$ to $|m\rangle$  \\
$\Gamma^{\rm rf}_{\rm h}$        & heating rate due to noise on RF drive\\
$\Gamma_{\rm hop}$        &              hopping rate between adsorption sites\\
$\Gamma_{0}$        &           characteristic transition rate of adatom fluctuation\\
$\gamma$               &             temperature-scaling exponent\\
\hline
$\Delta$                     &             energy difference between two wells \\
$\Delta_0$               &           tunneling coupling between two wells\\
$\Delta_{\rm max}$        &        maximum value for $\Delta$ \\
$\Delta_{\rm 0, max}$        &        maximum value for $\Delta_0$         \\
$\Delta_{\rm 0, min}$        &        minimum value for $\Delta_0$        \\
$\delta$                      &             frequency shift                                \\
$\delta_s$                      &           skin depth                                        \\
\hline
$\epsilon_0$              &              electric permittivity of free space\\
$\epsilon$              &              relative electric permittivity\\
\hline
$\zeta$                    &             a numerical factor in Eq.~\eqref{eq:AdatomPotentialEstimate}\\
\hline
$\eta$                        &         Lamb-Dicke parameter \\
\hline
$\Theta$                      &            unit step function                \\
tan $\theta$              &            dielectric loss tangent\\
$\theta_{\rm cc}$     &            angle of a cylindrical cone on a surface\\
\hline
$\kappa$                    &        dipole geometrical efficiency factor \\
\hline
$\lambda$                  &             tunneling parameter\\
$\lambda_0$             &             most-likely tunneling parameter\\
\hline
$\boldsymbol{\mu}$        &             dipole on a surface\\
$\mu$                     &             dipole moment \\
$\hat \mu$                     &             dipole operator of a TLF \\
$\mu_0$                    &        permeability of free space\\
$\mu_n$                  &            dipole moment of adatom in motional state $|n\rangle$ \\
$\delta \mu$            &        change in dipole moment on/off patch        \\
$\delta \mu_i$            &        fluctuating dipole moment $\perp$ to a surface \\
\hline
$\nu_n$                     &              vibrational freq. of adatom in motional state $|n\rangle$ \\
$\nu_{nm}$               &              $\nu_n - \nu_m$\\
$\nu_{10}$               &              characteristic vibrational frequency $= \nu_1-\nu_0$\\
\hline
$\Xi_{\rm l}$            &         longitudinal deformation potential constant \\
$\Xi_{\rm t}$            &        transverse deformation potential constant \\
\hline
$\rho$                      &                density of electrode material\\
$\rho_{\rm coh}$            &        coherence of a superposition state        \\
$\rho_{\rm e}$            &        electrical resistivity\\
$\rho_{\rm I}$             &               reduced ion density operator\\
\hline
$\sigma_{\rm d}$      &             areal density of surface dipoles\\
$\bar{\sigma}_{\rm d}$&     mean areal density of surface dipoles\\
$\delta \sigma_{\rm d}$&        adsorbate density fluctuations\\
$\sigma_{\rm p}$        &        patch coverage fraction \\
$\sigma_{x}$              &          Pauli operator\\
$\sigma_{z}$              &          Pauli operator\\
$\tilde{\sigma}_{z}$         &          Pauli operator in a rotated eigenbasis\\
\hline
$\tau_{\rm e}$        &        temporal width of electron pulse        \\
$\tau_0$                    &        characteristic time of TLF switching \\
&        \hspace{1cm} or adsorbate hopping\\
\hline
$\Phi$                        &        potential\\
$\Phi_0$                &        constant electrode potential\\
$\Phi_{\rm dc}$         &          DC potential\\
$\Phi_{\rm rf}$            &        RF potential\\
$\Phi_{\rm p}$            &           potential of a patch \\
$\phi$                 &          angle used in the diagonalization of $\hat H_{\rm TLF}$\\
$\varphi_0$                  &            initial phase of ion's motion\\
\hline
$\chi_i$                  &            step function defining patch $i$\\
\hline
$\vert \psi\rangle$          & wave function of the motional state of the ion\\
$\vert \psi_{\rm L}\rangle$  & wave function in the left well\\
$\vert\psi_{\rm R}\rangle$ & wave function in the right well\\
\hline
$\Omega_{\rm c}$        &         Rabi frequency on carrier transition\\
$\Omega_{\rm L}$        &         Rabi frequency\\
$\Omega_{n,m}$        &        coupling between states  $|g,n\rangle$ and $|e,m\rangle$ \\
$\Omega_{\rm rf}$         &        trap-drive frequency\\
$\omega$                    &         angular frequency\\
$\tilde{\omega}$        &         dimensionless frequency for adatom diffusion\\
$\omega_{\rm COM}$         &        frequency of COM mode\\
$\omega_{\rm d}$         &        characteristic frequency set by $d$        \\
$\omega_k$                 &        frequency of $k$-th motional mode        \\
$\omega_{\rm R}$          &         characteristic frequency for set by $R_{\rm el}$        \\
$\omega_{\rm rec}$        &         recoil frequency of a trapped ion\\
$\omega_{\rm t}$          &         ion motional frequency\\
$\omega_{\rm ax}$        &         ion motional frequency in the axial direction\\
$\omega_{\rm r}$          &         ion motional frequency in the radial direction\\
$\omega_{x}$          &         ion motional frequency in the $x$ direction\\
$\omega_{y}$          &         ion motional frequency in the $y$ direction\\
$\omega_{z}$          &         ion motional frequency in the $z$ direction\\
\end{longtable}

\section{Acronyms}
\label{Appendix_Acronyms}
\begin{longtable}{ll}
\caption[List of Abbreviations]
{Abbreviations used in text}
\label{tab:abbreviations}
\endfirsthead
\endhead
\endfoot
\endlastfoot

AC  \hspace{1cm}& Alternating Current               \\
AES             & Auger Electron Spectroscopy       \\
COM             & Center of Mass                    \\
DAC             & Digital-to-Analog Converter       \\
DC              & Direct Current                    \\
EPR             & Equivalent Parallel Resistance    \\
ESR             & Equivalent Series Resistance      \\
EM              & Electromagnetic                   \\
EMF             & Electro-Motive Force              \\
EMI             & Electro-Magnetic Interference     \\
FEM             & Finite-Element Method             \\
MF              & Medium Frequency                  \\
PDOS            & Partial Density of States         \\
RF              & Radio Frequency                   \\
RMS             & Root Mean Squared                 \\
SOI             & Silicon On Insulator              \\
TLF             & Two-Level Fluctuator              \\
\end{longtable}

\end{document}